# QCD corrections to the Golden decay channel of the Higgs boson

**Mandeep Kaur**

*A thesis submitted for the partial fulfillment of the degree of Doctor of Philosophy*

Department of Physical Sciences

Indian Institute of Science Education & Research (IISER) Mohali,
Sector 81, SAS Nagar, Manauli PO 140306 Punjab, India

July 2024

**Dedicated to**

*my papa, mumma, beloved brother and Jerry*

# Abstract

Precision studies within the Higgs sector of the Standard Model are crucial to deepen our understanding of the fundamental particle interactions and uncovering new physics phenomena beyond the Standard Model, thereby guiding us towards a more comprehensive understanding of the universe. In this context, future colliders aim to provide highly precise experimental measurements of the properties and couplings of the Higgs boson. To fully leverage the potential of these precision machines, it is crucial to minimize theoretical errors in the Higgs sector observables, such as production and decay rates, to at least match the level of experimental uncertainties.

Reducing theoretical uncertainties involves including higher-order terms in perturbative calculations of the observables. Central to these calculations are Feynman integrals, which form the backbone of the theoretical framework.

In this thesis, we focus on providing precise predictions for the observables associated with one of the most important decay channels of the Higgs boson, the **"Golden decay channel"**. Our work improves theoretical predictions for the partial decay width of this channel, specifically the $H \to Z^{(*)}Z^{(*)} \to e^+e^-\mu^+\mu^-$ channel, by incorporating NNLO mixed QCD-electroweak corrections in the perturbation theory. The entire calculation of the decay amplitude at $\mathcal{O}(\alpha\alpha_s)$ is performed systematically following the conventional workflow of multi-loop calculations. This approach involves the application of the Feynman rules, the tensor decomposition of contributing amplitudes, and the utilization of integration-by-parts identities to express the amplitudes in terms of master integrals.

We describe the full analytical evaluation of these two-loop master integrals, specifically those appearing in the $\mathcal{O}(\alpha\alpha_s)$ corrections to the $HZZ$ vertex with both $Z$-bosons being off-shell, employing the *method of differential equations*. The calculation retains full dependence on the masses of all the particles, including those in the loop. Despite encountering non-rationalizable square roots, we transform the system of differential equations into the *canonical dlog-form* with a minimal set of independent one-forms. The analytic results for all the master integrals are obtained in terms of *Chen's iterated integrals* with logarithmic kernels order-by-order in the dimensional regularization parameter $\epsilon$, along with analytic boundary constants.

After the analytic evaluation of relevant two-loop master integrals, this thesis also presents a phenomenological study of the $H \to e^+e^-\mu^+\mu^-$ decay at $\mathcal{O}(\alpha\alpha_s)$. The entire calculation of the decay amplitude is implemented in the public code **Hto4l** to perform the phase space integration over final state leptons and to obtain improved predictions for the partial decay width with an accuracy of $\mathcal{O}(\alpha\alpha_s)$. We study the impact of these mixed QCD-electroweak corrections on the invariant mass distribution of the final-state leptons and angular variables, specifically focusing on the angle between the decay planes of the intermediate $Z$-bosons relative to leading-order predictions and NLO electroweak corrections. Our study shows that there are certain kinematic and angular bins in which the mixed QCD-electroweak corrections dominate the NLO EW corrections, highlighting the importance of these corrections in the context of data analysis aimed at probing new physics in the Higgs sector. Finally, the thesis concludes with a summary of the main results and outlines possible future directions.

# List of Publications/Preprints/Proceedings

## Publications based on thesis work

1. Ekta Chaubey, **Mandeep Kaur**, and Ambresh Shivaji, "Master integrals for $\mathcal{O}(\alpha\alpha_s)$ corrections to $H \to ZZ^*$", JHEP **10**, 056 (2022), arXiv:2205.06339 [hep-ph].

2. **Mandeep Kaur**, Maguni Mahakhud, Ambresh Shivaji, and Xiaoran Zhao, "QCD corrections to the Golden decay channel of the Higgs boson", JHEP **04**, 069 (2024), arXiv:2307.16063 [hep-ph].

## Other Publications/Proceedings

1. K. M. Black, S. Jindariani et al., "Muon Collider Forum report", JINST **19** T02015, arXiv:2209.01318 [hep-ex].

2. Taushif Ahmed, Ekta Chaubey, **Mandeep Kaur**, and Sara Maggio, "Two-loop non-planar four-point topology with massive internal loop", JHEP **05**, 064 (2024), arXiv:2402.07311 [hep-th].

3. **Mandeep Kaur**, Maguni Mahakhud, Ambresh Shivaji, and Xiaoran Zhao, "QCD corrections to $H \to e^+e^-\mu^+\mu^-$", Contribution to: 25th DAE-BRNS High Energy Physics Symposium, Springer Proc. Phys. **304** (2024) 866-868.

# List of Figures

# List of Tables

# Contents

# 1

# Introduction

Two fundamental observations common to our everyday experience are: things exist (i.e. there is matter) and things happen (i.e. interactions occur between particles). Particle physics aims to simplify our understanding of these observations to their most elementary level. This field of physics studies the *fundamental particles* – the primary constituents of matter – and their interactions, often achieved by colliding particles at high energies in particle colliders. The Standard Model (SM)[1] of particle physics, which is the current best theoretical framework, describes the outcomes of these collider experiments. The SM primarily consists of *Fermions*, categorized as *quarks* and *leptons*, which constitute matter; and the *force carriers* or *gauge bosons*, that mediate interactions between the matter particles.

The SM is formulated within the framework of *Quantum Field Theory (QFT)*, where the fundamental particles are interpreted as excitations of underlying *quantum fields*. These fields and their interactions are described by a mathematical object called the *Lagrangian*. The SM is a specific type of QFT known as a gauge theory, characterized by its invariance under certain symmetry transformations, collectively called the *symmetry group*. The symmetry group of the SM is $SU(3) \times SU(2)_L \times U(1)_Y$. The subscript $L$ indicates that the weak gauge group $SU(2)$ acts only on the left-handed fermions, and $Y$ is the hypercharge. Each of these gauge groups corresponds to a fundamental interaction: the strong interactions are characterized by $SU(3)$, the weak interactions by $SU(2)$, and the electromagnetic interactions by a $U(1)$ gauge symmetry. At high energies, the weak and electromagnetic interactions unify into the electroweak (EW) interaction, described by the EW gauge symmetry $SU(2)_L \times U(1)_Y$, with four gauge bosons as mediators – three of weak isospin from $SU(2)_L$ and one boson of weak hypercharge from $U(1)_Y$.

[1] Refer to [4] for a pedagogical introduction to the SM

The underlying symmetries of the SM Lagrangian initially forbid the inclusion of mass terms for the fermions and the gauge bosons. However, experimental observations demonstrated that both fermions and weak gauge bosons possess mass [5]. For many years, the origin of mass of these fundamental particles remained a puzzle for the particle physics community. In 1964, Peter Higgs, François Englert, and four other theorists proposed a groundbreaking solution to this problem in the form of the *Brout-Englert-Higgs-Kibble (BEHK)* mechanism, also known as the *Higgs mechanism* [6–10]. This mechanism introduces a background field – the Higgs field, which interacts with particles and gives them mass after the spontaneous breaking of $SU(2)_L \times U(1)_Y$ symmetry. The Higgs mechanism further inherently predicts the existence of the *Higgs boson*, a scalar particle corresponding to the Higgs field. I will briefly describe the Higgs mechanism in the following section; for further details, see [11–13].

## 1.1 The Higgs Mechanism

The Higgs mechanism extends the SM by introducing a color singlet, $SU(2)_L$ doublet complex scalar field,

$$\Phi = \frac{1}{\sqrt{2}} \begin{pmatrix} \phi^+ \\ \phi^0 \end{pmatrix} = \frac{1}{\sqrt{2}} \begin{pmatrix} \phi_1 + i\phi_2 \\ \phi_3 + i\phi_4 \end{pmatrix}, \tag{1.1}$$

possessing two complex or four real degrees of freedom and weak hypercharge $Y = +1$. Correspondingly, new terms introduced into the SM Lagrangian are:

$$\mathcal{L}_\Phi = (\mathcal{D}_\mu \Phi)^\dagger (\mathcal{D}^\mu \Phi) - V(\Phi) + \mathcal{L}_{\text{Yukawa}}, \tag{1.2}$$

where the first piece represents the kinetic term for $\Phi$, containing its interactions with gauge-bosons through the covariant derivative $\mathcal{D}^\mu$, defined as

$$\mathcal{D}_\mu \Phi = \left( \partial_\mu - i g_2 \frac{\sigma^a}{2} W_\mu^a - i \frac{g_1}{2} B_\mu \right) \Phi, \tag{1.3}$$

with three $SU(2)_L$ gauge fields $W_\mu^a$ $(a = 1, 2, 3)$, and one $U(1)_Y$ gauge field $B_\mu$. Here, $g_1$ and $g_2$ are the coupling constants of $U(1)_Y$ and $SU(2)_L$ gauge groups, respectively, and the $\sigma_i$ are the $2 \times 2$ Pauli matrices. The term $V(\Phi)$ in equation (1.2) represents the potential energy function of the field $\Phi$:

$$V(\Phi) = -\mu^2 \Phi^\dagger \Phi + \lambda (\Phi^\dagger \Phi)^2, \tag{1.4}$$

where $\mu^2$ and $\lambda$ are real parameters. The last piece of $\mathcal{L}_\Phi$ accounts for the interactions between the scalar field $\Phi$ and the fermions.

The Lagrangian in equation (1.2) is in its unbroken phase of the EW symmetry $SU(2)_L \times U(1)_Y$. For $V(\Phi)$ to have a finite minimum, $\lambda$ must be positive, but there is no a priori preference for choice of the sign of $\mu^2$. If $\mu^2 < 0$, the potential has a minimum at $\Phi_0^2 = |\langle 0| \Phi |0\rangle|^2 = 0$, and the EW symmetry remains unbroken in the vacuum. However, for $\mu^2 > 0$, the minimum of $V(\Phi)$ lies away from $\Phi_0^2 = 0$, and the field $\Phi$ develops a non-zero *vacuum expectation value* ($v$, VEV),

$$\Phi_0^2 = \frac{v^2}{2} = \frac{\mu^2}{2\lambda}. \tag{1.5}$$

In order to preserve the $U(1)_{em}$ symmetry of electromagnetism, the ground state is chosen as

$$\Phi_0 = \frac{1}{\sqrt{2}} \begin{pmatrix} 0 \\ v \end{pmatrix}. \tag{1.6}$$

The choice of this particular ground state breaks the $SU(2)_L \times U(1)_Y$ symmetry spontaneously while preserving the gauge symmetry $U(1)_{em}$ in the vacuum. The small field excitations from this ground state can be parametrized as

$$\Phi = \frac{1}{\sqrt{2}} \exp\left( i \frac{\xi_a \sigma^a}{v} \right) \begin{pmatrix} 0 \\ v + H \end{pmatrix}, \tag{1.7}$$

where $H$ is the *Higgs field* and $\xi_a$ $(a = 1, 2, 3)$ are the three unphysical massless Goldstone bosons arising from the Goldstone theorem [14]. These unphysical fields can be eliminated by employing the unitary gauge, i.e. $\zeta_a = 0$, yielding

$$\Phi = \frac{1}{\sqrt{2}} \begin{pmatrix} 0 \\ v + H \end{pmatrix}, \tag{1.8}$$

leaving the massive scalar Higgs field with the quantum excitation known as the *Higgs boson*, with mass $M_H = \sqrt{2\lambda v^2}$.

### 1.1.1 Gauge Boson Masses and Their Couplings to the Higgs Boson

In order to understand how the weak gauge bosons acquire masses through their interaction with the Higgs field after symmetry breaking. Let us examine the covariant derivative term in equation (1.2)

$$\begin{aligned} |\mathcal{D}(\Phi)|^2 &= \frac{1}{2}\left| \begin{pmatrix} \partial_\mu + \frac{i}{2}(g_1 B_\mu + g_2 W_\mu^3) & \frac{i}{2} g_2 (W_\mu^1 - i W_\mu^2) \\ \frac{i}{2} g_2 (W_\mu^1 + i W_\mu^2) & \partial_\mu + \frac{i}{2}(g_1 B_\mu - g_2 W_\mu^3) \end{pmatrix} \begin{pmatrix} 0 \\ v + H \end{pmatrix} \right|^2 \\ &= \frac{1}{2}(\partial_\mu H)(\partial^\mu H) + \frac{1}{8}(v+H)^2 g_2^2 \left( (W_\mu^1)^2 + (W_\mu^2)^2 \right) + \frac{1}{8}(v+H)^2 (g_1 B_\mu - g_2 W_\mu^3)^2, \end{aligned} \tag{1.9}$$

where the first term $\frac{1}{2}(\partial_\mu H)(\partial^\mu H)$ is the kinetic term for the Higgs field $H$. Defining the physical charged $W$ bosons – mediators of weak interactions as,

$$W_\mu^\pm = \frac{1}{\sqrt{2}}(W_\mu^1 \mp i W_\mu^2), \tag{1.10}$$

the second term proportional to $g_2^2$ in equation (1.9) becomes

$$\begin{aligned} \mathcal{L}_\Phi &\supset \frac{1}{4}(v+H)^2 g_2^2 W_\mu^+ W^{-\mu}, \\ &= \frac{g_2^2 v^2}{4} W_\mu^+ W^{-\mu} + \frac{g_2^2 v}{2} H W_\mu^+ W^{-\mu} + \frac{g_2^2}{4} H H W_\mu^+ W^{-\mu}. \end{aligned} \tag{1.11}$$

In the above equation, first term yields the $W$ boson mass, with

$$M_W^2 = \frac{g_2^2 v^2}{4}. \tag{1.12}$$

Thus, the $W$ boson acquires a mass through the Higgs vacuum expectation value $v$. The second and third terms in equation (1.11) are the interaction terms, yielding interactions of $W^+W^-$ with one or two Higgs bosons, shown in figure 1.1. The corresponding Feynman rules utilizing

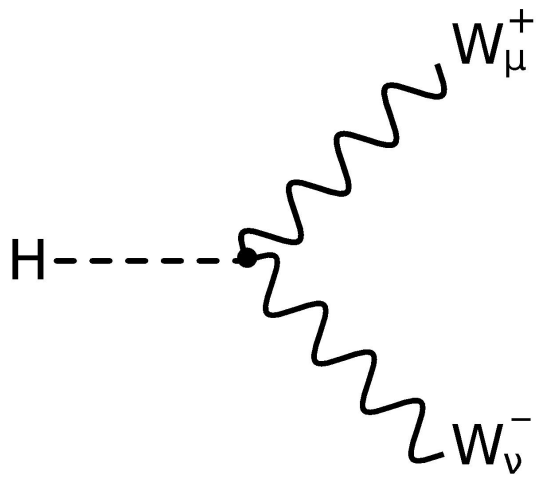

$$: \quad 2i\frac{M_W^2}{v}g_{\mu\nu}$$

(A)

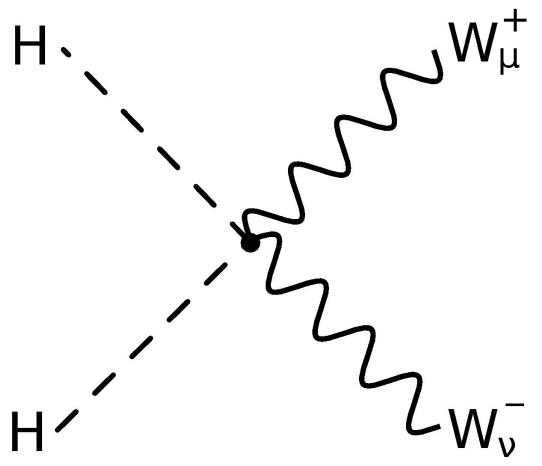

$$: \quad 2i\frac{M_W^2}{v^2}g_{\mu\nu}$$

(B)

FIGURE 1.1: Feynman rules for the $HWW$ and $HHWW$ vertices derived from the Lagrangian given in equation (1.11).

equation (1.12) are

$$\begin{aligned} HW_\mu^+W^{-\mu}: &\quad i\frac{g_2^2 v}{2}g_{\mu\nu} = 2i\frac{M_W^2}{v}g_{\mu\nu}, \\ HHW_\mu^+W^{-\mu}: &\quad i\frac{g_2^2}{4}\times 2!g_{\mu\nu} = 2i\frac{M_W^2}{v^2}g_{\mu\nu}, \end{aligned} \tag{1.13}$$

where in the second expression, 2! is a symmetry factor coming due to the presence of two identical Higgs bosons in the interaction term. Now returning back to the third term of equation (1.9). The gauge fields $W_\mu^3$ and $B_\mu$ can be transformed into the physical neutral weak field $Z_\mu$ and photon field $A_\mu$ through an orthogonal rotation as follows

$$\begin{pmatrix} Z_\mu \\ A_\mu \end{pmatrix} = \begin{pmatrix} \cos\theta_W & -\sin\theta_W \\ \sin\theta_W & \cos\theta_W \end{pmatrix} \begin{pmatrix} W_\mu^3 \\ B_\mu \end{pmatrix}, \tag{1.14}$$

where $\cos\theta_W$ and $\sin\theta_W$ are the sine and cosine of the Weinberg angle or weak mixing angle, $\theta_W$, defined as

$$\cos\theta_W = \frac{g_2}{\sqrt{g_1^2+g_2^2}}, \qquad \sin\theta_W = \frac{g_1}{\sqrt{g_1^2+g_2^2}}. \tag{1.15}$$

Using the equations (1.14) and (1.15), the third term in equation (1.9) becomes,

$$\begin{aligned} \mathcal{L}_\Phi \supset &\frac{1}{8}(v+H)^2(g_1B_\mu - g_2W_\mu^3)^2, \\ = &\frac{1}{8}(g_1^2+g_2^2)(v+H)^2 Z_\mu Z^\mu, \end{aligned}$$

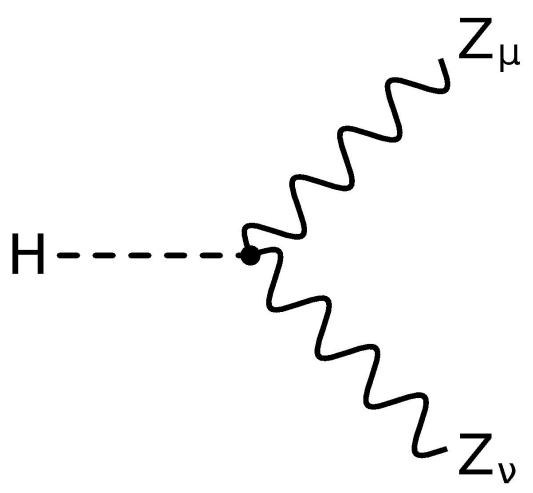

$$: \quad 2i\frac{M_Z^2}{v}g_{\mu\nu}$$

(A)

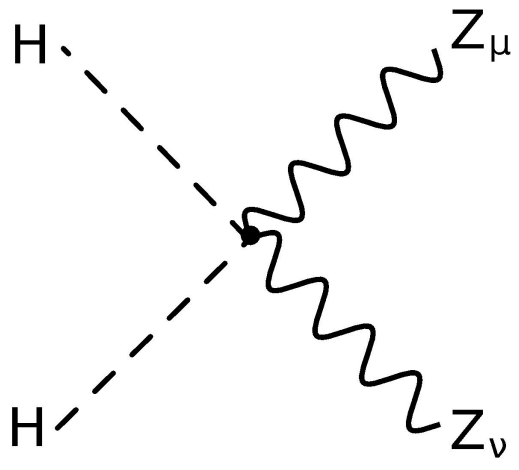

$$: \quad 2i\frac{M_Z^2}{v^2}g_{\mu\nu}$$

(B)

FIGURE 1.2: Feynman rules for the $HZZ$ and $HHZZ$ vertices derived from the Lagrangian given in equation (1.16).

$$= \frac{(g_1^2+g_2^2)v^2}{8}Z_\mu Z^\mu + \frac{(g_1^2+g_2^2)v}{4}HZ_\mu Z^\mu + \frac{(g_1^2+g_2^2)}{8}HHZ_\mu Z^\mu, \tag{1.16}$$

where the first term gives the mass of the $Z$ boson,

$$M_Z^2 = \frac{(g_1^2+g_2^2)v^2}{4}. \tag{1.17}$$

The subsequent second and third terms in equation (1.16) give interactions of two $Z$-bosons with the one or two Higgs bosons, shown in figure 1.2. The corresponding Feynman rules utilizing equation (1.17) are

$$\begin{aligned} HZ_\mu Z^\mu : &\quad i\frac{(g_1^2+g_2^2)v}{4} \times 2! g_{\mu\nu} = 2i\frac{M_Z^2}{v}g_{\mu\nu}, \\ HHZ_\mu Z^\mu : &\quad i\frac{(g_1^2+g_2^2)}{8} \times 2! \times 2! g_{\mu\nu} = 2i\frac{M_Z^2}{v^2}g_{\mu\nu}, \end{aligned} \tag{1.18}$$

where each interaction vertex contains a $2!$ factor for two identical $Z$ bosons, and in the second term, an extra factor of $2!$ is because of the two identical Higgs bosons in the interaction Lagrangian term.

Note that the photon field $A_\mu \equiv (\sin\theta_W W_\mu^3 + \cos\theta_W B_\mu)$ does not appear in equation (1.9), which means $A_\mu$ does not couple to the Higgs field and hence remains massless after the symmetry breaking, as it has to be.

### 1.1.2 Fermion Masses and Their Couplings to the Higgs Boson

As for the gauge bosons, the masses of the fermions can be generated through their interactions with the Higgs field described by the $\mathcal{L}_{\text{Yukawa}}$ term in the equation (1.2). The gauge-invariant renormalizable Yukawa Lagrangian ($\mathcal{L}_{\text{Yukawa}}$) is given as:

$$\mathcal{L}_{\text{Yukawa}} = \sum_{i,j=1}^{3} -y^{\ell}_{ij}\bar{L}_{i,L}\Phi\, e_{j,R} - y^{d}_{ij}\bar{Q}_{i,L}\Phi\, d_{j,R} - y^{u}_{ij}\bar{Q}_{i,L}\tilde{\Phi}\, u_{j,R} + h.c. \quad , \tag{1.19}$$

where indices $i, j$ represent the three generations of the SM fermions, the terms $L_{i,L} = (\nu_{i,L} \;\; e_{i,L})^T$, $Q_{i,L} = (u_{i,L} \;\; d_{i,L})^T$ denote left-handed lepton and quark doublets, respectively; and $e_{j,R}$, $d_{j,R}$, and $u_{j,R}$ represent individual right-handed leptons, down-type quarks, and up-type quarks, respectively. The $y_{i,j}$ are the $3 \times 3$ *complex Yukawa matrices* describing the interaction strength between the Higgs field and fermions, and $\tilde{\Phi} = i\sigma_2\Phi^*$ is the complex conjugate of the Higgs doublet ($\Phi$), with the second Pauli matrix $\sigma_2$. The term $h.c.$ stands for the Hermitian conjugate.

Now, after the EW symmetry breaking, setting the Higgs doublet to the unitary gauge as

$$\Phi = \frac{1}{\sqrt{2}}\begin{pmatrix} 0 \\ v+H \end{pmatrix}, \qquad \tilde{\Phi} = \frac{1}{\sqrt{2}}\begin{pmatrix} v+H \\ 0 \end{pmatrix}, \tag{1.20}$$

in equation (1.19), we get

$$\begin{aligned}\mathcal{L}_{\text{Yukawa}} = \sum_{i,j=1}^{3} \Bigg[ &- \frac{y^{\ell}_{ij}}{\sqrt{2}}(v+H)\bar{e}_{i,L}e_{j,R} - \frac{y^{d}_{ij}}{\sqrt{2}}(v+H)\bar{d}_{i,L}d_{j,R} \\ &- \frac{y^{u}_{ij}}{\sqrt{2}}(v+H)\bar{u}_{i,L}u_{j,R}\Bigg] + h.c. \end{aligned} \tag{1.21}$$

From the above equation, one can easily read off the mass matrices for the fermions of the form $\mathcal{M}^{f}_{i,j} = \frac{1}{\sqrt{2}} v y^{f}_{i,j}$. To obtain the mass eigenstates for fermions, we need to diagonalize these mass matrices with the help of unitary transformations (for details, see reference [11]). In terms of mass eigenstates denoted by a superscript $m$, the Yukawa Lagrangian given in equation (1.21) becomes

$$\begin{aligned}\mathcal{L}_{\text{Yukawa}} = \sum_{i=1}^{3} \Bigg[ &- \frac{y^{\ell}_{i}}{\sqrt{2}}(v+H)\bar{e}^{m}_{i,L}e^{m}_{i,R} - \frac{y^{d}_{i}}{\sqrt{2}}(v+H)\bar{d}^{m}_{i,L}d^{m}_{i,R} \\ &- \frac{y^{u}_{i}}{\sqrt{2}}(v+H)\bar{u}^{m}_{i,L}u^{m}_{i,R}\Bigg] + h.c. \end{aligned}$$

$$= \sum_{i=1}^{3} \Bigg[ -\frac{y_i^\ell}{\sqrt{2}} v\ \bar{e}_{i,L}^m e_{i,R}^m - \frac{y_i^d}{\sqrt{2}} v\ \bar{d}_{i,L}^m d_{i,R}^m - \frac{y_i^u}{\sqrt{2}} v\ \bar{u}_{i,L}^m u_{i,R}^m$$
$$- \frac{y_i^\ell}{\sqrt{2}} H \bar{e}_{i,L}^m e_{i,R}^m - \frac{y_i^d}{\sqrt{2}} H \bar{d}_{i,L}^m d_{i,R}^m - \frac{y_i^u}{\sqrt{2}} H \bar{u}_{i,L}^m u_{i,R}^m \Bigg] + h.c. \tag{1.22}$$

From the above equation, we can deduce the *Yukawa couplings* describing the interaction strength between the Higgs field and fermions denoted by $y_f$, which are related to the respective fermion masses $m_f$ as

$$y_f = \frac{\sqrt{2} m_f}{v}. \tag{1.23}$$

Additionally, in equation (1.22), the terms proportional to $H$ are the interactions terms, yielding the interactions of two fermions with one Higgs boson. The corresponding Feynman rule utilizing equation (1.23) (shown in figure 1.3) is

$$H f \bar{f} : \qquad -i \frac{y_f}{\sqrt{2}} = -i \frac{m_f}{v} \tag{1.24}$$

The above relation shows that the couplings of all the SM fermions (except neutrinos which are massless in the SM) with the Higgs boson are directly proportional to their masses.

Thus, the Higgs mechanism resolves the conundrum related to the origin of the masses for the weak gauge bosons and the fermions through the EW symmetry breaking in a consistent manner. It predicts the existence of the Higgs boson, a quantum particle associated with the Higgs field, whose discovery became a crucial test for both the mechanism and the SM.

After almost four decades since its postulation, in 2012, ATLAS [2] and CMS [3] experiments at the Large Hadron Collider (LHC) at CERN declared the discovery of a new scalar particle with a mass of $125.5 \pm 0.2(\text{stat})^{+0.5}_{-0.6}(\text{syst})$ GeV [15], exhibiting properties consistent with those anticipated for the SM Higgs boson [16, 17]. All the subsequent studies performed at particle colliders to date have consistently supported the association of this new scalar particle with the predicted SM Higgs boson [18–26]. The discovery of the Higgs boson was a landmark triumph

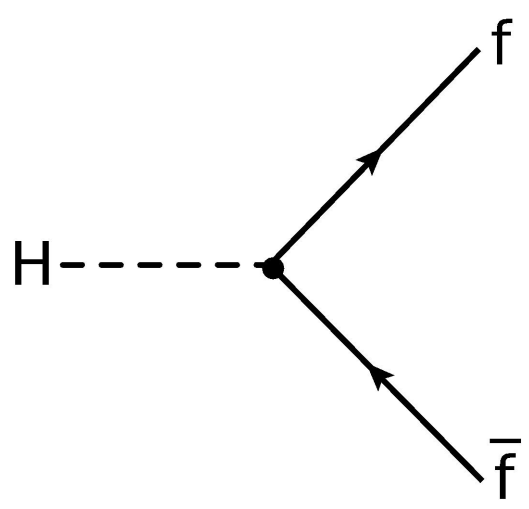

$$: \qquad -i \frac{y_f}{\sqrt{2}} = -i \frac{m_f}{v}$$

FIGURE 1.3: Feynman rule for the $H f \bar{f}$ vertex (see equation (1.24)).

in particle physics, leading to the Nobel Prize in Physics to François Englert and Peter Higgs in 2013. This discovery not only confirmed the Higgs mechanism but also completed the particle content of the SM and proved that the SM is a mathematically consistent renormalizable field theory to describe physics all the way up to the Planck scale.

Despite the remarkable precision with which SM predictions have been confirmed in countless experiments, it fails to provide a complete description of nature. It does not account for several phenomena such as dark matter, dark energy, observed matter-antimatter asymmetry in the universe, neutrino masses and their mixing, and gravity. Recent experimental measurements at Fermilab, such as the anomalous magnetic moment of the muon and the mass of the W boson [27, 28], appear to be inconsistent with the SM predictions. Additionally, the SM offers no explanation for the light mass of the Higgs boson, which is subject to large quantum corrections known as the *hierarchy problem*. All these and many other persistent shortcomings of the SM compel us to look for physics beyond the SM (BSM), and naturally raises the question of how to probe this new physics.

One approach to explore the BSM physics involves directly producing and detecting new particles at colliders. However, despite over a decade since the discovery of the Higgs boson, no new fundamental particles have been observed in colliders or other terrestrial experiments. This observation indicates the association of the potential new physics sector with an energy scale beyond the current energy capabilities of existing colliders. The absence of new particle discoveries has prompted increased interest in alternative methods and theoretical frameworks for exploring physics beyond the SM.

An alternative approach involves performing stringent tests of the SM predictions themselves. The existence of new particles can be inferred through their participation in virtual interactions with the SM particles. These interactions could leave subtle imprints on the SM observables, such as cross-sections and decay rates, near the electroweak scale, leading to small deviations from their predicted values. These deviations can be detected by comparing the precise theoretical calculations of these SM observables against the precise experimental measurements. Significant discrepancies between the two could serve as telltale signs of new physics beyond the SM. To ensure a fairer comparison, obtaining theoretical predictions for the SM observables with extremely high accuracy becomes crucial.

The backbone of theoretical calculations within the SM framework relies on perturbative methods due to our inability to solve the theory exactly. This approach involves expressing the observables as a series expansion in terms of small coupling constants of the underlying theory. Given the typical energy scales at which the Large Hadron Collider (LHC) operates, the SM couplings are sufficiently small for the perturbative calculations to yield reliable results. The first term in the expansion gives the leading order (LO) contribution, followed by the next-to-leading-order (NLO) term and so on. One main source of theoretical uncertainties in these calculations stems from the truncation of the perturbation series at finite orders. Therefore, to achieve the desired precision in theoretical predictions for SM observables, including higher-order terms in their calculations is essential.

The indirect approach to probe new physics signals raises an important question: Within the SM framework, where should we focus our efforts to maximize the chances of success? Different SM processes may exhibit varying sensitivities to BSM physics, so developing a strategy requires considering both theoretical and experimental aspects. Among the SM particles, the Higgs sector stands out due to its unique scalar nature and central role in the electroweak symmetry breaking. Additionally, the Higgs sector is the least constrained part of the SM to date. Furthermore, its properties, such as coupling strength being proportional to particle masses, make it a promising candidate for probing the landscape of heavy new physics. Quantum interactions with heavy new particles could significantly modify its production and decay rates compared to the SM predictions. Therefore, precision studies within the Higgs sector represent an important milestone in the journey beyond the SM. To achieve this, both present and future colliders, such as the Large Hadron Collider (LHC) [29], High-Luminosity LHC (HL-LHC) [30], Future Circular Collider (FCC-ee) [31], Circular Electron Positron Collider (CEPC) [32, 33], and the International Linear Collider (ILC) [34] aim to measure various Higgs properties with higher statistics. However, to fully exploit the experimental precision, equally precise theoretical predictions for the Higgs production and decay rates are essential.

Next, I will briefly provide an overview of the main production and decay mechanisms of the SM Higgs boson at the LHC, along with the current state-of-the-art theoretical predictions for various Higgs boson production cross-sections and decay rates.

## 1.2 Higgs Boson Physics At the LHC

The SM Higgs boson with a mass of 125 GeV is produced through a number of mechanisms by smashing proton beams together at the LHC. The LHC is the world's largest and most powerful particle accelerator, which facilitates proton-proton collisions with a center-of-mass energy $\sqrt{s} = 13.6$ TeV. ATLAS and CMS are two general-purpose detectors at the LHC, where the Higgs physics is primarily explored.

### 1.2.1 Main Production Mechanisms

The main single Higgs production channels at the LHC, in order of their significance, are: gluon fusion $(ggF)$, vector-boson fusion $(VBF)$, associated production with an electroweak gauge boson $(VH)$, and associated production with a top quark-antiquark pair $(t\bar{t}H)$ or with a single top quark $(tqH)$. The representative lowest-order Feynman diagrams contributing to these production modes are shown in figure 1.4.

Among all these production mechanisms, gluon fusion, $gg \to H$, is the dominant production channel for the SM Higgs boson at the LHC, with the largest cross-section. In this process, the Higgs is produced via the interaction of gluons through the exchange of a heavy top-quark loop [35], as shown in figure 1.4 (A). The contributions from light quark loops are suppressed due to small Yukawa couplings. The cross-section for ggF channel has been evaluated at next-to-next-to-next-to leading order (N3LO) QCD and next-to-leading order (NLO) EW accuracy [1, 36]. The total production cross-section for ggF process in $pp$ collisions at the LHC for $\sqrt{s} = 13$ TeV is approximately 49 pb [1, 37]. The major contribution to the radiative corrections to the cross-section comes from the QCD corrections at NLO and next-to-next-to-leading order (NNLO).

The vector-boson fusion (VBF) mechanism, depicted by the process $qq \to qqH$, provides the second most significant contribution to the Higgs production at the LHC. At the tree-level, it is a quark-initiated process that proceeds through the u or t-channel exchange of virtual $W$ or $Z$ bosons, followed by the emission of the Higgs boson from the weak gauge-boson propagator, as shown in figure 1.4 (B). VBF accounts for approximately $7\%$ of the total Higgs bosons produced at the LHC with a production cross-section of 3.8 pb [1, 37]. The inclusive cross-section for this process is now known with QCD corrections at N3LO [38]. Additionally, the EW corrections to the VBF process are calculated at NLO QCD+EW order [39]. Compared to other production mechanisms, this channel offers a relatively clean environment for Higgs boson searches at the LHC and for studying the Higgs sector couplings, particularly after specific selection cuts are applied.

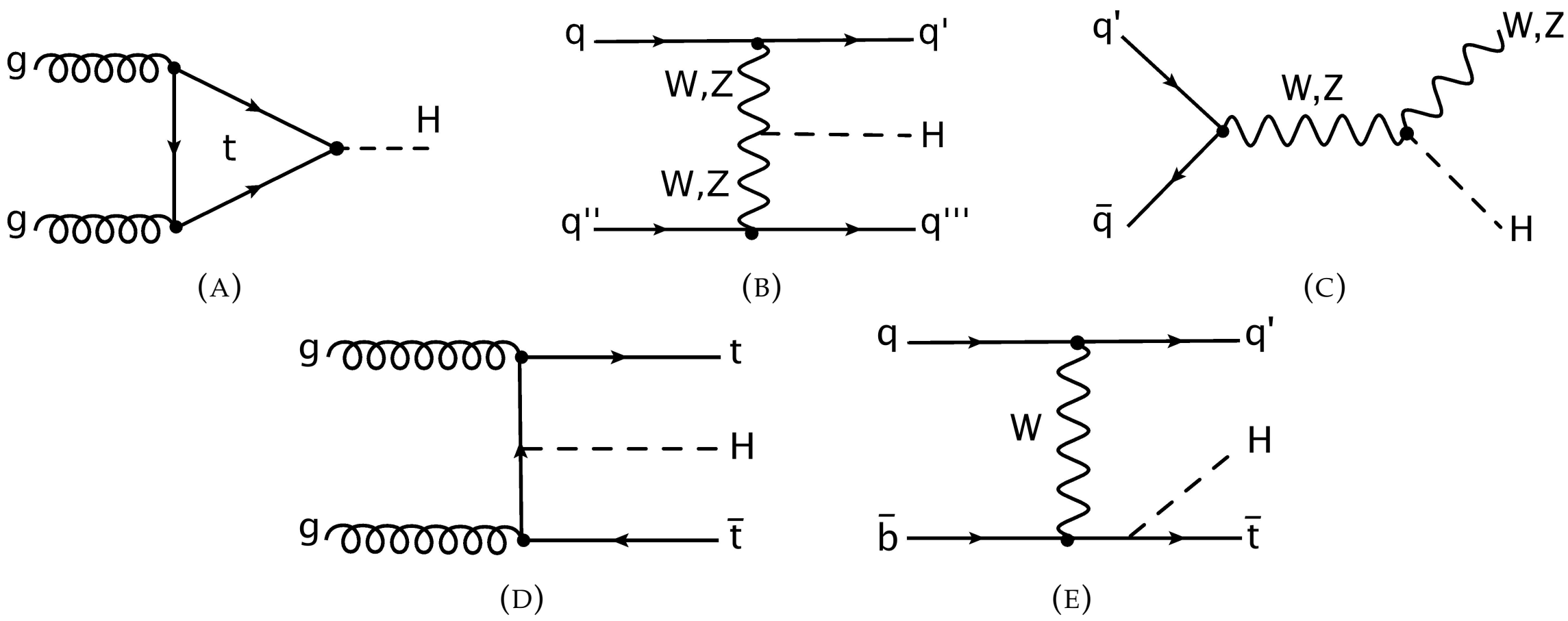

FIGURE 1.4: Representative leading order Feynman diagrams for single Higgs production in (A) gluon-gluon fusion, (B) vector-boson fusion, (C) association with an electroweak boson, (D) associated production with a top quark-antiquark pair, (E) production in association with a single top-quark.

The next most significant Higgs production channel at the LHC is its associated production with an electroweak gauge boson ($W$ or $Z$) known as the Higgs-strahlung process, shown in figure 1.4 (C). This production mechanism contributes approximately $4\%$ of the total Higgs bosons produced by the LHC. The inclusive cross-section for this production mechanism has recently been evaluated at N3LO in QCD [40].

Higgs production in association with a $t\bar{t}$ pair (shown in figure 1.4 (D)) or a single top quark (shown in figure 1.4 (E)) are less probable but still relevant production channels for the Higgs at the LHC, as they allow for the direct measurement of the Higgs coupling with the top-quark. The cross-section for the $t\bar{t}H$ process is calculated with NNLO (QCD) and NLO (EW) accuracy [1, 37, 41], while the cross-section for the $tqH$ process has been computed at NLO QCD in a five-flavour scheme, considering the bottom quark as a massless parton in the initial hadrons [37, 42].

The inclusive production cross-sections for the SM Higgs boson with $M_H = 125$ GeV as a function of center-of-mass energy ($\sqrt{s}$) in $pp$ collisions are shown in figure 1.5 (A) [37, 43]. A detailed discussion of the theoretical calculations for various Higgs production mechanisms and the impact of higher-order loop calculations on their predictions can be found in [1, 37] and references therein.

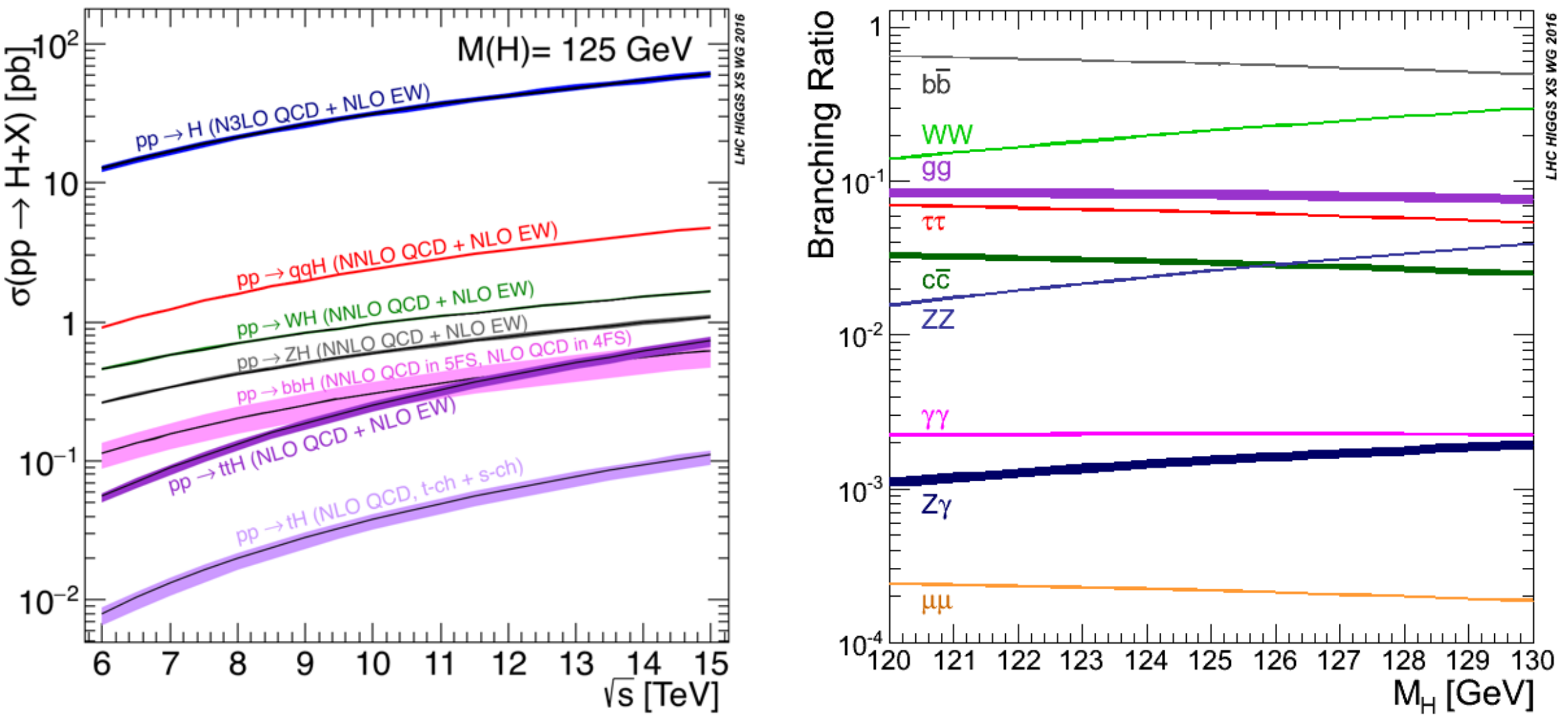

(A) (B)

FIGURE 1.5: (Left) Standard Model Higgs production cross-sections as a function of LHC center-of-mass energies, $\sqrt{s}$ (taken from reference [1]). (Right) The branching ratios for the main decay modes of the SM Higgs boson (taken from reference [1]). The theoretical uncertainties are indicated as bands.

### 1.2.2 Main Decay Mechanisms

The SM Higgs boson is an unstable particle with a lifetime of about $10^{-22}$ seconds and decays promptly after its production at the LHC. Precise calculations of all its relevant decay widths are crucial for fully understanding and analyzing the Higgs data at the LHC. The interaction of the Higgs boson with all massive SM particles results in a wide variety of decay modes, providing a direct probe of various Higgs sector couplings. In addition to decaying into massive particles, the Higgs boson can also decay into massless particles, but only via loops involving massive particles. The significant decay mechanisms of the SM Higgs boson can be categorized as follows:

- Decay into a fermion-antifermion pair ($f\bar{f}$)
- Decay into a pair of weak gauge-bosons ($W$ or $Z$)
- Loop-induced decays

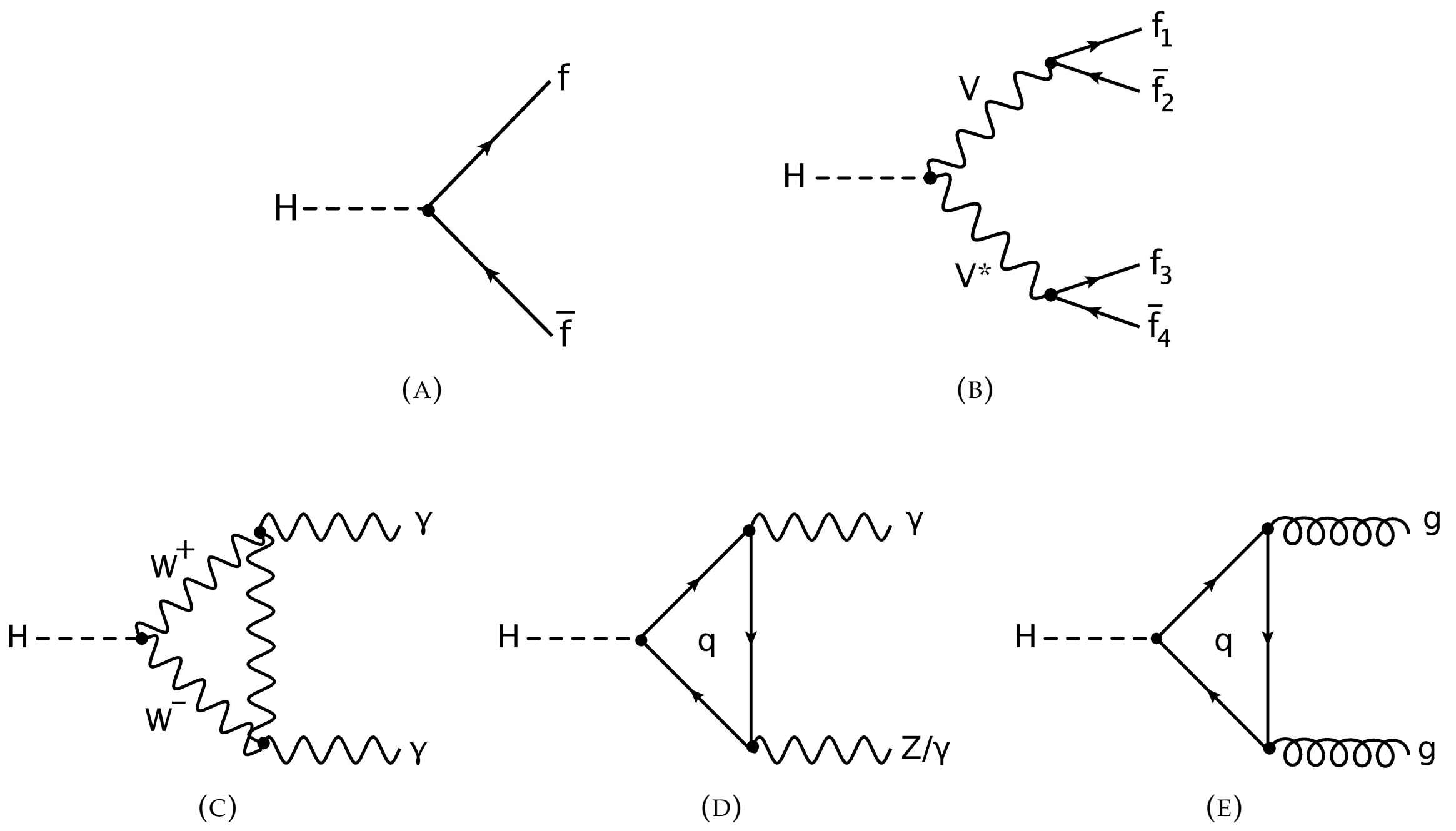

FIGURE 1.6: Representative leading-order Feynman diagrams for the Standard Model Higgs boson decaying into (A) a fermion-antifermion pair, (B) a pair of weak vector bosons ($W$ or $Z$), with each vector boson further decaying into a pair of fermions, (C) a pair of photons through a massive $W$-boson loop, (D) $\gamma\gamma$ or $Z\gamma$ through a heavy quark loop, and (E) a pair of gluons.

#### 1.2.2.1 $H \to f\bar{f}$

At the tree-level, the Higgs boson decays into a fermion-antifermion pair, as shown in figure 1.6 (A). Since the strength of the Higgs-fermion interaction depends on the fermion's mass (refer to equation (1.23)), therefore, the Higgs boson preferentially decays into the heaviest kinematically allowed fermionic final states, with the fermion mass close to $M_H/2$. With a mass of 125 GeV, the dominant fermionic decay modes are into a bottom quark-antiquark pair ($b\bar{b}$), tau lepton-antilepton pair ($\tau^+\tau^-$), and charm quark-antiquark pair ($c\bar{c}$). Among these, the dominant decay mode is $H \to b\bar{b}$ with the largest branching ratio (BR= $58\%$) as shown in figure 1.5 (B).

The partial decay widths of $H \to b\bar{b}$ and $H \to c\bar{c}$ have been computed, including the QCD corrections up to next-to-next-to-next-to-leading order (N4LO) [37]. The corresponding EW corrections to $H \to f\bar{f}$ ($f\bar{f} = b\bar{b},\ c\bar{c},\ \tau^+\tau^-$) are known up to NLO [37]. The evaluation of branching ratios and partial decay widths of various fermionic decay modes of the SM Higgs boson, including higher-order radiative corrections, is implemented in the publicly available **HDECAY** code [44].

#### 1.2.2.2 $H \to WW^*$ and $H \to ZZ^*$

The tree-level decays of the SM Higgs boson, $H \to WW^*$ and $H \to ZZ^*$, are crucial as they directly probe the $HVV$ $(V = W, Z)$ couplings, offering valuable insights into the electroweak symmetry breaking mechanism. Since the SM Higgs mass is less than twice the W or Z boson masses, at least one of the final state bosons must be off-shell (denoted by a superscript $*$) during on-shell Higgs decay. In the doubly off-shell decay, $H \to V^*V^*$, each gauge boson further decays into leptons or quark-antiquark pairs, leading to fully leptonic (4 leptons), semi-leptonic (2 leptons, 2 quarks), or hadronic (4 quarks) final states (depicted in figure 1.6 (B)). Among these final states, fully leptonic final states are promising for precision studies in the Higgs sector due to their easy identification, good measurement capabilities, and large signal-over-background ratio. The $H \to WW \to (\ell^+\nu_\ell)(\ell^-\bar{\nu}_\ell)$ $(\ell = e, \mu)$ channel, with a branching fraction of $1.1\%$, is important, but the presence of undetected neutrinos hinders precise W boson reconstruction. In contrast, the decay channel $H \to ZZ \to 4\ell$ $(\ell = e, \mu)$, with four well-identified charged leptons in the final state, provides a distinctive and clean signature among all possible Higgs decays. This decay mode, dubbed as the **"Golden decay channel"**, has a branching fraction of $\sim 1.3 \times 10^{-4}$ for a Higgs mass of 125 GeV. Despite its rarity, it was one of the key discovery channels of the Higgs boson in 2012 [2, 3], as it provides a large peak around the Higgs mass in the invariant mass distribution of final state leptons (as shown in figure 1.7 (A)) and is highly utilized for precision measurements in the Higgs sector.

Theoretical predictions for Higgs decay into any possible four-fermion final state (i.e. leptonic, semi-leptonic, and hadronic), including full NLO electroweak and NLO QCD corrections are implemented in **`PROPHECY4F`** [45] and predictions with NLO EW accuracy matched with QED Parton Shower for the Higgs decay to four charged leptons are implemented in a public code **`Hto4l`** [46].

#### 1.2.2.3 Loop-induced Decays

The loop-induced decays of the SM Higgs boson involve $H \to gg$, $H \to \gamma\gamma$, and $H \to Z\gamma$. These decays occur via loops involving massive $W$-bosons and top (bottom) quarks and are suppressed compared to direct decays into fermions and gauge-bosons due to the loop effect. The representative LO Feynman diagrams for these decay mechanisms are shown in figure 1.6 (C-E). Among these decay modes, the decay $H \to \gamma\gamma$ is rare, with a branching ratio of $\mathcal{O}(10^{-3})$.

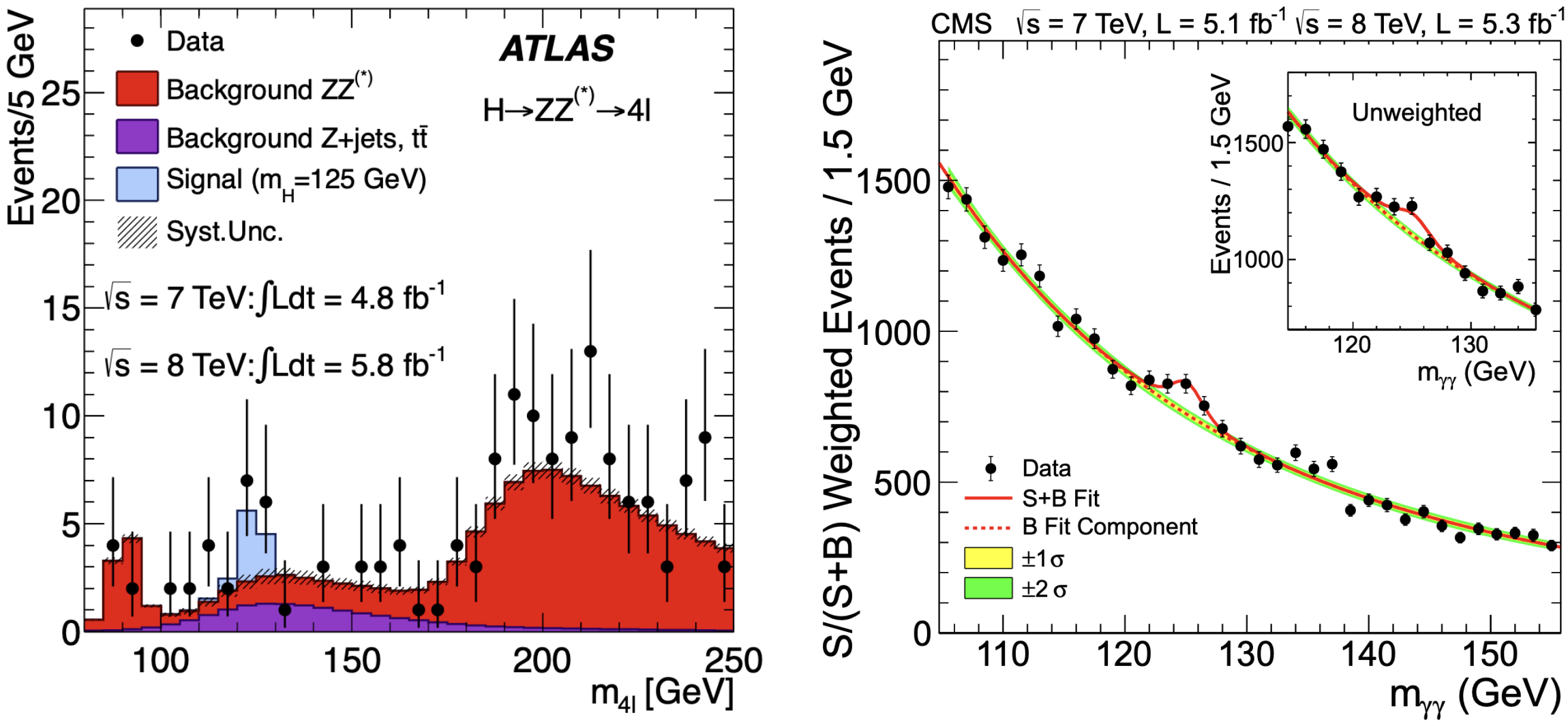

(A) (B)

FIGURE 1.7: (A) The distribution of the four-lepton invariant mass, $m_{4\ell}$, for the selected candidates, compared to the background expectation in the 80–250 GeV mass range, for the combination of the $\sqrt{s} = 7$ TeV and $\sqrt{s} = 8$ TeV data. The signal expectation for the SM Higgs with $M_H = 125$ GeV is also shown (image taken from reference [2]). (B) The diphoton invariant mass distribution with each event weighted by the S/(S+B) value of its category. The lines represent the fitted background and signal, and the coloured bands represent the $\pm 1$ and $\pm 2$ standard deviation uncertainties in the background estimate. The inset shows the central part of the unweighted invariant mass distribution (image taken from reference [3]).

However, due to its clean final-state topology, it possesses high sensitivity for the Higgs detection in the low $M_H$ region. This channel, along with the golden decay channel, played a crucial role in the Higgs discovery in 2012 [2, 3] by providing a narrow peak on top of a small background around 125 GeV in the invariant mass spectrum of diphoton as shown in figure 1.7 (B). The radiative corrections to the $H \to gg$ decay are known at N3LO QCD [47–49] and NLO EW [37] accuracy. For the $H \to \gamma\gamma$ decay, NLO QCD and EW corrections are available in references [37, 47], NNLO QCD corrections have also been computed in [50] and are implemented in the code **HDECAY** [44]. The partial decay width for $H \to Z\gamma$ channel is implemented in **HDECAY** at LO only, and NLO QCD corrections to this channel are available in [51].

The SM Higgs boson with a mass of 125 GeV has a total decay width, $\Gamma_H = 4.07 \times 10^{-3}$ GeV, with a relative uncertainty of $^{+4.0\%}_{-3.9\%}$. The branching ratios for the main decay modes of 125 GeV Higgs boson, including theoretical uncertainties (due to missing higher-order corrections and errors in SM input parameters) are shown in figure 1.5 (B) and are summarized in Table 1.1 taken from [37]. Further details on the theoretical calculations of all relevant branching ratios and corresponding uncertainties in their predictions can be found in the review articles [1, 52–55]

| Decay channel | Branching ratio | Rel. uncertainty |
|---|---|---|
| $H \to b\bar{b}$ | $5.82 \times 10^{-1}$ | ${}^{+1.2\%}_{-1.3\%}$ |
| $H \to \tau^+\tau^-$ | $6.27 \times 10^{-2}$ | $\pm 1.6\%$ |
| $H \to c\bar{c}$ | $2.89 \times 10^{-2}$ | ${}^{+5.5\%}_{-2.0\%}$ |
| $H \to W^+W^-$ | $2.14 \times 10^{-1}$ | $\pm 1.5\%$ |
| $H \to ZZ^*$ | $2.62 \times 10^{-2}$ | $\pm 1.5\%$ |
| $H \to \gamma\gamma$ | $2.27 \times 10^{-3}$ | $\pm 2.1\%$ |

TABLE 1.1: SM Higgs boson branching ratios for $M_H = 125$ GeV and related uncertainties.

and references therein.

## 1.3 What This Thesis Is About

The precision studies of Higgs production cross-sections and decay rates offer a promising avenue for indirect searches of the BSM physics. Given that the Higgs boson can only be observed at the LHC through its decay products, detailed investigations of its various decay channels are crucial. Among the SM Higgs boson decays discussed earlier, the golden channel, $H \to ZZ^* \to 4\ell$ ($\ell = e, \mu$), is particularly intriguing due to its exceptionally clean experimental signature and central role in the Higgs discovery. This channel with well-identified charged leptonic final states allows for the precise reconstruction of the Higgs bosons produced at the LHC. The differential distributions of the final state leptons in this decay mode serve as a powerful tool for precise Higgs mass measurements [56], the study of the spin and CP properties of the Higgs boson [19, 20, 57, 58], and precision tests of the SM [59]. Moreover, as a direct probe of the $HZZ$ coupling, this channel deepens our understanding of the mass generation mechanism. The data collected in the off-shell production and decay of Higgs to four leptons via Z-boson pairs can constrain its total decay width [60–62]. These unique features of the $H \to ZZ^* \to 4\ell$ channel motivate our work to provide precise predictions for observables associated with this channel to have a better understanding of the Higgs properties.

The context of this thesis mainly lies in providing precise theoretical predictions for the partial decay width of the **Golden decay channel** of the Higgs boson by incorporating higher-order terms in perturbation theory. The thesis work focuses on computing QCD corrections to the golden decay channel, specifically targeting $H \to Z^*Z^* \to e^+e^-\mu^+\mu^-$ channel, on top of pure electroweak corrections. A crucial challenge in this computation lies in evaluating non-trivial two-loop Feynman integrals arising from the QCD corrections to the $HZZ$ vertex. We employ the *method of differential equations* [63–65] to analytically solve these integrals, keeping the full dependence on all particle masses, including those running in the loop. This thesis presents the first-ever full analytical results for these two-loop master integrals in terms of *Chen's iterated integrals* [66]. Finally, to provide improved predictions for the partial decay width of $H \to e^+e^-\mu^+\mu^-$, we implement our whole calculation within **Hto4l** [46], a public code that simulates $H \to 4\ell$ $(\ell = e, \mu)$ decay events. This code allows us to quantify the QCD corrections, simulate decay events, and provide improved numerical predictions for the partial decay width and kinematic distributions for $H \to e^+e^-\mu^+\mu^-$ with an accuracy of $\mathcal{O}(\alpha\alpha_s)$ [67].

## 1.4 Outline of This Thesis

The thesis is organized as follows:

**Chapter 2:** We begin with a comprehensive overview of *scalar Feynman integrals*, as they are an indispensable part of loop calculations. In this chapter, we introduce standard notations and terminology used in Feynman integral evaluation. We then discuss the origin of ultraviolet (UV) and infrared (IR) divergences within Feynman integrals and outline methods to regularize these divergences. We focus specifically on *dimensional regularization* [68, 69], the technique employed throughout the thesis to handle divergences. Finally, we explore the different types of relations between Feynman integrals – symmetry relations (SR), Lorentz invariance (LI), and integration-by-parts (IBP) identities [70, 71] – and explain how to use these relations to reduce a set of scalar Feynman integrals belonging to a family to a minimal set of independent integrals called *master integrals*.

**Chapter 3:** After obtaining a set of master integrals, the next step is their evaluation. This chapter explores the tools and methods used for their numerical and analytical evaluation, highlighting the significant advantages of analytical results over purely numerical approaches. We then delve into the powerful *method of differential equations* [63–65] for analytical evaluation of Feynman integrals. The chapter provides a detailed explanation of how to derive the system of differential equations for a given integral family and further discusses the concept of a *good basis choice*, which transforms the differential equation system into a simplified *canonical form* [72]. We discuss guidelines for selecting an appropriate integral basis and outline how to solve the differential equation system using suitable boundary conditions. Furthermore, we explore special functions and their properties that frequently appear in the solutions of integral differential equation systems. Finally, we address the crucial task of determining boundary conditions once a general solution for the differential equation system, including integration constants, is obtained.

**Chapter 4:** Following the discussion of introductory chapters, in this chapter, we describe the key components for calculating mixed QCD-electroweak corrections to the partial decay width of $H \to ZZ^* \to e^+e^-\mu^+\mu^-$ channel. Beyond the LO, the decay of Higgs into $e^+e^-\mu^+\mu^-$, in addition to $Z^{(*)}Z^{(*)}$ channel, receives contributions from additional $Z^{(*)}\gamma^{(*)}$ and $\gamma^{(*)}\gamma^{(*)}$ channels.

To systematically compute the two-loop amplitudes, we classify the contributing diagrams at $\mathcal{O}(\alpha\alpha_s)$ into three categories: diagrams with correction to $HV_1V_2$ $(V_1, V_2 = Z, \gamma)$ vertex, diagrams involving $\mathcal{O}(\alpha\alpha_s)$ self-energy corrections to the $Z$-boson propagators, and tree-level diagrams with $\mathcal{O}(\alpha\alpha_s)$ counter-term insertion on the $Z$-lepton-antilepton vertex. The computation of two-loop amplitudes, particularly those involving the $HV_1V_2$ vertex correction, presents a significant challenge. To address this, we employ the *projector technique*, which allows us to organize the amplitudes in terms of scalar functions called *form factors*, obtained using our in-house codes and routines. Finally, publicly available codes based on *integration-by-parts (IBP)* and *Lorentz invariance (LI) identities* are employed to express these two-loop form factors as linear combinations of an independent set of 41 *master integrals*.

**Chapter 5:** In this chapter, we present the analytic computation of 41 master integrals appearing in $\mathcal{O}(\alpha\alpha_s)$ corrections to the $HV_1V_2$ vertex. The calculation is performed using the *method of differential equations* while retaining the full dependence on the masses of all the particles involved, including the one running in the loop. With suitable basis choice for integrals, we transform the system of differential equations into the *canonical form* in $d = 4 - 2\epsilon$ dimensions. The presence of multiple kinematic scales poses significant challenges in the computation and introduces several square roots in the differential equation system after performing a basis transformation. Despite the simultaneous non-rationalizability of all the occurring square roots, we construct an ansatz to rewrite the dependence on kinematic variables in differential equation system entirely in terms of differentials of logarithms, the so-called *dlog-forms*. The final analytic results for all the master integrals are expressed in terms of *Chen's iterated integrals* with algebraic kernels order-by-order in dimensional regularization parameter $\epsilon$, along with boundary constants determined using the **PSLQ** [73] algorithm. Furthermore, we perform numerical evaluations of the analytical results using in-house **`Mathematica`** code. The obtained results are verified by comparing them with numerical values from publicly available tools.

**Chapter 6:** The two-loop amplitudes for $HV_1V_2$ vertex corrections and self-energy corrections discussed in chapter 4 can potentially develop divergences, classified as ultraviolet and infrared divergences, arising from loop integrations. To handle these infinities, we employ *dimensional regularization*, where we work in $d = 4 - 2\epsilon$ space-time dimensions. Fortunately, in our case, the absence of real emission diagrams at $\mathcal{O}(\alpha\alpha_s)$ leads to an IR finite two-loop amplitude, as the virtual infrared contributions cancel out among the contributing diagrams. To eliminate the remaining UV divergences from the virtual two-loop amplitudes, we utilize the *counterterm approach*, evaluating all relevant counterterm diagrams. The counterterm constants are then fixed using the standard *on-shell renormalization scheme* [74, 75]. Furthermore, we discuss the effects of introducing finite widths for unstable internal particles, which can lead to several issues like violation of gauge invariance [76, 77] and gauge dependence in fixed-order calculations. To ensure gauge independence of our results, we employ the *complex-mass scheme* [78–80] throughout this work. Finally, we discuss the input parameter scheme adopted for the numerical calculations within the context of this thesis.

**Chapter 7:** In this chapter, we detail the implementation of our UV finite matrix elements for $H \to e^+e^-\mu^+\mu^-$ contributing at $\mathcal{O}(\alpha\alpha_s)$ into the public Monte-Carlo code **`Hto4l`** [46] to perform the phase-space integration over final state leptons. We also detail the various checks performed to ensure the accuracy of this implementation. Furthermore, we present improved predictions for the partial decay width of the golden decay channel of the Higgs boson, achieved with an accuracy of $\mathcal{O}(\alpha\alpha_s)$. Additionally, we present and discuss in detail the impact of these $\mathcal{O}(\alpha\alpha_s)$ corrections on the invariant mass distribution of the final state leptons and the angular distribution $\phi$, where $\phi$ is the angle between the decay planes of the $Z$-bosons.

**Chapter 8:** This chapter provides a brief summary of the results obtained in the thesis and discusses the outlook for future work based on these results.

All the Feynman diagrams presented in this thesis are drawn with the help of **`Jaxodraw`** [81] based on **`Axodraw`** [82].

# 2

# Feynman Integrals to Master Integrals

FEYNMAN integrals, representing *ill-defined* integrations over loop momenta, are the cornerstone of precision calculations within pQFT, contributing to scattering amplitudes at one-loop and beyond. Depending on the nature of interacting fields, these integrals, in general, may involve a tensorial structure i.e., the numerator of the integrand can have a tensor structure carrying Lorentz indices and/or indices associated with the gauge groups. However, it is well known that these tensor integrals can always be expressed by means of *tensor reduction* in terms of integrals known as *scalar Feynman integrals*, which are free from any such indices. The method for this tensor reduction was pioneered by Passarino and Veltman for the one-loop case [83], with Tarasov later extending it to the more general case [84, 85]. Therefore, in this thesis, we restrict our study of Feynman integrals to the scalar ones only.

In this chapter, we discuss basic notations, definitions and properties of the scalar Feynman integrals. After introducing the concept of the *family* of Feynman integrals, we explore the various relations that connect different integrals belonging to a given integral family. We also describe their reduction into a linearly independent set of integrals called *master integrals*.

## 2.1 Scalar Feynman Integrals and Their Properties

Consider a Feynman diagram with $l$ number of loops, $n$ internal, and $e+1$ external lines. Using momentum conservation, let $P = \{p_1, \ldots, p_e\}$ indicates the set of independent external momenta. The general expression for a scalar Feynman integral associated with this Feynman diagram in space-time dimension $d$ can be given as

$$I_{\nu_1,\cdots,\nu_n}(d;\{p_i^2\};\{s_i\};\{m_j^2\};\mu^2) = (\mu^2)^{\nu-\frac{ld}{2}} \int \prod_{j=1}^{l} \frac{d^d k_j}{i\pi^{d/2}} \prod_{j=1}^{n} \frac{\mathcal{N}(K,P)}{D_j^{\nu_j}}, \tag{2.1}$$

where the Feynman integral is a function of the set of squared independent external momenta $\{p_i^2\}$, Mandelstam variables $\{s_i\}$, and the masses of internal particles $\{m_j^2\}$. The inverse propagators $D_j$ are given by

$$D_j = -q_j^2 + m_j^2. \tag{2.2}$$

The number of inverse propagators present in the integral depends on the structure of the Feynman graph under consideration. In equation (2.2), $q_j$ are a linear combination of $K = \{k_1, k_2, \ldots, k_l\}$ loop momenta and $P = \{p_1, \ldots, p_e\}$ external momenta, $m_j$ are the masses of particles running in the loops. The scale $\mu$ is introduced to make the integral dimensionless. The exponents $\nu_j$ are the powers of individual inverse propagators $D_j$ and are positive integers. The quantity $\nu$ is given as

$$\nu = \sum_{j=1}^{n} \nu_j \tag{2.3}$$

The numerator $\mathcal{N}$ is a polynomial containing scalar dot products of the form $k_i.k_j$ and $k_i.p_j$ between independent loop momenta and external momenta. The number of possible independent scalar products which can appear in $\mathcal{N}$ can be evaluated as

$$\text{no. of scalar products} = \frac{l(l+1)}{2} + e \cdot l, \tag{2.4}$$

where the first term counts all the possible scalar products between two loop momenta, while the second term describes the number of scalar products between a loop momenta and independent external momenta. Generally, one can write the scalar products in $\mathcal{N}$ as linear combinations of inverse propagators $D_j$. However, on going beyond the one-loop, we often encounter more independent scalar products than the number of internal propagators. Consequently, it is not always possible beyond the one-loop to express all scalar products in the numerator using the inverse propagators. These scalar products that cannot be expressed using the given set of inverse propagators are called *irreducible scalar products* (ISPs) [86]. One way to handle these ISPs is to enlarge our basis set of propagators by introducing some propagators artificially called *auxiliary propagators*, which appear only with negative exponents. This enlarged set allows us to express each independent scalar product in terms of inverse propagators. As a result, our scalar Feynman integral with unit numerator took the following form.

$$I_{\nu_1,\cdots,\nu_r}(d;\{p_i^2\};\{s_i\};\{m_j^2\};\mu^2) = (\mu^2)^{\nu-\frac{ld}{2}} \int \prod_{j=1}^{l} \frac{d^d k_j}{i\pi^{d/2}} \prod_{j=1}^{r} \frac{1}{D_j^{\nu_j}}, \tag{2.5}$$

where the exponents $\nu_j$ satisfy

$$\nu_j \begin{cases} \geq 0 & j = 1,\ldots,n \\ \leq 0 & j = n+1,\ldots,r \end{cases} \tag{2.6}$$

and the propagators with exponents $\nu_j$ for $j = n+1,\ldots,r$ are considered as auxiliary propagators. To understand this whole concept, let us consider the example of a massless two-loop double box diagram [87] shown in figure 2.1. For this diagram, we have

- Two loops with loop momenta $(k_1, k_2)$, i.e. $l = 2$ and four external lines after applying momentum conservation $(p_4 = -p_1 - p_2 - p_3)$ giving 3 independent external momenta $(p_1, p_2, p_3)$, therefore, $e = 3$. Thus, the number of possible independent scalar products using equation (2.4) is 9.

$$k_1^2, k_2^2, k_1 \cdot k_2, k_1 \cdot p_1, k_1 \cdot p_2, k_1 \cdot p_3, k_2 \cdot p_1, k_2 \cdot p_2, k_2 \cdot p_3 \tag{2.7}$$

- Seven internal lines, therefore, correspondingly 7 propagators,

$$\begin{aligned} &D_1 = -q_1^2 = -(k_1 - p_1)^2, && D_2 = -q_2^2 = -(k_1 - p_1 - p_2)^2, && D_3 = -q_3^2 = -k_1^2, \\ &D_4 = -q_4^2 = -\left(k_1 + k_2\right)^2, && D_5 = -q_5^2 = -\left(k_2 + p_1 + p_2\right)^2, && D_6 = -q_6^2 = -k_2^2, \\ &D_7 = -q_7^2 = -\left(k_2 + p_1 + p_2 + p_3\right)^2. \end{aligned} \tag{2.8}$$

Out of 9 independent scalar products, we can write 7 in terms of the inverse propagators given in equation (2.8). So, we will be left with two irreducible scalar products ($k_1 \cdot p_3$, $k_2 \cdot p_2$), which we cannot express in terms of these inverse propagators. Thus, we will enlarge the set of propagators by introducing two auxiliary propagators, $D_8 = -q_8^2 = -(q_3 + p_2 + p_4)^2$ and $D_9 = -q_9^2 = -(q_6 + p_1 + p_3)^2$. The corresponding auxiliary diagram for a massless double box with auxiliary propagators is shown in figure 2.2. Now, we will be able to write all the scalar products in terms of inverse propagators.

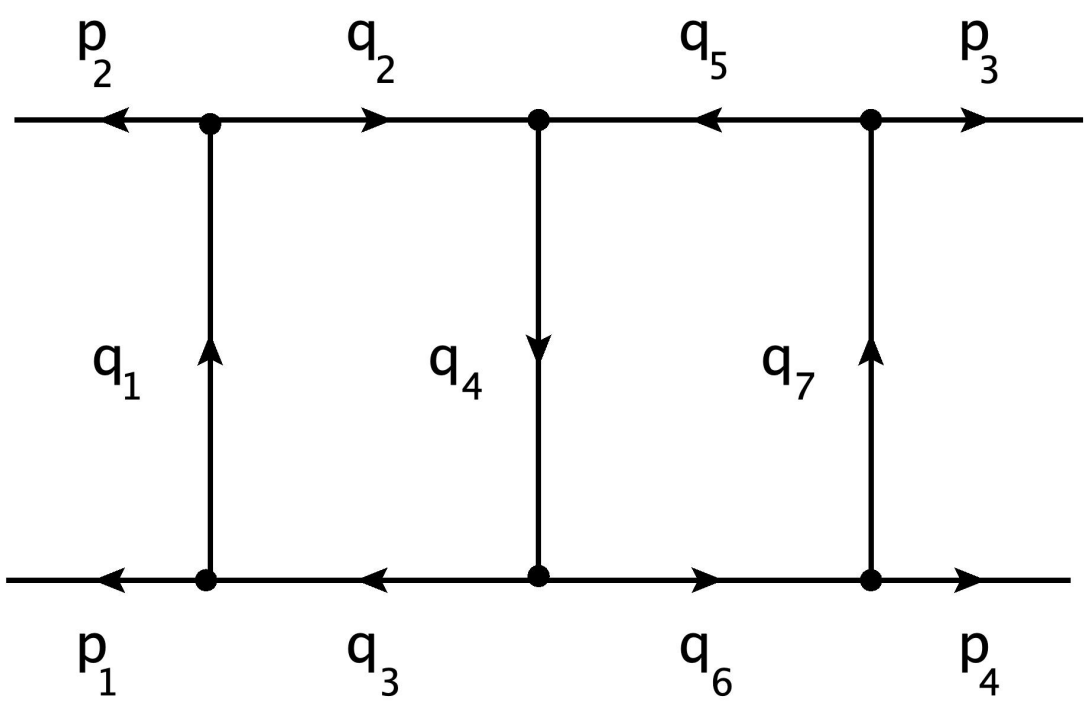

FIGURE 2.1: A massless two-loop double box diagram

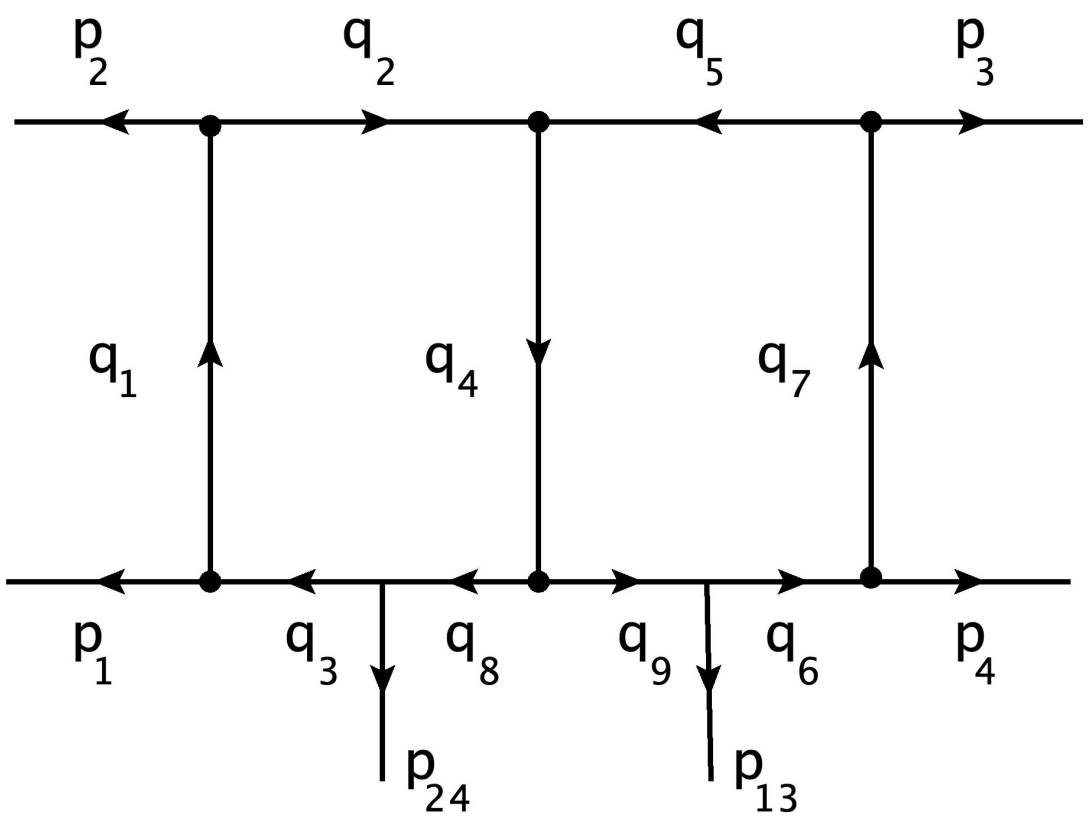

FIGURE 2.2: An auxiliary graph for two-loop double box diagram with nine propagators. We use the notation $p_{mn} = p_m + p_n$.

Now, we will introduce some "vocabulary" often used in the calculation of Feynman integrals.

**Definition. 1**

**Family:** Feynman integrals sharing an identical complete set of propagators $D_i$, which encompass all independent scalar products, yet may differ in the powers of the exponents $\nu_i$, are classified as members of the same integral family.

For example, the one-loop triangle integral $I_{1,1,1}$ and the bubble integral $I_{0,1,1}$, shown in figure 2.3, belong to the same family. The bubble integral is essentially obtained from the triangle diagram by removing one of its propagators, thereby changing the exponent of the corresponding propagator from 1 to 0.

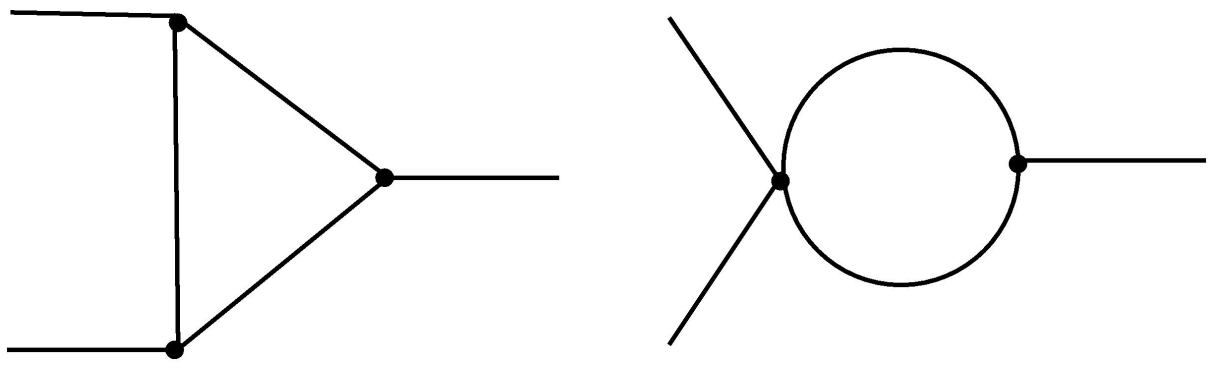

FIGURE 2.3: An example of triangle and bubble diagrams belonging to the same family – sharing the same set of propagators but differing in their individual exponents.

**Definition. 2**

**Topology:** A topology is a subset of a given integral family defined for a set of propagators with only positive exponents. It can be drawn as a graph for which momentum conservation holds at each vertex.

**Definition. 3**

**Sub-topology:** A sub-topology refers to a subset of a topology, where some of the propagators are missing. It can also be depicted as a diagram with momentum conservation at each vertex.

**Definition. 4**

**Sector:** Within a given integral family, Feynman integrals belonging to the same sector share an identical set of propagators with positive exponents. However, they may still differ in the powers of common propagators or in the powers of auxiliary propagators.

For example, one-loop box integrals with four propagators $I_{1,2,0,2}$ and $I_{2,1,-1,3}$ belong to the same sector, but $I_{1,1,0,2}$ and $I_{3,2,4,1}$ belong to the different sectors. We can define a *sector id* ($S_{\text{id}}$) to label each sector uniquely, which helps in organizing the calculation of these integrals. For a given integral family with $r$ inverse propagators,

$$I_{\nu_1,\ldots,\nu_n,\nu_{n+1},\ldots,\nu_r}, \quad \text{where} \quad \nu_i \begin{cases} \geq 0, & \text{for } 1 \leq i \leq n, \\ \leq 0, & \text{for } n+1 \leq i \leq r. \end{cases} \tag{2.9}$$

sector id is defined as

$$S_{\text{id}} = \sum_{i=1}^{r} 2^{i-1}\Theta\left(\nu_i - \frac{1}{2}\right), \tag{2.10}$$

where $\Theta(p)$ is the Heaviside step function, which will be equal to one for positive arguments,

i.e., $p > 0$ and zero otherwise. Within an integral family, the integrals with a smaller sector id are considered to be simpler. For $i \leq n$, integrals with the highest sector id or equivalently, with the maximum number of positive indices $\nu_i$ are classified as being in the *top sector* of the family, while integrals for which one or more indices satisfy $\nu_i < 1$ are classified as belonging to the *sub-sectors*.

**Definition. 5**

**Cutting a propagator and maximal cut:** Cutting an internal line in a Feynman graph or an internal propagator with an exponent equal to one is equivalent to putting the corresponding propagating particle on-shell. In the momentum space, it corresponds to the following replacement

$$\frac{1}{-q_i^2 + m_i^2} \rightarrow 2\pi i\, \delta(-q_i^2 + m_i^2) \tag{2.11}$$

Similarly, for a given Feynman integral $I_{\nu_1,\cdots,\nu_n}$, if we cut all the internal propagators having indices $\nu_i > 0$ simultaneously, it corresponds to the *maximal cut*.

## 2.2 Divergences and Regularization

The Feynman integrals defined in equation (2.5), in general, suffer from the divergences in different limits of loop momenta. To understand this concept, let us consider the simplest example of a one-loop massive tadpole integral $I(m^2)$ associated with a diagram shown in figure 2.4 in $d = 4$ space-time dimension

$$I(m^2) = \int d^4k \frac{1}{(-k^2 + m^2)}, \tag{2.12}$$

where $k$ is the loop momenta and $m$ is the mass of the particle running in the loop. To keep things simple, we have omitted the prefactors introduced in the equation (2.5) for a scalar

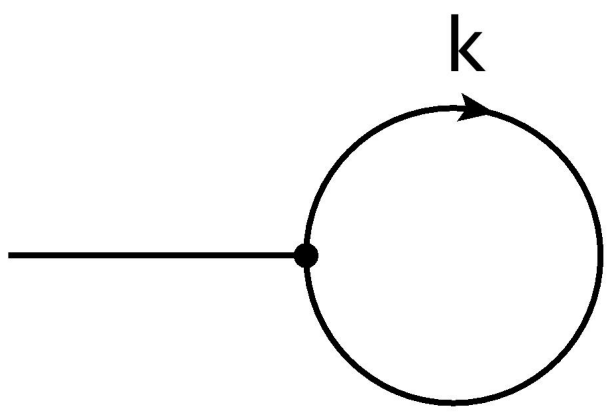

FIGURE 2.4: one-loop massive tadpole.

Feynman integral. In the region where loop momentum $k$ becomes much larger compared to the other scales, such as $m$, the integrand of the integral in equation (2.12) behaves as $1/k^2$. By employing a Wick rotation (for details refer to [4]) for Lorentz vector $k$ to transform it into an Euclidean vector $K$, where

$$k^0 = iK^0, \tag{2.13}$$

$$k^j = K^j \text{ for } j = 1, 2, 3\ , \tag{2.14}$$

and using spherical coordinates in four dimensions and integrating over the angles, we can express the tadpole integral as

$$\int d^4k \frac{1}{k^2} = i \int d^4K \frac{1}{K^2} = i2\pi^2 \int_0^\infty dKK. \tag{2.15}$$

In the limit of large $K$, this integral is ill-defined and quadratically divergent in the upper limit of integration, as shown below

$$\lim_{A\to\infty} \int_\Lambda^A dKK = \lim_{A\to\infty} \left(\frac{A^2}{2} + C_1\right),\ \Lambda > 0\ , \tag{2.16}$$

where $C_1$ is some constant. These types of divergences arising in the limit of large loop momenta, are known as *ultraviolet (UV) divergences*. In contrast, for massless particles running inside the loops, another type of singularities, known as *infrared (IR) divergences*, arise in the limit of small loop momenta. For example, in the massless case with the second power of the propagator, the integrand of the one-loop tadpole integral will have a factor of $1/(k^2)^2$. In the small loop momenta region, using Wick rotation and spherical coordinates, we obtain

$$\lim_{A\to 0} \int_A^\Lambda dK \frac{1}{K} = \lim_{A\to 0} \Big(\ln(A) + C_2\Big),\ \Lambda < \infty,\ C_2 = \text{constant}, \tag{2.17}$$

a logarithmically divergent integral.

These divergent integrals lead to infinite results in the calculation of physical observables in the perturbative quantum field theory. Therefore, these divergences must be removed from these integrals to obtain well-defined results for physical observables. The first step in dealing with these divergent integrals is to separate the divergences from the finite parts, a procedure known as *regularization*. Regularization is based on the following basic idea:
we introduce an arbitrary parameter called the *regularization parameter*, say $\lambda$, in our divergent

Feynman integral ($I$), such that the integral $I(\lambda)$ becomes well-defined in some range of $\lambda$. Then, during physical calculations, we expand our integrals with respect to $\lambda$ and maintain a clear separation between the terms involving $\lambda$ and the other terms. At the end of the calculation, we take the limit $\lambda \to 0$, and then our well-defined integral ($I(\lambda)$) coincides with the previous ill-defined integral ($I$), i.e.,

$$\lim_{\lambda \to 0} I(\lambda) = I. \tag{2.18}$$

The integral $I(\lambda)$ is called the regularized integral. By regularization, we mean to confine all divergences to the terms involving the regulator $\lambda$. Different schemes are available in the literature to employ the regularization procedure, such as, *momentum cutoff regularization* [4], *Pauli-Villars regularization* [88], *dimensional regularization* [68, 69] and *analytic regularization* [89]. Among these schemes, dimensional regularization is the most commonly used and efficient method to perform the regularization as it preserves the symmetries of the underlying theory, such as gauge and Lorentz invariance.

### 2.2.1 Dimensional Regularization

The idea of dimensional regularization in QFT was first proposed by t'Hooft and Veltman [68], and Bollini and Giambiagi [69]. It is based on the observation of convergence of the UV and IR divergences in the space-time dimensions lower and higher than 4, respectively. This observation suggests that the integrals which diverge for $d = 4$ may give finite results when computed in other dimensionalities. Therefore, the computation of integrals is performed in $d$-dimensions, where $d$ is considered as a continuous variable. As this procedure involves the analytic continuation of the space-time dimension, it is referred to as *dimensional regularization.*

We generally work in $d = 4 - 2\epsilon$ dimensions, where $\epsilon$ is the dimensional regularization parameter. To regularize UV divergences, we typically assume $\epsilon > 0$; for IR divergences, we take $\epsilon < 0$. Thus, dimensionally regulated integrals become functions of $\epsilon$, denoted as I($\epsilon$). In the limit $d \to 4$, divergences appear as poles in $\epsilon$ in the Laurent series expansion of these dimensionally regulated integrals. To understand this, let us evaluate the massive tadpole integral in $d$-dimensions, which we previously identified as quadratically UV divergent in equation (2.16).

$$I(d, m^2) = \int d^d k \frac{1}{(-k^2 + m^2)} = i \int_0^\infty d^d K \frac{1}{(K^2 + m^2)}. \tag{2.19}$$

In the above equation, we perform a Wick rotation of the integral to Euclidean space in $d$ dimensions by changing the time component to $k_0 = iK_0$ and leaving the space components unchanged i.e., $k_j = K_j$, for $0 < j \leq d-1$. Since the integration variable appears only squared, we can use generalized spherical coordinates (see [87] for details) to split the integration into radial and angular parts. The integration measure gives $d^dK = K^{d-1}dK d\Omega_d$. Performing the angular integration using $\int d\Omega_d = \frac{2\pi^{d/2}}{\Gamma(d/2)}$, the tadpole integral in equation (2.19) gives

$$I(d, m^2) = i\frac{2\pi^{d/2}}{\Gamma\left(\frac{d}{2}\right)} \int_0^\infty dK \frac{K^{d-1}}{(K^2+m^2)}. \tag{2.20}$$

Here $\Gamma(x)$ denotes the Euler's gamma function. In the next step, we will manipulate the integrand to put the integral in the form of the beta function to perform the integration over $K$.

$$I(d, m^2) = i\frac{2\pi^{d/2}}{\Gamma\left(\frac{d}{2}\right)} \int_0^\infty \frac{1}{2} dK^2 \frac{(K^2)^{d/2-1}}{(K^2+m^2)}. \tag{2.21}$$

Taking the mass dependence outside the integral and putting $K^2/m^2 = x$, we will get

$$I(d, m^2) = i(m^2)^{d/2-1}\frac{\pi^{d/2}}{\Gamma\left(\frac{d}{2}\right)} \int_0^\infty dx\, x^{d/2-1}(x+1)^{-1}. \tag{2.22}$$

Using the property of the beta function,

$$B(z_1, z_2) = \int_0^\infty dt \frac{t^{z_1-1}}{(1+t)^{z_1+z_2}} = \frac{\Gamma(z_1)\Gamma(z_2)}{\Gamma(z_1+z_2)}. \tag{2.23}$$

We get,

$$I(d, m^2) = i(m^2)^{d/2-1}\frac{\pi^{d/2}}{\Gamma\left(\frac{d}{2}\right)} \frac{\Gamma\left(\frac{d}{2}\right)\Gamma\left(1-\frac{d}{2}\right)}{\Gamma(1)} = i(m^2)^{d/2-1}\pi^{d/2}\Gamma\left(1-\frac{d}{2}\right). \tag{2.24}$$

Taking $d = 4-2\epsilon$, and performing the Laurent expansion with respect to $\epsilon$, we get

$$I(\epsilon, m^2) = -i(m\pi)^2\left(\frac{1}{\epsilon} + 1 - \gamma_E - \ln\left(m^2\pi\right)\right) + \mathcal{O}(\epsilon), \tag{2.25}$$

where $\gamma_E$ denotes the Euler-Mascheroni constant. The divergence of massive tadpole integral now appears as a simple pole in $\epsilon$. The integral will diverge for $\epsilon \to 0$ or as $d \to 4$. The insertion of suitable counter terms cancels these UV poles through the procedure called *renormalization*.

The regulated integrals in $d$-dimensions satisfy the following standard properties: Let $f(k)$ and $g(k)$ be any functions depending on the $d$-dimensional vector $k$. Then, for any complex numbers $a, b$

- **Linearity and Additivity:**

$$\int d^d k \; (a \; f(k) + b \; g(k)) = a \int d^d k \; f(k) + b \int d^d k \; g(k). \tag{2.26}$$

- **Scaling:** For any $\lambda \in \mathbb{C}$

$$\int d^d k \; f(\lambda k) = \lambda^{-d} \int d^d k \; f(k). \tag{2.27}$$

- **Translation invariance:** For any $d$-dimensional vector $q$

$$\int d^d k \; f(k+q) = \int d^d k \; f(k). \tag{2.28}$$

In addition to the above three axioms, some other useful properties of these dimensionally regulated integrals are described below:

- Order of integrations is interchangeable, i.e.,

$$\int d^d k \int d^d l \; f(k,l) = \int d^d l \int d^d k \; f(k,l), \tag{2.29}$$

where, $f$ is any function of two vectors $k$ and $l$.

- Order of integration and differentiation with respect to two different vectors can be interchanged, i.e.,

$$\frac{\partial}{\partial l} \int d^d k \; f(k,l) = \int d^d k \frac{\partial}{\partial l} \; f(k,l). \tag{2.30}$$

- For differentiating $f(k,q)$ with respect to the components of vector $k$, we have

$$\int d^d k \; \frac{\partial}{\partial k^\mu} q^\mu f(k,q) = 0. \tag{2.31}$$

This property is useful for deriving the integration-by-parts relations among Feynman integrals with different indices, as we will discuss in section 2.3.

- For $f$ be any function of $k^2$, we have

$$\int d^dk\ k^\mu f(k^2) = 0. \tag{2.32}$$

As we now have all the necessary tools to handle divergent integrals, in the next section, we will explore how to leverage the linear relations among these dimensionally regulated integrals to decompose them into a smaller set of independent integrals known as *master integrals*.

## 2.3 Linear relations among Feynman integrals

In perturbative calculations of physical observables in QFT, the number of Feynman integrals that need to be evaluated increases drastically with each successive higher order. For example, one-loop calculations require evaluation of only a few integrals, but the precision calculations at two-loop and beyond may require consideration of hundreds or even thousands of Feynman integrals. Evaluating such a huge number of integrals at any given order poses a significant challenge. Therefore, at a fixed order, it becomes necessary to utilize relations among these integrals to reduce them into a minimal set of independent integrals. Thus, our task will be reduced to evaluating only these independent sets of integrals. The procedure to obtain this minimal set of integrals is called *reduction to master integrals*. The integrals in this minimal set are referred to as the *master integrals*. The identities that relate the Feynman integrals within a family can be classified as

- Symmetry relations (SR)
- Lorentz invariance (LI) identities
- Integration-by-parts (IBP) identities

These identities are valid for any space-time dimension and are derived from the symmetry properties of dimensionally regularized Feynman integrals. Now, we will explain each of them briefly in the following sections.

### 2.3.1 Symmetry Relations

The discrete shifts in the loop momentum, which leave the integrals unchanged, give the first set of relations among the dimensionally regularized Feynman integrals. Let us consider the example of one-loop massive bubble integral shown in figure 2.5

$$I_{\nu_1\nu_2}(d,p^2,m^2) = \int d^dk \frac{1}{(-k^2+m^2)^{\nu_1}(-(k-p)^2+m^2)^{\nu_2}}, \tag{2.33}$$

where $k$, $p$ are the loop and external momenta respectively and $m$ is the mass of the internal particle. This integral exhibits the symmetry

$$I_{\nu_1\nu_2} = I_{\nu_2\nu_1}, \tag{2.34}$$

for the shift in the loop momentum $k$

$$k \to k' = -k+p. \tag{2.35}$$

Symmetry relations can generally be obtained from shifts in loop momentum

$$k'_m = \sum_{n=1}^{l} M_{mn}k_n + \sum_{n=1}^{e} c_n^{(m)} p_n, \quad m = (1,\cdots,l), \qquad M_{mn}, c_n^{(m)} \in \{-1,0,1\}, \tag{2.36}$$

and from the permutations of external momenta. These symmetry relations not only relate the integrals belonging to the same sector but can also establish a relation between the integrals belonging to different sectors. The symmetries that establish the connection between the integrals belonging to the same sector are called *sector symmetries* and those which establish relations among different sectors of the same or of different families are known as *sector mappings*.

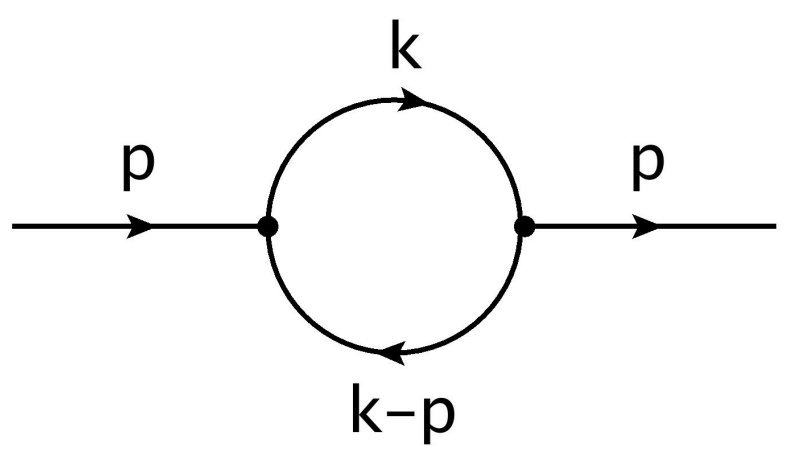

FIGURE 2.5: one-loop bubble.

### 2.3.2 Lorentz Invariance (LI) Identities

The Lorentz scalar nature of Feynman integrals under consideration given in equation (2.5) allows us to derive another set of identities called Lorentz invariance (LI) identities [65]. These identities are based on the fact that the integrals remain invariant under the Lorentz transformations of external momenta. Let us consider an infinitesimal Lorentz transformation of external momenta,

$$p_i^\mu \to p_i^\mu + \delta p_i^\mu = p_i^\mu + \omega^{\mu\nu} p_{i,\nu} \ , \tag{2.37}$$

where $\omega_{\mu\nu}$ is a totally anti-symmetric tensor, i.e. $\omega_{\mu\nu} = -\omega_{\nu\mu}$. The scalar Feynman integral should remain unchanged under this transformation, i.e.,

$$I(p_i + \delta p_i) = I(p_i). \tag{2.38}$$

For any infinitesimal transformation, we can expand the integral $I(p_i + \delta p_i)$ as

$$I(p_i + \delta p_i) = I(p_i) + \sum_{j=1}^{e} \delta p_j^\mu \frac{\partial}{\partial p_j^\mu} I(p_i) = I(p_i) + \omega^{\mu\nu} \sum_{j=1}^{e} p_{j,\nu} \frac{\partial}{\partial p_j^\mu} I(p_i). \tag{2.39}$$

For equation (2.38) to be true, we will have

$$\omega^{\mu\nu} \sum_{j=1}^{e} p_{j,\nu} \frac{\partial}{\partial p_j^\mu} I(p_i) = 0. \tag{2.40}$$

Using the anti-symmetric property of $\omega^{\mu\nu}$, one arrives at

$$\sum_{j=1}^{e} \left( p_{j,\nu} \frac{\partial}{\partial p_j^\mu} - p_{j,\mu} \frac{\partial}{\partial p_j^\nu} \right) I(p_i) = 0. \tag{2.41}$$

By contracting the above equation with anti-symmetric tensors built by exploiting all possible combinations of external momenta, we get the required LI identities for the integrals under consideration. For $e$ independent external momenta, total $e(e-1)/2$ LI identities are possible for a given set of integrals.

Consider the example of a one-loop triangle topology shown in figure 2.3. Employing momentum conservation, we have two independent external momenta, say $p_1$ and $p_2$. As only one anti-symmetric combination can be constructed using two external momenta. Thus, in this case, only one LI identity is possible, given as

$$(p_1^\mu p_2^\nu - p_2^\mu p_1^\nu)\sum_{j=1}^{2}\left(p_{j,\nu}\frac{\partial}{\partial p_j^\mu} - p_{j,\mu}\frac{\partial}{\partial p_j^\nu}\right)I(p_1,p_2) = 0. \tag{2.42}$$

### 2.3.3 Integration-by-parts (IBP) Identities

The integration-by-parts (IBP) identities, first introduced by Chetyrkin and Tkachov [70, 71], are one of the most important classes of identities established for dimensionally regularized Feynman integrals for their reduction to a minimal set of independent integrals. These identities are based on the idea that for dimensionally regularized Feynman integrals, we can always find a value of the space-time dimension $d$ for which the integral gives finite results and will converge[1]. The condition for the convergence of a dimensionally regularized integral at the boundaries is that the total derivative of the integrand must vanish at the boundaries, which can be seen as a $d$-dimensional Gauss theorem. Such a condition for the scalar Feynman integral $I_{\nu_1,\cdots,\nu_r}$ given in (2.5) with exponent set $(\nu_1,\cdots,\nu_r)$ can be given as:

$$\int\prod_{i=1}^{l} d^d k_i\,\frac{\partial}{\partial k_m^\mu}\left(v_n^\mu\prod_{j=1}^{r}\frac{1}{D_j^{\nu_j}}\right) = 0, \qquad 1\le m\le l\,, \tag{2.43}$$

where $v_n^\mu$ is one of the internal $(k_1,\cdots,k_l)$ or independent external $(p_1,\cdots,p_e)$ momentum vectors. Such an identity is known as integration-by-parts identity or IBP.

For a given integral family, after differentiating an integral of a given sector with respect to one of the loop momentum and contracting with $v_n^\mu$, we will get scalar products among the loop momenta, and/or between a loop and external momenta in the numerator. Again, expressing these scalar products in terms of inverse propagators $D_i$, we will get linear IBP relations among different integrals of the type

$$\sum_i c_i I_{\nu_{i1},\cdots,\nu_{ir}} = 0, \tag{2.44}$$

where the coefficients $c_i$ are rational functions of space-time dimension $d$, masses of the internal particles, kinematic invariants and the exponents $\nu_i$. The integrals $I_{\nu_{i1},\cdots,\nu_{ir}}$ are the scalar integrals of the same sector or of subsectors, with the same exponents or with the set of exponents in which some of the exponents are possibly decreased and increased by one. Thus, IBP identities relate the integrals with different set of exponents. In total, $l(l+e)$ IBP identities can be established for each integral. By solving these identities, we can express a generic dimensionally regularized Feynman integral of the given family as a linear combination of some preferred independent integrals called master integrals.
A simple example to understand the idea is of a one-loop massive tadpole integral

$$I_\nu(d, m^2) = \int d^d k \, \frac{1}{(-k^2 + m^2)^\nu}. \tag{2.45}$$

[1] All scaleless integrals vanish in dimensional regularization.

Since there is no external momentum involved, so with loop momentum $k$, the only possible IBP relation can be given as

$$\int d^d k \, \frac{\partial}{\partial k^\mu} \frac{k^\mu}{(-k^2 + m^2)^\nu} = 0. \tag{2.46}$$

Taking the derivative with respect to loop momenta gives

$$\int d^d k \left( \frac{1}{(-k^2 + m^2)^\nu} \frac{\partial}{\partial k^\mu} k^\mu + k^\mu \frac{\partial}{\partial k^\mu} \frac{1}{(-k^2 + m^2)^\nu} \right) = 0. \tag{2.47}$$

$$\int d^d k \left( \frac{1}{(-k^2 + m^2)^\nu} . \, d + k^\mu \, (-\nu) \frac{1}{(-k^2 + m^2)^{\nu+1}} \, (-2k_\mu) \right) = 0. \tag{2.48}$$

Here, in the last equation, we have used the $d$-dimensional algebra $\frac{\partial}{\partial k^\mu} k^\mu = d$. Simplifying the integrand using $k^\mu k_\mu = k^2$ and after some modifications, we arrive at

$$d \int d^d k \, \frac{1}{(-k^2 + m^2)^\nu} - 2\nu \int d^d k \, \frac{1}{(-k^2 + m^2)^\nu} + 2\nu m^2 \int d^d k \, \frac{1}{(-k^2 + m^2)^{\nu+1}} = 0. \tag{2.49}$$

Writing back the above equation in terms of tadpole integral notation $I_\nu$, we obtain

$$(d - 2\nu) I_\nu(d, m^2) + 2\nu m^2 I_{\nu+1}(d, m^2) = 0. \tag{2.50}$$

The equation (2.50) will lead to the recurrence relation

$$I_\nu(d, m^2) = -\frac{(d - 2\nu + 2)}{2(\nu - 1) m^2} I_{\nu-1}(d, m^2), \;\; \nu \geq 2 \tag{2.51}$$

Thus, for the tadpole topology, any integral with exponent $\nu > 1$ can be expressed in terms of one integral $I_1(d, m^2)$, called the master integral of this family.

Therefore, by utilizing the symmetry properties of dimensionally regulated Feynman integrals, we derive a linear system of equations that relate different integrals within an integral family. Solving these equations allows us to express integrals from the family in terms of a "finite" set of master integrals, forming a basis [90]. However, as with any algebraic basis, there is considerable freedom in choosing the basis integrals for the complete family. In applications, the main bottleneck lies in solving the system of equations resulting from these symmetry properties. Depending on the problem, this can lead to systems with hundreds, thousands, or even millions of linear equations. Hence, using computer algebra systems to handle and solve such a huge system of equations becomes crucial. One of the most commonly used algorithms to efficiently solve the system of identities is the *Laporta algorithm* [91]. In this algorithm, we basically order the integrals based on their complexity, and then the system of equations is solved using *Gauss's substitution rule*. Examples of computer programs that implement the Laporta algorithm to reduce Feynman integrals to master integrals are: **FIRE** [92–94], **Kira** [95] and **Reduze** [96]. Another widely used computer package based on constructing explicit symbolic rules to solve the system of identities to obtain master integrals is **LiteRed** [97, 98].

## 2.4 Other Relations among Feynman integrals

The linear relations discussed in the previous section focus on the connections among Feynman integrals with varying exponents $\nu_i$ but within a fixed space-time dimension $d$. Tarasov introduced a systematic algorithm in [84] for deriving recurrence relations among dimensionally regularized Feynman integrals with different values of space-time dimension $d$. These relations, known as *dimensional shift relations*, primarily relate $d$ and $d-2$ dimensional integrals in terms of a differential operator. Using these relations along with IBP identities, the master integrals of a given integral family (denoted by say a vector $\vec{g}$) in $d-2$ dimensions can be expressed as linear combinations of the same master integrals in $d$ dimensions:

$$g_k(d, \{p_i^2\}, \{s_i\}; \{m_j^2\}) = \sum_l B_{kl}(d, \{p_i^2\}, \{s_i\}; \{m_j^2\})\; g_l(d-2, \{p_i^2\}, \{s_i\}; \{m_j^2\}). \tag{2.52}$$

Here, the coefficients $B_{kl}(d, \{p_i^2\}, \{s_i\}; \{m_j^2\})$ are rational functions of scalar products of external independent momenta, Mandelstam variables, internal masses, and the space-time dimension $d$. These relations are highly valuable for practical calculations of multi-loop and multi-leg integrals. To obtain these dimensional recurrence relations among the given set of integrals, one can utilize the publicly available **Mathematica** package **LiteRed** [97, 98].

After getting the set of master integrals, the next crucial step will be their computation using various techniques, which we will discuss in the next chapter.

# 3

# Evaluation of master integrals

IN the previous chapter, we highlighted the properties of the scalar Feynman integrals and discussed how one can obtain the set of independent integrals i.e. master integrals of an integral family, for the problem under consideration. In this chapter, we will focus on the methods to obtain solutions for these master integrals.

In general, two different approaches can be employed for evaluating Feynman integrals: the numerical approach and the analytic methods (for a summary, see [99]). Since high school, we have been well aware of the fact that integration is objectively a complex operation to perform. As a result, numerical methods will appear more promising to us for the evaluation of Feynman integrals. There are numerous public numerical tools based on the various algorithms available for the computation of Feynman integrals.

## 3.1 Numerical Tools

In this section, we will discuss some aspects of numerical computation; the tools and techniques used in the numerical evaluation of Feynman integrals.

**Sector decomposition:** As discussed in the last chapter, Feynman integrals are generally divergent in nature. Therefore, these divergences must be adequately taken care of and regularized before employing the numerical algorithms for their evaluation to obtain reliable results. To constructively isolate these divergences from the integrals *sector decomposition approach* is commonly employed. It is based on the simple idea of dividing the integration region into sectors, isolating the singularities in terms of a Laurent series in dimensional regularization parameter $\epsilon$ with finite coefficients, and finally integrating these coefficients numerically. The famous Monte-Carlo integrators based on this approach often used for Feynman integral evaluation are **`FIESTA`** [100] and **`pySecDec`** [101].

In addition to sector decomposition, alternative numerical methods have been developed to handle the evaluation of Feynman integrals. One such method is implemented in the **`AMFlow`** package.

**AMFlow:** **`AMFlow`** [102] is a publicly available **`Mathematica`** package used for automated, numerical computations of dimensionally regularized Feynman integrals. It utilizes the *auxiliary mass flow method* [103–106], where the integrals are treated as functions of an auxiliary mass parameter, say $m_{\text{aux}}$. The method involves setting up and solving linear differential equations for these integrals with respect to $m_{\text{aux}}$. **`AMFlow`** is particularly useful for high-performance numerical evaluation of Feynman integrals, making it suitable for high-precision phenomenological studies.

**PSLQ algorithm:** The PSLQ algorithm [73, 107–109] is an effective integer relation detection algorithm. It is a numerical algorithm, which for a given n-long vector $x = (x_1, x_2, \cdots, x_n)$ with high-precision floating point numbers, can find integer coefficients $a_1, a_2, \cdots, a_n$ such that the linear combination with these integer coefficients is close to zero within numerical precision, i.e.,

$$a_1 x_1 + a_2 x_2 + \cdots + a_n x_n = 0. \tag{3.1}$$

Otherwise, depending on the numerical precision chosen, it finds the bound within which no such relation can exist. This technique can be efficiently used if one knows the set of numbers by which the result of a calculation is finally spanned or if one wants to find a linear combination of such numbers. For example,

$$I = \int_0^1 dz \, \frac{Li_3(z)}{1+z}. \tag{3.2}$$

Calculating this integral with high precision up to 40 digits gives,

$$I = 0.3395454690873598695906678484608602061388\ .$$

The integral $I$ has *transcendentality* 4, further clarification on this term can be found in section 3.3.3. Therefore, we employ the **PSLQ** algorithm to fit this numerical value using the expected basis spanned by

$$\ln^4(2),\ \ln(2)\zeta_3,\ \ln^2(2)\zeta_2,\ \zeta_2^2,\ Li_4(1/2).$$

Which yields

$$I = -\frac{1}{12}\ln^4(2) + \frac{\pi^4}{60} + \frac{3}{4}\ln(2)\zeta_3 + \frac{1}{12}\ln^2(2)\pi^2 - 2Li_4\left(\frac{1}{2}\right). \tag{3.3}$$

While evaluating Feynman integrals as a Laurent expansion in dimensional regulator $\epsilon$, this technique is widely used to fit the numerical values of the $\epsilon$-coefficients in terms of well-known mathematical constants. Despite various private implementations of **PSLQ**, it is now implemented within **Mathematica**. The above-mentioned techniques and implementations are used in performing numerical checks for the multi-scale Feynman integrals evaluated in chapter 5.

Numerical methods provide fast and promising results for many multi-loop Feynman integrals, but they have some limitations. For instance, in the presence of endpoint singularities or spurious poles, they provide results with lower convergence and stability. Moreover, computing observables such as cross-sections and decay rates requires integrating the corresponding amplitude at hundreds or even thousands of phase-space points. However, for each point, the integrator must be rerun from the beginning, making numerical results for Feynman integrals more susceptible to errors and resource-intensive in terms of time and memory. Additionally, in some cases, we might have access to the numerical results for a given integral topology only within a limited phase-space region.

Therefore, analytical solutions for the Feynman integrals, whenever obtainable, are often preferred over numerical ones for several reasons, as discussed below:

1. **Precision and reliability:** The analytical solutions are exact and free of any approxima-

tions. Thus, they can be evaluated numerically with very high precision, and the results are much more reliable than the numerical ones.

2. **Fast and error-free evaluation:** The availability of compact analytical solutions of Feynman integrals in terms of known mathematical functions makes them suitable for fast and error-free numerical evaluations of physical observables.

3. **Insights into the theory:** The analytical solutions of the Feynman integrals are expressed in terms of special mathematical functions [110] and thus are helpful to obtain the analytical expressions for respective scattering amplitudes in pQFT. The detailed study of the mathematical structure of these special functions in certain limits or regions provides deeper insights into the underlying physics (symmetries, threshold regions, singularity structure, etc.) of the theory under consideration.

4. **Interconnecting different theories:** The mathematical functions appearing in the analytical solutions of Feynman integrals in QFT also emerge in other formal theories. Thus, exploiting the geometrical structure of these functions helps to understand the connection between various theories, such as QCD, electroweak theory and super Yang-Mills (SYM) theories.

Another motivation to pursue analytical approaches comes from the complex geometrical structure of functions appearing in the solutions of multi-loop Feynman integrals. The unitarity properties of the scattering amplitudes of which these integrals are an essential part, necessitate functions with intricate branch cuts to appear in the solutions of integrals. The geometrical study of these functions helps to establish a direct connection between Physics and Mathematics and is thus, currently a hot topic of research in physics. A more detailed examination of the functional basis of Feynman integrals and their properties will be presented in the upcoming sections.

## 3.2 Art of Analytical Evaluation of Feynman Integrals

As previously discussed, the analytical solutions for Feynman integrals hold significance not only from a physics perspective but also from a mathematical point of view. For the analytical computation of Feynman integrals, two approaches are mainly employed. One approach is to perform the direct integration, and the other is to do the integration indirectly.

Prominent examples of methods for direct integration include *Feynman parametrization* and *Mellin-Barnes representation* [111–116], which efficiently work for one-loop integrals and for multi-loop integrals with a few mass scales. Among the indirect integration methods, the most commonly utilized are the *difference equation method* [91, 117] and *differential equations method* [63–65]. The difference equation method establishes functional relations similar to the recurrence relations between different integrals by shifting their exponents by integers. However, in the method of differential equations, Feynman integrals are considered to be the function of kinematic invariants, and this method describes how the integrals behave under the continuous change of these invariants. Over the past few years, the method of differential equations has emerged as a powerful tool for analytically evaluating multi-loop Feynman integrals. In the context of this thesis, we will discuss the method of differential equations in detail in the following sections.

## 3.3 Differential Equation Method

Using the IBP identities, the method of solving Feynman integrals by differentiating them with respect to the masses of internal particles was first introduced by Kotikov in 1990 [63] and later Remiddi and Gehrmann [64, 65] extended and generalized this method with respect to all external kinematic invariants. The main idea of this method involves taking the derivative of a given family of master integrals with respect to both, the kinematic invariants and internal masses and utilizing the IBP relations to write the result of differentiation in terms of the same set of master integrals. As a result, we will get a linear system of differential equations, a solution to which can be obtained using appropriate *boundary conditions*.

### 3.3.1 Derivation of System of Differential Equations

In order to derive the system of differential equations, let us consider a family of Feynman integrals, belonging to a Feynman graph having '$l$' internal and '$e+1$' external lines, with '$m$' master integrals denoted by vector $\vec{f}(d,x) = (f_1(d,x), \cdots, f_m(d,x))$, depending on '$n+1$' number of dimensionless kinematic variables $x = (x_1, \cdots, x_{n+1})$ and phase-space dimension $d$. The kinematic variables consist of Lorentz invariant products of '$e$' linearly independent external momenta and '$l$' internal masses of the form

$$\frac{p_i \cdot p_j}{\mu^2}, \quad 1 \le i \le j \le e, \tag{3.4}$$

$$\frac{m_i^2}{\mu^2}, \quad 1 \le i \le l, \tag{3.5}$$

where $\mu$ is the regularization scale. For a given scalar master integral in the family, if we rescale all the kinematic variables by a factor $\lambda$, we will have

$$f(\lambda x_1, \cdots, \lambda x_{n+1}) = \lambda^{\frac{lD}{2}-\nu} f(x_1, \cdots, x_{n+1}). \tag{3.6}$$

This can be easily seen by replacing $\mu^2 \to \mu^2/\lambda$ in equation (2.5). Thus, we may set one of the kinematic variables to 1 (say $x_{n+1} = 1$)) or change the variables set from $(x_1, \cdots, x_n, x_{n+1})$ to $(y_1, \cdots, y_n) = (x_1/x_{n+1}, \cdots, x_n/x_{n+1})$ and the full dependence on all kinematic variables can be recovered later using scaling relation given in equation (3.6). Therefore, we can view the Feynman integrals in vector $\vec{f}$ as functions of $x_1, \cdots, x_n$ kinematic variables (and dimension $d$). By taking the derivative of each integral in $\vec{f}$ with respect to each of the kinematic variables, on the right-hand side, we will get a linear combination of Feynman integrals belonging to the same family but with shifted exponents (all of them may not be the master integrals belonging to $\vec{f}$). Using IBP identities, we may reduce the right-hand side of each differentiated integral in terms of a linear combination of the chosen master integrals belonging to $\vec{f}$. Thus, we will get a system of first-order differential equations for the integrals $f_1, \cdots, f_m$ compactly written as

$$d\vec{f}(\epsilon, x) = A(\epsilon, x)\vec{f}(\epsilon, x), \tag{3.7}$$

where, $d = \sum_{j=1}^{n} dx_j \partial_{x_j}$ denotes the total derivative with respect to $x_1, \cdots, x_n$ and $A$ is a matrix-valued one-form

$$A = \sum_{j=1}^{n} A_{x_j} dx_j, \tag{3.8}$$

having $m \times m$ dimension with entries depending on the rational functions of kinematic variables $x$ and dimensional regulator $\epsilon$ (for space-time dimension $d = 4 - 2\epsilon$).

**Note:** As the total differential must satisfy $d^2 = 0$, the matrices $A_{x_j}$ are not all independent. We have

$$0 = d^2\vec{f}(\epsilon, x) = d\Big[A(\epsilon, x)\vec{f}(\epsilon, x)\Big] = \Big[dA(\epsilon, x) - A(\epsilon, x) \wedge A(\epsilon, x)\Big]\vec{f}(\epsilon, x). \tag{3.9}$$

This gives the *integrability condition* for $A(\epsilon, x)$

$$dA(\epsilon, x) - A(\epsilon, x) \wedge A(\epsilon, x) = 0. \tag{3.10}$$

Here, the symbol '$\wedge$' denotes the wedge product. The equation (3.10) provides a set of relations among the matrices $A_{x_j}(\epsilon, x)$, which can be used to cross-check the correctness of the system of differential equations.

**Example:** Let us derive the system of the differential equation for the one-loop bubble integral with two equal massive propagators shown in figure 2.5.

$$I_{\nu_1\nu_2}(d, p^2, m^2) = e^{\epsilon\gamma_E}(\mu^2)^{\nu_{12}-d/2} \int \frac{d^d k}{i\pi^{d/2}} \frac{1}{(-k^2+m^2)^{\nu_1}(-(k-p)^2+m^2)^{\nu_2}}, \tag{3.11}$$

where $\nu_{12} = \nu_1 + \nu_2$, and the factor $e^{\epsilon\gamma_E}$ is introduced to ensure the final result is independent of $\gamma_E$. Next, we set $\mu^2 = m^2$ for convenience. Utilizing the scaling relation and defining $x = p^2/m^2$ as the only scaleless variable, we rewrite the integral as

$$I_{\nu_1\nu_2}(d, x) = e^{\epsilon\gamma_E}(m^2)^{\nu_{12}-d/2} \int \frac{d^d k}{i\pi^{d/2}} \frac{1}{(-k^2+m^2)^{\nu_1}(-(k-p)^2+m^2)^{\nu_2}} \tag{3.12}$$

Utilizing IBP identities, one can show that there are two master integrals associated with this bubble topology (for details refer to [87]). We select

$$\vec{I}(d, x) = (I_{20}, I_{21})^T \tag{3.13}$$

as the basis of master integrals. In $d = 4 - 2\epsilon$ phase-space dimensions, after differentiating the basis integrals with respect to $x$ and applying IBP identities, we obtain the system of differential equations

$$\begin{aligned} &\frac{d}{dx} I_{20}(\epsilon, x) = 0, \\ &\frac{d}{dx} I_{21}(\epsilon, x) = \frac{\epsilon}{x(4-x)} I_{20}(\epsilon, x) - \frac{2-(1+\epsilon)x}{x(4-x)} I_{21}(\epsilon, x). \end{aligned} \tag{3.14}$$

Expressing this in the form of equation (3.7), i.e., $d\vec{I} = B(\epsilon, x)\vec{I}$ gives us the matrix one-form $B(\epsilon, x)$ as

$$B(\epsilon, x) = \begin{pmatrix} 0 & 0 \\ \frac{\epsilon}{4x} - \frac{\epsilon}{4(x-4)} & -\frac{1}{2x} - \frac{1+2\epsilon}{2(x-4)} \end{pmatrix} dx \tag{3.15}$$

To derive this differential equation, various computer programs, such as **LiteRed** [97, 98], can be used. It is worth noting that all entries of the matrix $B$ are rational functions in $x$.

### 3.3.2 Change of Basis

As the choice for the basis integrals $\vec{f}(\epsilon, x)$ is arbitrary, we always have the freedom to switch to a different basis $\vec{g}(\epsilon, x)$. As $\vec{f}(\epsilon, x)$ and $\vec{g}(\epsilon, x)$ are bases belonging to the same vector space, they must be related to each other via an invertible matrix $T(\epsilon, x)$ as follows

$$\vec{g}(\epsilon, x) = T(\epsilon, x)\vec{f}(\epsilon, x). \tag{3.16}$$

Under this integral basis change $f \to g$, the system of differential equations (3.7) transforms as follows

$$d\vec{g}(\epsilon, x) = A'(\epsilon, x)\vec{g}(\epsilon, x), \tag{3.17}$$

where the matrix $A'(\epsilon, x)$ is related to the old matrix $A(\epsilon, x)$ as follows

$$A'(\epsilon, x) = T(\epsilon, x)\ A(\epsilon, x)\ T^{-1}(\epsilon, x) - T(\epsilon, x)\ dT^{-1}(\epsilon, x). \tag{3.18}$$

One important point to note here is that while IBP identities restrict entries in the original matrices $A_{x_j}$ to be rational in variables $x_j$, there are no such restrictions for the entries of the matrix $T(\epsilon, x)$. In general, we consider the change of bases such that the transformation matrix $T(\epsilon, x)$ depends rationally on $\epsilon$, with entries that can be rational, algebraic, or transcendental functions of kinematic variables $x$.

In the next section, we will explore what kind of basis choice can dramatically simplify the system of differential equations.

### 3.3.3 Canonical Form of Differential Equations

In applications, we are usually interested in the Laurent expansion of the master integrals up to some finite order in $\epsilon$. However, solving the full differential equation system (3.7) with complete dependence on the dimensional regulator $\epsilon$ is often challenging. So, we are often interested in a basis choice that results in a simplified system of differential equations that can be solved easily. In reference [72], it is conjectured that with a suitable "*good*" basis choice $\vec{g}(\epsilon, x)$, the differential equation system can be put into the form

$$d\vec{g}(\epsilon, x) = \epsilon \tilde{A}(x)\vec{g}(\epsilon, x), \tag{3.19}$$

with $A' = \epsilon\tilde{A}$, such that the dependence on the dimensional regulator $\epsilon$ gets completely factorized out from the transformed matrix $\tilde{A}$. This factorization significantly simplifies the analysis of the system of differential equations. Furthermore, the transformed system is *fuchsian* in the sense that it only exhibits regular singularities in the kinematic variables [118]. The new basis $\vec{g}(\epsilon, x)$ that satisfies equation (3.19) is referred to as the *canonical basis*. The corresponding system of differential equations is said to be in the *canonical form* or *$\epsilon$-form*. In the $\epsilon$-form, the solutions for basis integrals can be easily obtained by solving the differential equation system iteratively, order by order in $\epsilon$, as will be shown in the section 3.4. Furthermore, this form helps in identifying the class of functions involved in their solutions. Another fundamental property of the canonical form, as evident from equation (3.19), is that in the limit $\epsilon \to 0$, the *homogeneous part* of the differential equation system becomes trivial, i.e.,

$$d\vec{g}(\epsilon, x) = 0. \tag{3.20}$$

This implies that the solution of the homogeneous part is constant for $\epsilon = 0$. In the context of this thesis, we focus on the canonical form of the differential equation system, obtained only through rational or algebraic transformations of the original basis $\vec{f}(\epsilon, x)$. For this case, the canonical differential equation system (3.19) can be cast into the *dlog-form*. In this form the matrix $\tilde{A}$ takes the form

$$\tilde{A}(x) = \sum_{r=1}^{k} C_r d\log p_r(x) = \sum_{r=1}^{k} C_r \omega_r, \tag{3.21}$$

with $C_r$ being $m \times m$-matrices with constant entries, and $p_r$ denoting algebraic functions of kinematic variables $x_i$ referred as the *letters*, their independent set is termed as the *alphabet*, and $\omega_r$ denote the independent differential one-forms. In this form the integration of the differential equation system becomes trivial and can be carried out algebraically.

An important question remains to be investigated: How do we choose a "good" basis that transforms the differential equation system into the canonical dlog-form? In this context, Henn [72] proposed a guiding principle based on the concept of *transcendentality*. The degree of transcendentality $\mathcal{T}(F)$ of a function $F$ is defined as the number of iterated integrals needed to define $F$ (a detailed explanation of iterated integrals will be provided in section 3.5). For instance, $\mathcal{T}(\log(x)) = 1$, $\mathcal{T}(\mathrm{Li}_k(x)) = k$, and constants like $\zeta_n$ and $\pi$ are assigned the same transcendentality value as the functions from which they are derived. For example, $\zeta_n = \mathrm{Li}_n(1)$, so $\mathcal{T}(\zeta_n) = n$, and $\zeta_2 = \pi^2/6$ implies $\mathcal{T}(\pi) = 1$ due to the property $\mathcal{T}(F_1 F_2) = \mathcal{T}(F_1) + \mathcal{T}(F_2)$. Algebraic functions have a transcendentality degree of zero. Furthermore, we introduce two additional concepts to understand how the notion of transcendentality is related to the canonical form of the differential equation system. A function $F$ is said to have a *uniform transcendental (UT) degree* if, when expressed as a sum of terms, all its terms have the same degree of transcendentality. Additionally, a function $F$ is considered as *pure* if its degree of transcendentality decreases by one by taking its derivative, i.e., $\mathcal{T}(dF) = \mathcal{T}(F) - 1$. Moreover, assigning the degree of transcendentality $-1$ to the dimensional regularization parameter $\epsilon$ is conventional. Then, for the transcendentality of the canonical differential equation system given in (3.19), we obtain

$$\mathcal{T}(d\vec{g}) = \mathcal{T}(\tilde{A}) + \mathcal{T}(\vec{g}) - 1. \tag{3.22}$$

For the dlog-form of the differential equation, $\tilde{A}$ is a logarithmic matrix one-form (3.21), and thus has a degree of transcendentality equal to 0. Therefore, it results in $\mathcal{T}(d\vec{g}) = \mathcal{T}(\vec{g}) - 1$, which means that the basis integrals that cast the differential equation system into canonical dlog-form are pure functions. Thus, the choice of the basis integrals is made in such a way that integrals are UT functions having the following expansion in $\epsilon$

$$\vec{g} = \sum_{j=0}^{\infty} \vec{g}^{(j)} \epsilon^j, \tag{3.23}$$

where $g_i^{(j)}$ represents a pure function with transcendentality $j$, ensuring that the entire function $g_i$ has a uniform zero transcendental degree at each expansion order. Furthermore, the differential equation system remains unchanged under the rescaling of basis integrals by any kinematic-independent factor. Hence, we can always renormalize our basis integrals such that their expansion starts at $\epsilon = 0$.

A significant question that remains is: how does one select these UT candidates for the canonical integral basis? One approach is the study of generalized unitarity cuts of the master integrals, which serve as an important tool in this regard [118]. Cutting a propagator involves replacing it with a Dirac $\delta$-function, as discussed earlier in chapter 2. Importantly, cuts of an integral do not affect the derivation of the IBP identities, so the cut integral satisfies the same set of differential equations as the original uncut integral. Additionally, the maximal cut of a Feynman integral corresponds to taking the n-fold residue of the integrand in the complex plane. *A crucial observation is that the maximal cuts provide solutions to the homogeneous system of differential equations involving integrals of the highest topologies, as integrals of sub-topologies drop out* [119]. Furthermore, it is worth noting that the property of uniform transcendentality of an integral remains preserved under the maximal cut. Therefore, the uniform transcendental degree of candidates for the canonical basis can be inferred from studying their simpler maximally cut integrals. This maximal cut information offers a standard approach to determine the transformation matrix, up to an $\epsilon$-dependent prefactor, for the integral basis, which transforms the homogeneous part of the differential equation system into the canonical form. Moreover, an interesting thing to note is that the maximal cut computed in any number of space-time dimensions $d$ provides a solution to its $d$-dimensional differential equation. The simplified analysis of maximal cuts is achievable through the utilization of the so-called Baikov representation [120, 121].

Furthermore, in the Laurent expansion of maximal cut of an integral with respect to dimensional regularization parameter $\epsilon$, if the first term comes out to be a constant, i.e., independent of any kinematic variables, the integral is said to have a *constant leading singularity*. According to a conjecture in reference [122], integrals with leading singularities $\pm 1$ always evaluate to pure functions. Therefore, in computing the maximal cut of integrals, we can normalize the integrand or modify the integration contour so that the leading term of its $\epsilon$ expansion evaluates to $\pm 1$. Correspondingly, normalizing the integrand of the original integral in a similar manner gives us a candidate for our integral basis with a uniform transcendental degree. Thus, the crucial strategy for obtaining a canonical dlog-form for the differential equation system is related to finding the integral basis with a unit leading singularity [118].

It is important to note that the differential equation system for cases with integrals involving transcendental functions such as periods of an elliptic curve in their leading singularities cannot be cast into the canonical form. However, several examples exist where obtaining the canonical form is still possible [123–126].

Several public computer codes are available to obtain a canonical basis, each employing a different algorithm and working mostly for problems involving a few kinematic variables. The examples include **`Fuchsia`** [127], **`Canonica`** [128], **`Libra`** [129] and **`Initial`** [130]. In addition to these codes, alternative methods based on the computation of intersection numbers have been put fourth [131–136]. For problems involving many mass scales, a combination of the above-mentioned approaches may be needed to effectively find the canonical basis.

Let us return to the example of bubble topology for which we have obtained the differential equation system in equation (3.15). Now, we switch to a different basis of master integrals to simplify the differential equation system, utilizing the maximal cut information. In the Baikov representation, the leading term in the expansion of the maximal cut (i.e., the leading singularity) of integral $I_{21}$, up to a constant prefactor in $d = 4 - 2\epsilon$, is given by (see [87] for details)

$$\text{MaxCut } I_{21} \propto \frac{1}{\sqrt{-x(4-x)}}. \tag{3.24}$$

Using this information, let us switch the basis from $\vec{I} \to \vec{J}$, as

$$J_{20}(\epsilon, x) = \epsilon I_{20}(\epsilon, x), \quad J_{21}(\epsilon, x) = \epsilon\sqrt{-x(4-x)} I_{21}(\epsilon, x). \tag{3.25}$$

The basis integrals $\vec{J}$ are multiplied with the $\epsilon$ factor to ensure that their $\epsilon$ expansion starts at a non-negative order. This new choice of basis integrals satisfies the following differential equation system,

$$\frac{d}{dx}\vec{J} = \epsilon \begin{pmatrix} 0 & 0 \\ -\frac{1}{\sqrt{-x(4-x)}} & -\frac{1}{x-4} \end{pmatrix} \vec{J}, \tag{3.26}$$

ensuring the complete factorization of the $\epsilon$-dependence from transformed matrix $\tilde{B}$ above. However, this simplification of the system of differential equations comes with a penalty; note that the transformation introduces a square root $\sqrt{-x(4-x)}$ in our system. The presence of these square roots may pose challenges in solving the corresponding differential equation systems, which we will discuss in the upcoming sections.

## 3.4 Solving the Differential Equations

After casting the differential equation system into the canonical form, given in equation (3.19), the solution for the basis integrals can be straightforwardly written as the path-ordered exponential expansion

$$\vec{g}(\epsilon, x) = \mathbb{P} \exp\left(\epsilon \int_\gamma \tilde{A}(x')\right) \cdot \vec{g}_0(\epsilon). \tag{3.27}$$

Here $\gamma$ is a path in the kinematic space from $x_0$ to $x$, and $\vec{g}_0(\epsilon) = \vec{g}(\epsilon, x_0)$ denotes the boundary values of the basis integrals at a specified point $x = x_0$. The expansion of path-ordered exponential in powers of $\epsilon$ gives

$$\vec{g}(\epsilon,x) = \left(\mathbf{1} + \epsilon \int_\gamma \tilde{A}(x') + \epsilon^2 \int_\gamma \tilde{A}(x')\tilde{A}(x'') + \ldots\right)\vec{g}_0(\epsilon). \tag{3.28}$$

It is evident that the solution at each order in $\epsilon$ takes the form of the *Chen's iterated integral* [137] (see section 3.5 for details). Additionally, each term in the series expansion consists of iterated integrals of the same depth, i.e., the number of integrations that need to be performed iteratively matches with the given $\epsilon$ order. The solution (3.28) for the basis integrals can be expressed as a Taylor expansion around $\epsilon = 0$

$$\vec{g}(\epsilon,x) = \sum_{i=0}^{\infty} \vec{g}^{(i)}(x)\epsilon^i. \tag{3.29}$$

Substituting this expansion back into the equation (3.19), we observe that due to the complete $\epsilon$-factorization on the right-hand side of equation (3.19), the solution for each basis integral at any $k^{th}$ order of $\epsilon$-expansion depends only on the solution at the $(k-1)^{th}$ order up to an integration constant,

$$\vec{g}^{(k)}(x) = \vec{g}^{(k)}(x_0) + \int_\gamma \tilde{A}(x')\vec{g}^{(k-1)}(x'), \tag{3.30}$$

$$\frac{\partial \vec{g}^{(0)}(x)}{\partial x} = 0 \quad \Leftrightarrow \quad \vec{g}^{(0)}(x) \equiv \vec{g}_0(x_0), \tag{3.31}$$

with the leading term of each basis integral a constant, determined by the boundary conditions. Thus, reformulating the differential equations into the canonical form simplifies the task of solving the basis integrals up to determining the boundary constants. The discussion on finding these boundary constants for a given differential equation system is postponed to section 3.7. Furthermore, to solve the system of differential equations, integrals are organized sector-wise, resulting in the lower block-triangular form of the matrices. A bottom-up approach is then employed afterwards to solve the differential equation system. The strategy involves starting with the equations for the master integrals of simpler topologies (those with fewer propagators) and utilizing these results to solve the equations for integrals of complicated topologies (those with more propagators).

In the solution for basis integrals given in equation (3.28), it is essential to note that the kernels of the iterated integrals constitute the one-forms appearing in the matrix $\tilde{A}$. Thus, the entries of matrix $\tilde{A}$ carry crucial information about the types of functions that can emerge in the solutions. For example, for the dlog-form of the differential equation system given in equation (3.21), the iterative kernels constitute algebraic logarithmic one-forms. In the cases when the logarithmic one-forms are the functions of rational letters only, the results for the basis integrals can always be expressed in terms of *Multiple Polylogarithms* (described in section 3.5), which are special cases of iterated integrals. However, in the cases where the entries of the matrix $\tilde{A}$ involve algebraic functions incorporating square roots, as observed in the case of the bubble integral (3.26), one seeks a change of variables $x_i \to y_i$ that simultaneously rationalizes all the occurring square roots. This coordinate transformation of variables, commonly known as a *base transformation*, proves instrumental in writing the results for the corresponding differential equation system in terms of polylogarithmic functions depending on the new variables $y_i$. In the literature, one can find numerous papers [138–145] that analyze systematic variable transformations to rationalize the occurring square roots, or to demonstrate the non-existence of such simultaneous rationalization. In Ref. [141], the authors provide a Mathematica package **`RationalizeRoots`** which can find a change of variables under certain conditions to rationalize the square roots. For example, in the case of the bubble integral given in equation (3.26), we can perform the coordinate transformation $x \to x'$ such that

$$x = -\frac{(1-x')^2}{x'}. \tag{3.32}$$

Putting it back into equation (3.26), we obtain the transformed system of differential equation

$$\frac{d}{dx'}\vec{J} = \epsilon \begin{pmatrix} 0 & 0 \\ \frac{1}{x'} & \frac{1}{x'} - \frac{2}{1+x'} \end{pmatrix} \vec{J}. \tag{3.33}$$

With $\epsilon$-independent matrix $\tilde{B}$ having all rational entries in new variable $x'$. In this form, we can write the differential equation system into *dlog-form* as

$$d\vec{J} = \epsilon \begin{pmatrix} 0 & 0 \\ d\log x' & d\log x' - 2\, d\log(x'+1) \end{pmatrix} \vec{J}. \tag{3.34}$$

with matrix $\tilde{B}$

$$\tilde{B}(x) = \begin{pmatrix} 0 & 0 \\ 1 & 1 \end{pmatrix} d\log x' + \begin{pmatrix} 0 & 0 \\ 0 & -2 \end{pmatrix} d\log\big(x'+1\big) \tag{3.35}$$

having alphabet consisting of two independent letters $\{x', (x'+1)\}$. Now we can easily solve this differential equation system order by order in $\epsilon$ in terms of Multiple polylogaritms (shown in section 3.6).

The presence of non-rationalizable square roots in the dlog-form of a differential equation system makes the general approach to integrate the system of differential equations in terms of MPLs impossible. However, it is often possible to express the results of the iterated integrals with dlog-forms containing algebraic arguments in terms of MPLs [143, 146–148], although it is not valid in general. For instance, in [149], the authors have provided a specific example of an iterated integral with dlog-forms which is non-expressible in terms of MPLs. In such scenarios, as well as the cases where the integration kernels in equation (3.28) are not logarithmic one-forms, a more general class of functions may be required to express the results. Moving beyond the one-loop level and with an increase in the number of mass scales, numerous examples emerge where in the analytic evaluation of Feynman integrals involving non-rationalizable square roots with degree $\geq 3$, results are expressed in terms of complicated functions such as *elliptic multiple-polylogarithms (eMPLs)* (iterated integrals over moduli space of torus) [123–126, 148, 150–169], and the iterated integrals with one-forms defined over complicated geometries such as *K3 surfaces* [139, 140, 144, 170, 171] and *Calabi-Yau geometries* [172–176]. The characteristics of these functions are not thoroughly understood as those in the context of MPLs.

In the context of the work presented in this thesis, our focus will be confined to solving the differential equation system in dlog-form, specifically involving algebraic functions with square roots. As mentioned earlier in this section, for the dlog-form of the differential equation system, the entries of the transformed matrix $\tilde{A}$, and hence the alphabet, determine the class of special functions needed for writing the results. Thus, in the next section, we will briefly discuss the definitions and properties of these special functions.

## 3.5 Special Functions and Their Properties

The standard dlog-form of the system of differential equations, as represented in equation (3.21), guarantees that the solutions for the basis integrals can be systematically expressed in terms of iterated integrals involving logarithmic kernels. In the subsequent section 3.5.1, a mathematical description of these iterated integrals is provided, along with an exploration of their basic characteristics. Due to the many useful properties of iterated integrals, working with them proves to be highly convenient in certain cases. However, for phenomenological applications, it is often desired to express computational results (whenever possible) in terms of special functions, which have been extensively studied in the literature. The unitarity property of the S-matrix demands the presence of special functions with branch cuts. The simplest example of a function with a branch cut is the logarithm; however, more complicated functions can also emerge in scattering amplitudes. In section 3.5.2, a discussion on the definitions and properties of multiple polylogarithms, which represent generalizations of logarithmic functions and play a crucial role in the context of scattering amplitudes, is presented.

### 3.5.1 Chen's Iterated Integrals

Let $\omega_1, \ldots, \omega_r$ denote a set of independent differential one-forms on a smooth $n$-dimensional manifold $M$. Consider a piecewise smooth path $\gamma : [0,1] \to M$ on the manifold $M$, where $\gamma(0)$ and $\gamma(1)$ represent the base point and end point, respectively. In the context of Feynman integrals, the manifold $M$ corresponds to the space of independent kinematic variables. Moreover, the pull-back of the differential one-forms $\omega_j$ to the interval [0,1] is defined as

$$\gamma^* \omega_j = f_j(\lambda) d\lambda. \tag{3.36}$$

For $\lambda \in [0,1]$, the $r$-fold (Chen's) iterated integral over the one-forms $\omega_j$ is defined as

$$I_\gamma(\omega_1, \ldots, \omega_r; \lambda) = \int_\gamma \omega_1 \ldots \omega_r = \int_0^\lambda d\lambda_1 \, f_1(\lambda_1) \int_0^{\lambda_1} d\lambda_2 \, f_2(\lambda_2) \cdots \int_0^{\lambda_{r-1}} d\lambda_r \, f_r(\lambda_r). \tag{3.37}$$

The 0-fold or an empty iterated integral is defined as $I_\gamma(;\lambda) = 1$. In the context of this thesis, the logarithmic one-forms are of particular interest, specifically denoted as $\omega_j = d \log p_j$.

Let us look at some basic properties satisfied by the iterated integrals [137]:

- The iterated integral $I_\gamma(\omega_1, \ldots, \omega_r; \lambda)$ is invariant under the parametrization of the path $\gamma$.
- Consider two paths $\gamma_1, \gamma_2 : [0,1] \to M$, such that $\gamma_1(0) = 0$, $\gamma_1(1) = \gamma_2(0)$, and $\gamma_2(1) = 1$. Then the integral over composed path $\gamma \equiv \gamma_1\gamma_2$ is obtained by sequentially integrating along $\gamma_1$ and then along $\gamma_2$, i.e.,

$$I_{\gamma_1\gamma_2}(\omega_1, \ldots, \omega_r; \lambda) = \sum_{i=0}^{r} I_{\gamma_1}(\omega_1, \ldots, \omega_i; \lambda) I_{\gamma_2}(\omega_{i+1}, \ldots, \omega_r; \lambda). \tag{3.38}$$

- If $\gamma^{-1} : [0,1] \to M$ denotes the path $\gamma$ traversed in reverse direction such that $\gamma^{-1}(\lambda) =$

$\gamma(1-\lambda)$, then the iterated integral along $\gamma^{-1}$ is given as

$$I_{\gamma^{-1}}(\omega_1,\ldots,\omega_r;1) = (-1)^r I_\gamma(\omega_r,\ldots,\omega_1;1). \tag{3.39}$$

- Chen's iterated integrals satisfy shuffle algebra relations

$$I_\gamma(\omega_1,\ldots,\omega_r;\lambda)I_\gamma(\omega_{r+1},\ldots,\omega_{r+s};\lambda) = \sum_{\sigma\in\sum(r,s)} I_\gamma(\omega_{\sigma(1)},\ldots,\omega_{\sigma(r+s)};\lambda), \tag{3.40}$$

where
$\sum(r,s) = \{\sigma \in \sum(r+s) : \sigma^{-1}(1) < \cdots < \sigma^{-1}(r)$ and $\sigma^{-1}(r+1) < \cdots < \sigma^{-1}(r+s)\}$,
and $\sum(r+s)$ is the set of permutations on $\{1,\ldots,r+s\}$.

### 3.5.2 Multiple Polylogarithms (MPLs)

In the context of Feynman integrals, multiple polylogarithms (MPLs) [177–181] constitute an important class of functions, as many Feynman integrals can be evaluated in terms of them. Moreover, there are many publicly available computer tools [182–184] for their manipulation and numerical evaluation with very high precision. The MPLs exhibit mainly two representations: iterated integral representation and nested sum representation, which we will be studying briefly in this section. Let us start with defining an ordinary logarithm,

$$Li_1(x) = -\ln(1-x) = \sum_{k=1}^{\infty}\frac{x^k}{k}, \tag{3.41}$$

and a dilogarithm as

$$Li_2(x) = \sum_{k=1}^{\infty}\frac{x^k}{k^2}. \tag{3.42}$$

The classical polylogarithms $Li_m(x)$, for any positive integer $m$ is defined as

$$Li_m(x) = \sum_{k=1}^{\infty}\frac{x^k}{k^m}. \tag{3.43}$$

This series representation converges for complex arguments $|x| \leq 1$ for $m \geq 2$, and for $|x| < 1$ when $m = 1$. The multiple polylogarithms, a generalization of classical polylogarithms, can be represented in the nested sum form as

$$Li_{m_1,\dots,m_k}(x_1,\dots,x_k) = \sum_{n_1>n_2>\dots>n_k>0}^{\infty} \frac{x_1^{n_1} x_2^{n_2} \dots x_k^{n_k}}{n_1^{m_1} n_2^{m_2} \dots n_k^{m_k}}, \tag{3.44}$$

which converges for

$$|x_1 \dots x_j| \leq 1 \;\; \forall\, j \in 1,\dots,k \;\; \text{and} \;\; (x_1, m_1) \neq (1,1). \tag{3.45}$$

Multiple zeta values are defined as a specific subclass of MPLs

$$\zeta_{m_1,\dots,m_k} = Li_{m_1,\dots,m_k}(1,\dots,1) = \sum_{n_1>n_2>\dots>n_k>0}^{\infty} \frac{1}{n_1^{m_1} n_2^{m_2} \dots n_k^{m_k}}. \tag{3.46}$$

The nested sum representation of MLPs can be highly useful for evaluating Feynman integrals using various techniques. However, when applying the differential equation method to solve Feynman integrals, the integral representation of MLPs proves to be more convenient. In the dlog-form of the differential equation system, given in equation (3.21), with all rational letters, it is always possible to express the integrable iterated integrals in terms of *Goncharov Polylogarithms* (GPLs). The integral representation of GPLs was first introduced by Goncharov in [178], and is defined recursively for $x_k \neq 0$ as:

$$G(x_1,\dots,x_k;y) = \int_0^y \frac{dt_1}{t_1 - x_1} \int_0^{t_1} \frac{dt_2}{t_2 - x_2} \cdots \int_0^{t_{k-1}} \frac{dt_k}{t_k - x_k}, \tag{3.47}$$

$$= \int_0^y \frac{dt_1}{t_1 - x_1} G(x_2,\dots,x_k;t_1), \tag{3.48}$$

with $G(;y) = 1$. In the context of multiple polylogarithms, $k$ denotes the transcendental degree of the GPL. The simplest examples of GPLs of transcendental degree 1 are:

$$G(-1;y) = \int_0^y \frac{dt}{t+1} = \int_0^y d\log(t+1), \qquad G(1;y) = \int_0^y \frac{dt}{t-1} = \int_0^y d\log(t-1). \tag{3.49}$$

The GPLs with trailing zeros, i.e., those having $x_k = 0$, are known to be divergent and can be defined after regularization through the following definition

$$G(\underbrace{0,\dots,0}_{k};y) = \frac{1}{k!}\log^k y. \tag{3.50}$$

The nested sum representation of MPLs is related to its integral representation as follows

$$Li_{m_1,\ldots,m_k}(x_1,\ldots,x_k) = (-1)^k G_{m_1,\ldots,m_k}\left(\frac{1}{x_1},\frac{1}{x_1x_2},\ldots,\frac{1}{x_1\ldots x_k};1\right). \tag{3.51}$$

The GPLs are well studied mathematical objects that exhibit a variety of useful mathematical properties, which we will be discussing here:

- Without any trailing zeros, GPLs satisfy the following scaling relations

  $$G(x_1,\ldots,x_k;y) = G(zx_1,\ldots,zx_k;zy), \tag{3.52}$$

  for all $z$ elements of the set of complex numbers excluding 0.

- Inherited from the iterated integral representation, GPLs satisfy the *shuffle algebra* [185]. A simple example of this is

  $$G(x_1,x_2;y).G(x_3;y) = G(x_1,x_2,x_3;y) + G(x_1,x_3,x_2;y) + G(x_3,x_1,x_2;y). \tag{3.53}$$

All these properties of MPLs are useful for systematically computing Feynman integrals to all orders in $\epsilon$. Especially, the shuffle algebra satisfied by these MPLs is of great importance in their numerical evaluation.

Since the early days of quantum field theory, it has been known that not all Feynman integrals can be expressed solely in terms of MPLs. Beyond the one-loop, the solutions of Feynman integrals give rise to new functions characterized by intricate mathematical structures. These functions are often associated with elliptic curves and higher-dimensional manifolds; the mathematical study of these functions is beyond the scope of work presented in this thesis.

## 3.6 Solution in Terms of Multiple Polylogarithms

In this section, we will outline the integration procedure for the one-loop bubble integral discussed in the preceding sections to obtain its result in terms of GPLs. The differential equation for this integral is derived in the dlog-form in equation (3.34) as a linear combination of two dlog differential one-forms

$$\omega_1 = d\log x', \quad \omega_2 = d\log\bigl(x'+1\bigr). \tag{3.54}$$

The integration is performed order-by-order in $\epsilon$ along a linear path $\gamma$ from some point $x'_0 > 0$ to $x' > x'_0$. Starting at order $\epsilon^{(0)}$ of equation (3.34), we find

$$d\vec{J}^{(0)} = 0 \implies \vec{J}^{(0)} = \vec{c}^{\,(0)}, \tag{3.55}$$

where $\vec{c}^0$ denotes the integration constants to be determined using boundary conditions. Utilizing this result in the next order $\epsilon^{(1)}$, we obtain

$$\vec{J}^{(1)} = \int_\gamma \begin{pmatrix} 0 & 0 \\ 1 & 1 \end{pmatrix} \vec{c}^{(0)} d\log x' + \int_\gamma \begin{pmatrix} 0 & 0 \\ 0 & -2 \end{pmatrix} \vec{c}^{(0)} d\log\bigl(x'+1\bigr) + \vec{c}^{(1)} \tag{3.56}$$

$$= \begin{pmatrix} c_1^{(1)} \\ (c_1^{(0)} + c_2^{(0)})G(0;x') - 2c_2^{(0)}G(-1;x') + c_2^{(1)} \end{pmatrix} \tag{3.57}$$

Before proceeding further, let us determine the integration constants. The fact that the derivative of the first integral $J_1 \equiv J_{20}$ with respect to $x'$ is zero, as reflected in equation (3.34), implies that for all orders in the $\epsilon$ expansion, we have

$$J_{20}^{(n)} = c_1^{(n)}, \quad n \geq 0. \tag{3.58}$$

This implies that the integral $J_{20}$ is solely determined by the boundary conditions. As this integral is just a tadpole, we can easily solve it using other methods. It is calculated in appendix C via direct integration yielding the result

$$J_1 \equiv J_{20} = \epsilon e^{\epsilon\gamma_E}\Gamma(\epsilon), \tag{3.59}$$

which fixes all the constants $c_1^{(n)}$. For instance, integral $J_1 \equiv J_{20}$ up to order $\epsilon^{(2)}$ provides

$$c_1^{(0)} = 1, \quad c_1^{(1)} = 0, \quad c_1^{(2)} = \frac{1}{2}\zeta_2. \tag{3.60}$$

To determine the integration constants for $J_2$, we utilize the fact that $J_2 \equiv J_{21}$ becomes zero when $x' = 1$, as indicated by equation (3.25). This condition is connected to the regularity of the integral $J_{21}$ for $p^2 = 0$. Thus, we have

$$J_2(x' = 1) = 0, \tag{3.61}$$

to fix the boundary constants for the integral $J_2$. For the zeroth order in $\epsilon$, this condition (3.61) yields $c_2^{(0)} = 0$. Using this result, along with equation (3.60) in equation (3.57), we get

$$J_2^{(1)} = G(0; x') + c_2^{(1)}. \tag{3.62}$$

The integration constant $c_2^{(1)}$ is fixed by evaluating (3.62) at $x' = 1$ utilizing (3.61). This gives

$$0 = J_2^{(1)}(x' = 1) \equiv G(0; 1) + c_2^{(1)} = \log(1) + c_2^{(1)} \implies c_2^{(1)} = 0. \tag{3.63}$$

Thus, at $\epsilon^{(1)}$ order for $J_2$, we have

$$J_2^{(1)} = G(0; x'). \tag{3.64}$$

To determine the results for $J_2$ and the boundary constants for the next orders in $\epsilon$, all the steps remain the same as above. Thus, we can find the result of $J_2$ to any desired order in $\epsilon$. The procedure for integration illustrated in this section can be generalized to cases with multiple loops and kinematic variables.

## 3.7 Boundary Conditions

Integration of the differential equation system after bringing it into the canonical form (3.19), determines the answer up to an integration constant as illustrated in section 3.4. This section is devoted to the discussion of how to fix these boundary constants, for which we need to follow these specific steps:

- One approach involves the initial analytic integration of the system of differential equations, which gives solutions for the integral basis expressed in terms of iterated integrals or other functional classes up to a boundary constant. Subsequently, these analytical results are computed numerically at a specific kinematic point and matched against the numerical results of the corresponding integrals evaluated at the same kinematic point using publicly available tools or other high-precision methods. This comparison helps determine the numerical values for the boundary constants. The next step involves retrieving analytical expressions corresponding to these numerical boundary values. Since the boundary constants depend on the choice of contour and are related to the alphabet and the boundary point, it is possible to infer which constants might appear in the analytic solutions. Using this information, the numerical values of the boundary constants are correlated with their respective analytical constants using the **PSLQ** algorithm introduced in section 3.1.

- Very often, the differential equations themselves give insights regarding the choice of the boundary conditions. The study of the singularity structure of the basis of Feynman integrals, in many cases, poses some constraints on the singularities present in the differential equation system. This information can be utilized to determine the boundary conditions without any explicit calculation, as demonstrated in the case of bubble integral.

- At specific kinematic points, the basis integrals may vanish. Opting for these particular kinematic points as the boundaries provides us with the corresponding boundary constants for those integrals. Additionally, at a given kinematic point, the basis integrals may reduce into more simple integrals, such as products of one-loop integrals. These simplified integrals can be analytically computed with ease, employing techniques like *Feynman parametrization* or other methods, facilitating the determination of the boundary constants.

- Additionally, boundary constants for the system of differential equations can be determined by evaluating the corresponding integrals in specific asymptotic limits employing techniques like *method of expansion by regions* [186, 187]. This approach, formulated using *alpha parametrization*, is implemented in the publicly accessible codes **asy.m** [188, 189], which is a part of **FIESTA** [100] and **ASPIRE** [190]. It is extensively utilized for determining boundary constants.

This topic concludes this chapter, where I have presented all the necessary tools required for the analytic computation of multi-loop Feynman integrals using the method of differential equations. We will employ these techniques for the analytic evaluation of two-loop master integrals contributing to the mixed QCD-electroweak corrections to the $H \to ZZ^*$ decay in chapter 5.

4

# Precision Studies of $H \to 4\ell$ Decay

HAVING established the necessary theoretical framework, this chapter delves into the central theme of this thesis: the computation of mixed QCD-electroweak corrections to the partial decay width of the Golden decay channel of the Higgs boson, specifically $H \to e^+e^-\mu^+\mu^-$. This channel is of great phenomenological importance due to its clean experimental signature, providing a unique opportunity to study the properties of the Higgs boson and gain deeper insights into the mechanisms of the universe through electroweak symmetry breaking, particularly due to the direct measurements of the $HZZ$ coupling, as discussed in chapter 1. Here, we discuss a portion of the collaborative paper [67] with M. Mahakhud, A. Shivaji, and X. Zhao, focusing on key calculation ingredients for $\mathcal{O}(\alpha\alpha_s)$ mixed QCD-electroweak corrections to the partial decay width of $H \to e^+e^-\mu^+\mu^-$ channel. The chapter is structured as follows: Section 4.1 details the kinematics and notations relevant to the calculation, in the context of the leading-order contribution to the $H \to e^+e^-\mu^+\mu^-$ decay. Section 4.2, provides a brief overview of the state-of-the-art studies of $\mathcal{O}(\alpha)$ electroweak corrections at the next-to-leading order for the Higgs decay into four fermions. Subsequently, section 4.3, categorizes the $\mathcal{O}(\alpha\alpha_s)$ amplitude into its real and virtual parts, discussing how to organize the calculation of the corresponding matrix elements. Finally, we conclude this chapter by delving into the systematic calculation of the matrix elements for the $\mathcal{O}(\alpha\alpha_s)$ corrections to the $HV_1V_2$ $(V_1, V_2 = Z, \gamma)$ vertex, which is the key bottleneck of the presented calculations. We express these matrix elements in terms of scalar functions known as form-factors and discuss their reduction into a linear combination of a set of 41 master integrals.

## 4.1 Leading Order Contribution

In the SM, the leading order (LO) contribution to the on-shell decay of the Higgs boson into two pairs of charged leptons with different flavors, i.e., $H \to e^+e^-\mu^+\mu^-$, mediated by a pair of $Z$-bosons, comes from a single tree-level diagram, as shown in figure 4.1. Due to energy conservation, at least one of the two $Z$-bosons has to be off-shell, depicted by $Z^*$. We consider a more general case by treating both the $Z$-bosons off-shell. The following momenta assignments for the particles in the decay are chosen.

$$H(q) \to Z^{(*)}(p_1)Z^{(*)}(p_2) \to e^+(q_1)e^-(q_2)\mu^+(q_3)\mu^-(q_4), \tag{4.1}$$

where momentum conservation requires $q = (p_1 + p_2)$, $p_1 = (q_1 + q_2)$ and $p_2 = (q_3 + q_4)$. In our calculation, we are neglecting the masses of the final-state leptons. Various scalar products of interest are given by,

$$q_1^2 = q_2^2 = q_3^2 = q_4^2 = 0, \quad p_1^2 = 2q_1.q_2, \quad p_2^2 = 2q_3.q_4, \quad q^2 = M_H^2, \tag{4.2}$$

where $M_H$ is the Higgs mass. The leading order matrix element squared, $|\overline{M_0}|^2$, averaged over initial and summed over final state spin configurations for the decay $H \to Z^{(*)}Z^{(*)} \to e^+e^-\mu^+\mu^-$, is given as

$$|\overline{M_0}|^2 = e^4 \times \frac{4M_Z^4}{v^2} \times \frac{1}{(s_{12}^2 - M_Z^2)^2 + M_Z^2\Gamma_Z^2} \times \frac{1}{(s_{34}^2 - M_Z^2)^2 + M_Z^2\Gamma_Z^2} \times$$
$$\left\{ \left[(g_V^\ell)^4 + (g_A^\ell)^4\right] (32\, s_{14}s_{23} + 32\, s_{13}s_{24}) + (g_V^\ell g_A^\ell)^2 \, (-64\, s_{14}s_{23} + 192\, s_{13}s_{24}) \right\}, \quad (4.3)$$

where $e$ is the electromagnetic charge, the term $\frac{4M_Z^4}{v^2}$ corresponds to the $HZZ$ vertex as derived in chapter 1, third and fourth terms correspond to the squared $Z$-boson propagators, $\Gamma_Z$ is the decay width for the $Z$-boson, and $s_{ij} = q_i.q_j$ denotes the dot product of the momenta of the final state particles $q_i$ and $q_j$.

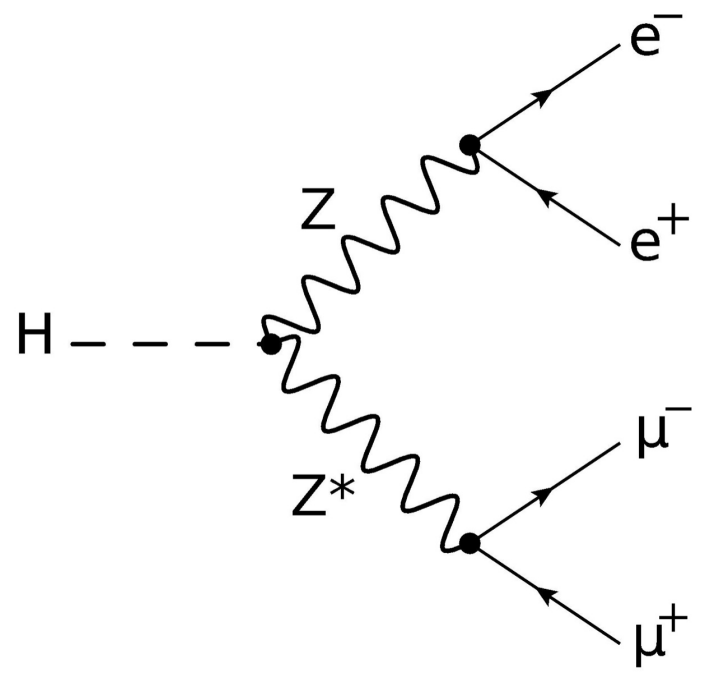

FIGURE 4.1: LO Feynman diagram contributing to $H \to e^+e^-\mu^+\mu^-$.

The terms $g_{V\ell}$ and $g_{A\ell}$ denote the vector and axial-vector couplings of the $Z$-boson to the leptons given as

$$g_V^\ell = \frac{1}{2\sin\theta_W \cos\theta_W}(I_{W,\ell}^3 - 2Q_\ell \sin^2\theta_W), \qquad g_A^\ell = \frac{1}{2\sin\theta_W \cos\theta_W} I_{W,\ell}^3. \quad (4.4)$$

Here $I_{W,\ell}^3$ is the third component of the weak isospin of left-handed leptons. For charged leptons, $I_{W,\ell}^3 = -1/2$, and $Q_\ell$ is the charge value relative to the electric charge of the proton, it is taken to be $-1$.

The corresponding leading-order partial decay width for $H \to e^+e^-\mu^+\mu^-$, denoted by $\Gamma^{\text{LO}}$, is calculated by integrating the squared LO matrix element over the four-body phase space of the final state leptons.

The expression for $\Gamma^{\text{LO}}$ is given by

$$\Gamma^{\text{LO}}(H \to e^+e^-\mu^+\mu^-) = \frac{1}{2M_H} \int d\Phi_4 \, |\overline{M_0}|^2, \quad (4.5)$$

with $d\Phi_4$ denoting the Lorentz-invariant four-body phase space factor for the decay, and is defined as

$$d\Phi_4 = \left( \prod_{i=1}^{4} \frac{d^3\vec{q}_i}{(2\pi)^3 2E_i} \right) (2\pi)^4 \delta^{(4)} \left( q - \sum_{i=1}^{4} q_i \right), \quad (4.6)$$

where $E_i$ denotes the energy of the final state leptons, and the four-dimensional delta function ensures the conservation of energy and momentum.

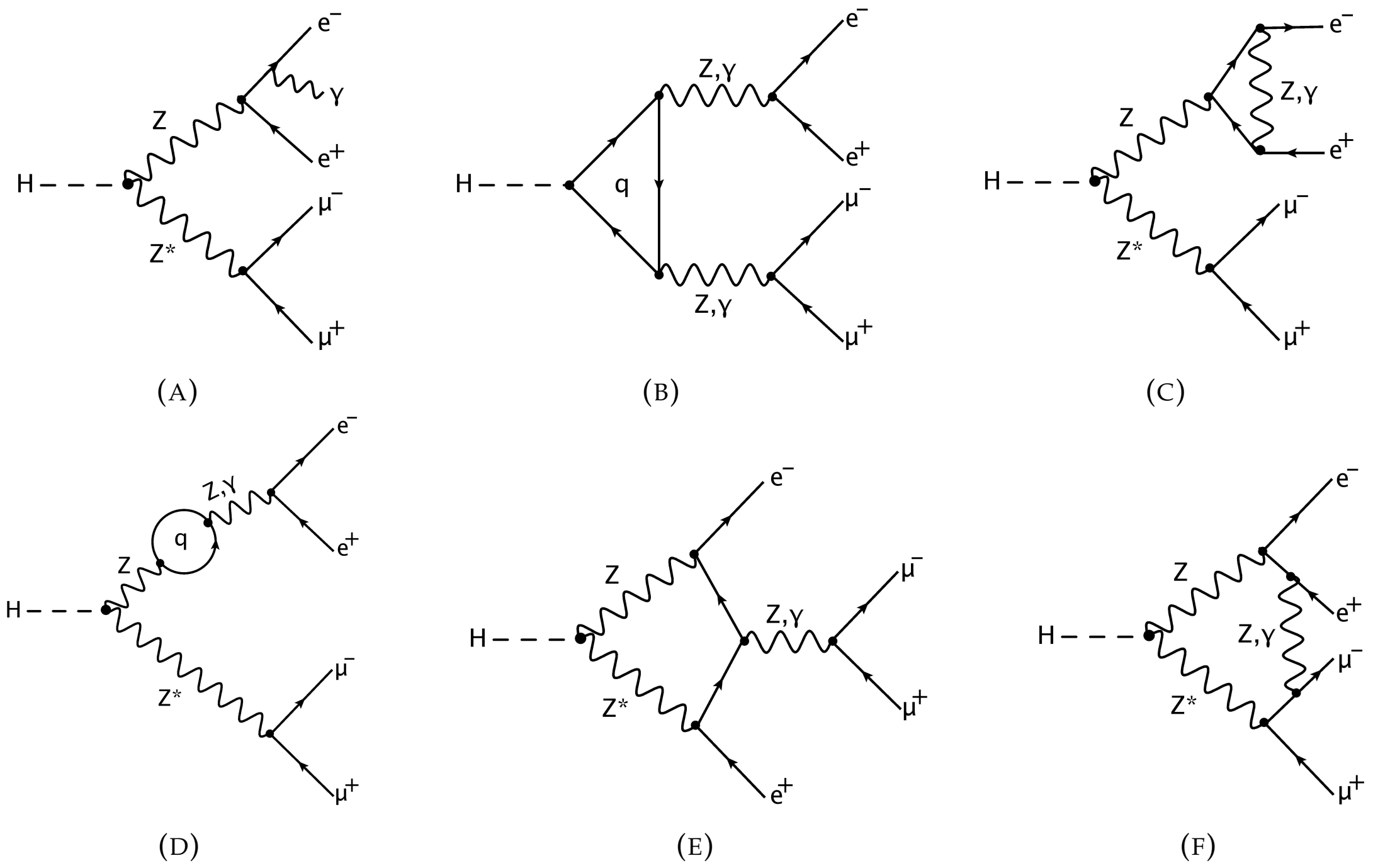

FIGURE 4.2: Representative real and virtual Feynman diagrams contributing to $\mathcal{O}(\alpha)$ corrections to $H \to 2e2\mu$. Diagrams include (A) real corrections, (B, C) vertex corrections, (D) self-energy corrections, (E) box corrections, and (F) pentagon corrections, where $q$ denotes the quark loop.

## 4.2 Next-to-Leading Order Corrections

In the context of $H \to Z^{(*)}Z^{(*)} \to 4\ell$ ($\ell = e, \mu$), the color-neutral nature of the particles involved at the leading order (LO) restricts the next-to-leading order (NLO) corrections to be purely electroweak (EW). These NLO corrections, being of $\mathcal{O}(\alpha)$ relative to the LO contribution, consist of two main components: virtual one-loop corrections and real corrections. The virtual part of the NLO amplitude denoted as $M_{virt.}^{\text{QED+weak}}$ receives contributions from the EW sector, i.e., both pure weak and QED sectors of the SM. Whereas the real correction part receive contributions only from the QED sector and is denoted as $M_{real}^{\text{QED}}$. The Feynman diagrams contributing to the $\mathcal{O}(\alpha)$ real corrections involve the inclusive emission of a photon from each final-state lepton leg (see figure 4.2 (A)). On the other hand, the diagrams contributing to the virtual part of the $\mathcal{O}(\alpha)$ EW amplitude can be classified based on their topology, including vertex, self-energy, box, and pentagon corrections at the one-loop level. The representative diagrams are shown in figure 4.2.

The $\mathcal{O}(\alpha)$ EW corrected partial decay width for the Higgs decay into four charged leptons

can be written as

$$\Gamma^{\rm NLO}(H \to 4\ell) = \int d\Gamma^{\rm LO} + \left\{ \int d\Gamma^{\rm QED+weak}_{virt.} + \int d\Gamma^{\rm QED}_{real} \right\}, \tag{4.7}$$

with

$$\int d\Gamma^{\rm QED+weak}_{virt.} = \frac{1}{2M_H} \int d\Phi_4 \sum_{s_1,s_2,s_3,s_4} 2\,{\rm Re}\Big[ M^{\rm QED+weak}_{virt.} (M_0)^* \Big], \tag{4.8}$$

$$\int d\Gamma^{\rm QED}_{real} = \frac{1}{2M_H} \int d\Phi_5 \sum_{s_1,s_2,s_3,s_4} \sum_{\rm helicity} |M^{\rm QED}_{real}|^2, \tag{4.9}$$

where the first sum in both the above equations is over the spins of the final state leptons, $d\Phi_5$ is the phase-space element corresponding to the four leptons and one photon in the final state, and the second sum in the expression for $\int d\Gamma^{\rm QED}_{real}$ is over the helicities of the emitted real photon. The detailed calculation of these $\mathcal{O}(\alpha)$ corrections to the partial decay width of the Higgs decay into four charged leptons can be found in reference [191].

The state-of-art of these $\mathcal{O}(\alpha)$ corrections to the partial decay width of the Higgs decay into four fermions is as follows. The exact one-loop QED corrections of $\mathcal{O}(\alpha)$ to the Higgs decay into four leptons with the off-shell $Z$-bosons have been evaluated in [192, 193]. The complete one-loop electroweak corrections for the leptonic, semi-leptonic, and hadronic final states, and the one-loop QCD corrections for the semi-leptonic and hadronic final states to the decay $H \to WW/ZZ \to 4f$ have already been evaluated in [194, 195] and are encoded in a Monte Carlo (MC) code **Prophecy4f** [45, 196]. The $\mathcal{O}(\alpha)$ corrections to the partial decay width, reported for the case of four charged leptons in the final state, are of the order of 2-4% for moderate Higgs masses ($M_H \leq 200$ GeV) and increase with the growing Higgs mass, reaching up to 14%. In addition to that, the one-loop electroweak and QCD corrections to the Higgs decay into four fermions in the context of a simple extension of the SM have also been studied and are implemented in the code **Prophecy4f** [197–199]. The next-to-leading order (NLO) electroweak corrections to the Higgs decay into charged leptonic final states $H \to Z^{(*)}Z^{(*)} \to 4\ell$ with $4\ell = 4e,\ 4\mu,\ 2e2\mu$ matched with QED Parton Shower (PS), have also been calculated, for which the results are available in a public event generator, **Hto4l** [46].

## 4.3 Next-to-Next-to Leading Order Corrections

In contrast to the NLO corrections to the amplitude, which receive contributions only from the EW sector, the amplitude for $H \to e^+e^-\mu^+\mu^-$ at the two-loop level receives contributions from both the EW and quantum chromodynamics (QCD) sectors of the SM. Thus, the amplitude for $H \to Z^{(*)}Z^{(*)} \to e^+e^-\mu^+\mu^-$ in the perturbative expansion up to two-loop order can be written as,

$$M_{\text{Total}} = M_0 + M_1^{(\alpha)} + (M_2^{(\alpha^2)} + M_2^{(\alpha\alpha_s)}) + \dots \tag{4.10}$$

Here $M_1$ and $M_2$ are the one-loop and two-loop amplitudes respectively. Among the contributions of $\mathcal{O}(\alpha^2)$ and $\mathcal{O}(\alpha\alpha_s)$ at the two-loop level, one can neglect the contributions of $\mathcal{O}(\alpha^2)$ due to the smallness of the EW coupling $\alpha$ in comparison to the strong coupling $\alpha_s$. Thus, we only consider the mixed QCD-electroweak corrections of $\mathcal{O}(\alpha\alpha_s)$ at the two-loop level and focus on the evaluation of $M_2^{\alpha\alpha_s}$.

Within all the EW particles participating in $\mathcal{O}(\alpha)$ corrections to the amplitude at NLO, only quarks are susceptible to couple with the gluons and take part in QCD corrections. Therefore, among the contributing diagrams at $\mathcal{O}(\alpha)$ (shown in figure 4.2), only the ones with quark loop will contribute at $\mathcal{O}(\alpha\alpha_s)$ along with the gluon dressing. Similar to the NLO corrections, these $\mathcal{O}(\alpha\alpha_s)$ corrections to the amplitude consist of two components: virtual two-loop corrections and real corrections. The following sections provide detailed descriptions of these two components.

### 4.3.1 Real Corrections

For the real part of the $\mathcal{O}(\alpha\alpha_s)$ amplitude, we need to consider the inclusive emission of a gluon. The possible Feynman diagrams for real emission can be obtained by attaching a gluon to the closed quark loop, as shown in figure 4.3. Due to a closed fermionic loop, the amplitudes of these diagrams are proportional to the trace over $T^a$, the generator of the $SU(3)$ gauge group. Since $\text{tr}(T^a) = 0$, diagrams for the real corrections give zero.

Furthermore, these real emission diagrams can only contribute at the amplitude-squared level, which corresponds to an $\mathcal{O}(\alpha^2\alpha_s)$ effect with respect to the LO. Therefore, the real emission diagrams do not contribute at $\mathcal{O}(\alpha\alpha_s)$.

### 4.3.2 Virtual Corrections

At $\mathcal{O}(\alpha\alpha_s)$, the virtual corrections to the amplitude can be divided into three categories as follows

$$M_2^{\alpha\alpha_s} = \delta M_{S.E.}^{\alpha\alpha_s} + \delta M_{Z\ell\bar{\ell}}^{\alpha\alpha_s} + \delta M_{HV_1V_2}^{\alpha\alpha_s} , \tag{4.11}$$

where $\delta M_{S.E.}^{\alpha\alpha_s}$ contains corrections due to the self-energy insertions on the vector-boson legs, $\delta M_{Z\ell\bar{\ell}}^{\alpha\alpha_s}$ appears due to $\mathcal{O}(\alpha\alpha_s)$ counter-term for the $Z\ell\bar{\ell}$ vertex, and $\delta M_{HV_1V_2}^{\alpha\alpha_s}$ consists of corrections to the $HV_1V_2$ vertex, respectively. The contributing virtual diagrams are generated with the help of the **QGRAF** [200] package. The amplitude for each of these diagrams is organized in **FORM** [201, 202] and is manipulated using **Mathematica**. We will describe the main features of the various virtual contributions below.

#### Self-energy corrections

The self-energy part of the two-loop amplitude, denoted by $\delta M_{S.E.}^{\alpha\alpha_s}$, receives contributions from $\mathcal{O}(\alpha\alpha_s)$ self-energy corrections to the $Z$-boson propagators. Since at the tree level, the process is mediated by $Z$-bosons, we must account for the $Z-\gamma$ transitions on going beyond the tree-level. Thus, the amplitude for self-energy corrections, $M_{S.E.}^{\alpha\alpha_s}$, can be expressed as

$$M_{S.E.}^{\alpha\alpha_s} = M_{e,ZZ}^{\alpha\alpha_s} + M_{e,Z\gamma}^{\alpha\alpha_s} + M_{\mu,ZZ}^{\alpha\alpha_s} + M_{\mu,Z\gamma}^{\alpha\alpha_s}. \tag{4.12}$$

The subscripts $e$ and $\mu$ are used to differentiate between the self-energy corrections of the two intermediate $Z$-bosons, where $e$ corresponds to the $Z$-boson decaying into $e^+e^-$ pairs, and $\mu$ corresponds to the $Z$-boson decaying into $\mu^+\mu^-$ pairs. Additionally, the other subscript distinguishes between the diagonal $(Z-Z)$ and non-diagonal $(Z-\gamma)$ transitions. A total of 72 self-energy diagrams contribute to the $\delta M_{S.E.}^{\alpha\alpha_s}$, some of which are depicted in figure 4.4. This part of the amplitude receives contributions from both the top quark and the light quarks. The two-point function $\Gamma^{VV'}_{\rho\sigma}$ $(VV' = ZZ, Z\gamma)$ corresponding to the gauge-boson self-energy can be decomposed into the transverse ($\Sigma^{VV'}_{\mathrm{T}}$) and longitudinal ($\Sigma^{VV'}_{\mathrm{L}}$) part as follows

$$\Gamma^{VV'}_{\rho\sigma}(q^2) = \left(g_{\rho\sigma} - \frac{q_\rho q_\sigma}{q^2}\right) \Sigma^{VV'}_{\mathrm{T}}(q^2) + \frac{q_\rho q_\sigma}{q^2} \Sigma^{VV'}_{\mathrm{L}}(q^2), \tag{4.13}$$

where $q^2$ represents the virtuality of the gauge-bosons $V$ and $V'$, while $\rho$ and $\sigma$ are the Lorentz indices associated with these gauge-bosons. Due to the presence of massless leptons in the final state, only the transverse part of those self-energies ($\Sigma^{V'V}_{\mathrm{T}}(q^2)$) contribute to the amplitude at $\mathcal{O}(\alpha\alpha_s)$. To compute $\delta M^{\alpha\alpha_s}_{S.E.}$, we need the $\mathcal{O}(\alpha\alpha_s)$ expressions for these transverse parts of the gauge-boson self-energies, which are available in reference [203][1], in terms of scalar functions denoted by $\Pi^{V,A}_T$ and are computed assuming zero mass for all quarks other than the top quark. For completeness, the necessary $\mathcal{O}(\alpha\alpha_s)$ expressions for the transverse part of the gauge-boson self-energies are provided in appendix B.

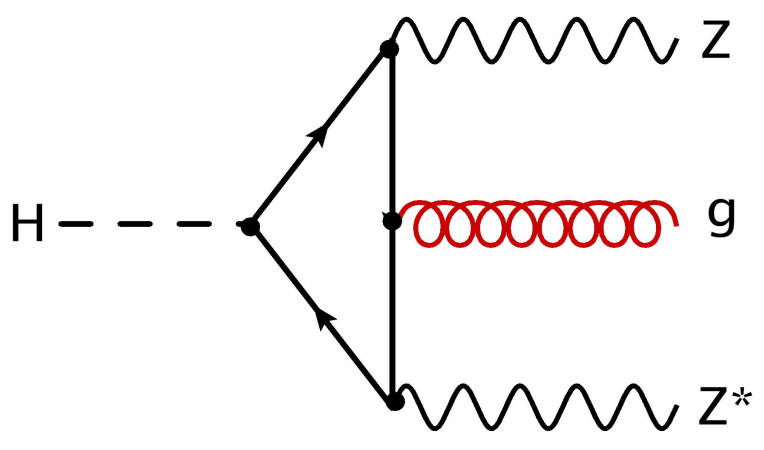

FIGURE 4.3: Representative diagram for real corrections to the amplitude for $H \to ZZ^*$.

### $Z\ell\bar{\ell}$ vertex correction

The two-loop amplitude also receives a contribution from the two diagrams shown in figure 4.5 involving a $\mathcal{O}(\alpha\alpha_s)$ counterterm at each $Z\ell\bar{\ell}$ vertex. The corresponding $\mathcal{O}(\alpha\alpha_s)$ amplitude can be written as

$$\delta M^{\alpha\alpha_s}_{Z\ell\bar{\ell}} = M^{\alpha\alpha_s}_{Ze^+e^-} + M^{\alpha\alpha_s}_{Z\mu^+\mu^-}\,. \tag{4.14}$$

The counterterm for the leptonic gauge-boson decay only receives contributions from the gauge-boson self-energies at $\mathcal{O}(\alpha\alpha_s)$. Thus, $Z\ell\bar{\ell}$ vertex counterterm depends on the gauge-boson wave-function renormalization constants and the renormalization constants of electromagnetic charge and the weak mixing angle [204]. The $\mathcal{O}(\alpha\alpha_s)$ expression for $Z\ell\bar{\ell}$ vertex counterterm, denoted as $\delta^{ct,(\alpha\alpha_s)}_{Z\ell\bar{\ell}}$, reads as

$$\delta^{ct,(\alpha\alpha_s)}_{Z\ell\bar{\ell}} = (g_{V\ell}\mathbb{1} - g_{A\ell}\gamma_5)\left[\delta Z_{e,(\alpha\alpha_s)} + \frac{s_W^2 - c_W^2}{c_W^2}\frac{\delta s_{W,(\alpha\alpha_s)}}{s_W} + \frac{1}{2}\delta Z_{ZZ,(\alpha\alpha_s)}\right] - \frac{2s_W}{c_W}Q_\ell\frac{\delta s_{W,(\alpha\alpha_s)}}{s_W} - \frac{Q_\ell}{2}\delta Z_{\gamma Z,(\alpha\alpha_s)} \tag{4.15}$$

where $s_W = \sin\theta_W$ and $c_W = \cos\theta_W$. Here, $\delta s_{W,(\alpha\alpha_s)}$ represents the renormalization constant for the weak mixing angle, $\delta Z_{ZZ,(\alpha\alpha_s)}$ and $\delta Z_{\gamma Z,(\alpha\alpha_s)}$ are the wave-function renormalization constants for the $Z$-boson and the photon-$Z$-boson mixing, respectively, and $\delta Z_{e,(\alpha\alpha_s)}$ is the charge renormalization constant, all computed at $\mathcal{O}(\alpha\alpha_s)$.

[1]The expression for $\Pi^{V,A}_T$ in the special case of one vanishing fermion mass has a typo, which is corrected in [204]

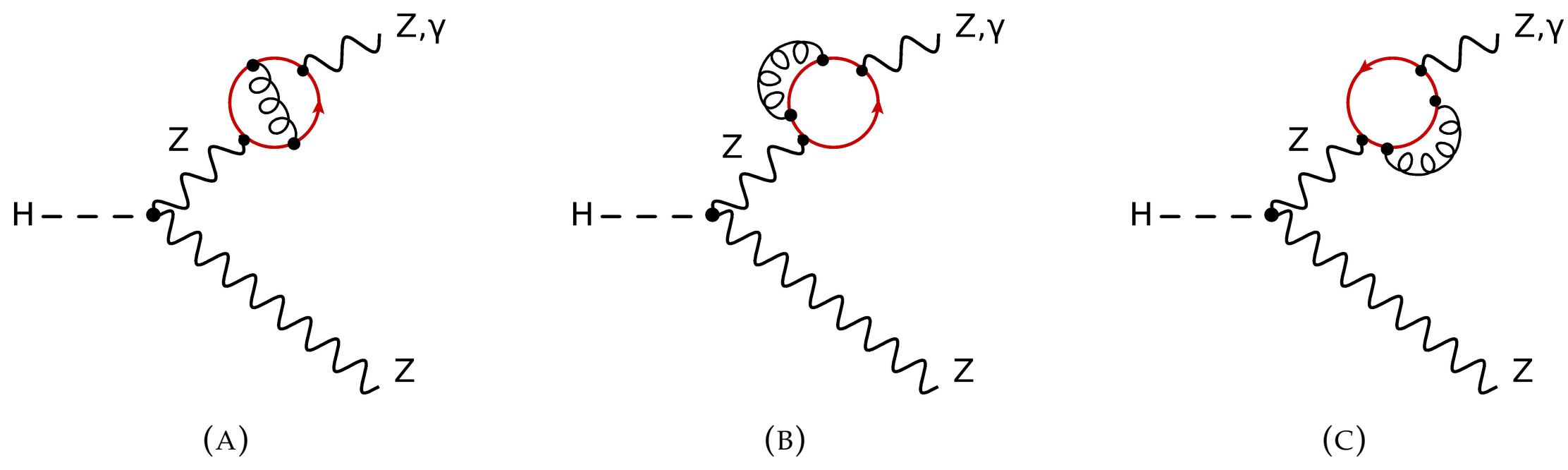

FIGURE 4.4: Representative diagrams contributing to $\delta M^{\alpha\alpha_s}_{S.E.}$ with a quark running in the loop. In these diagrams, five light quark flavors ($u, d, c, s$ and $b$) and the top quark contribute. The light quarks are taken to be massless.

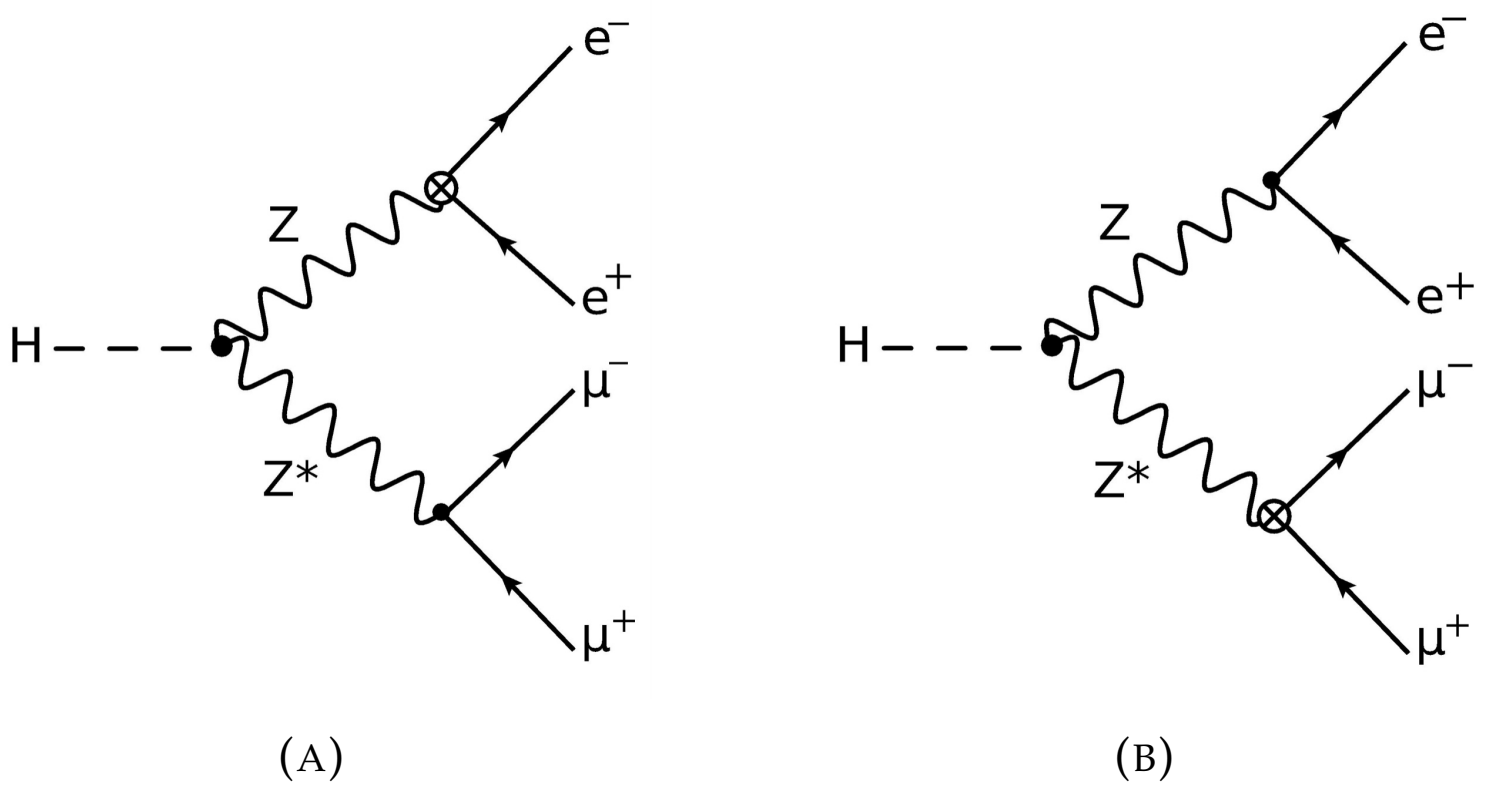

FIGURE 4.5: Tree level diagrams involving $\mathcal{O}(\alpha\alpha_s)$ $Z\ell\bar{\ell}$ vertex counterterm denoted by a crossed circle.

These various renormalization constants can be expressed in terms of gauge-boson self-energies and are discussed in chapter 6.

The pure electroweak nature of the $Z\ell\bar{\ell}$ vertex does not allow any kind of QCD corrections of $\mathcal{O}(\alpha\alpha_s)$; therefore, this counterterm for the $Z\ell\bar{\ell}$ vertex is non-divergent in nature. One must be careful about the renormalization scheme (discussed in chapter 6) in which the vertex correction is computed.

### $HV_1V_2$ vertex corrections

While at the LO, the decay of Higgs into $e^+e^-\mu^+\mu^-$ proceeds through the two intermediate $Z$-bosons, beyond the LO, in addition to the $Z^{(*)}Z^{(*)}$ channel, we also have to consider the loop induced contributions coming from the $Z^{(*)}\gamma^{(*)}$, and $\gamma^{(*)}\gamma^{(*)}$ channels. Thus, the most general amplitude for the $HV_1V_2$ vertex corrections denoted by $\delta M^{\alpha\alpha_s}_{HV_1V_2}(V_1, V_2 = Z, \gamma)$, where both $V_1$

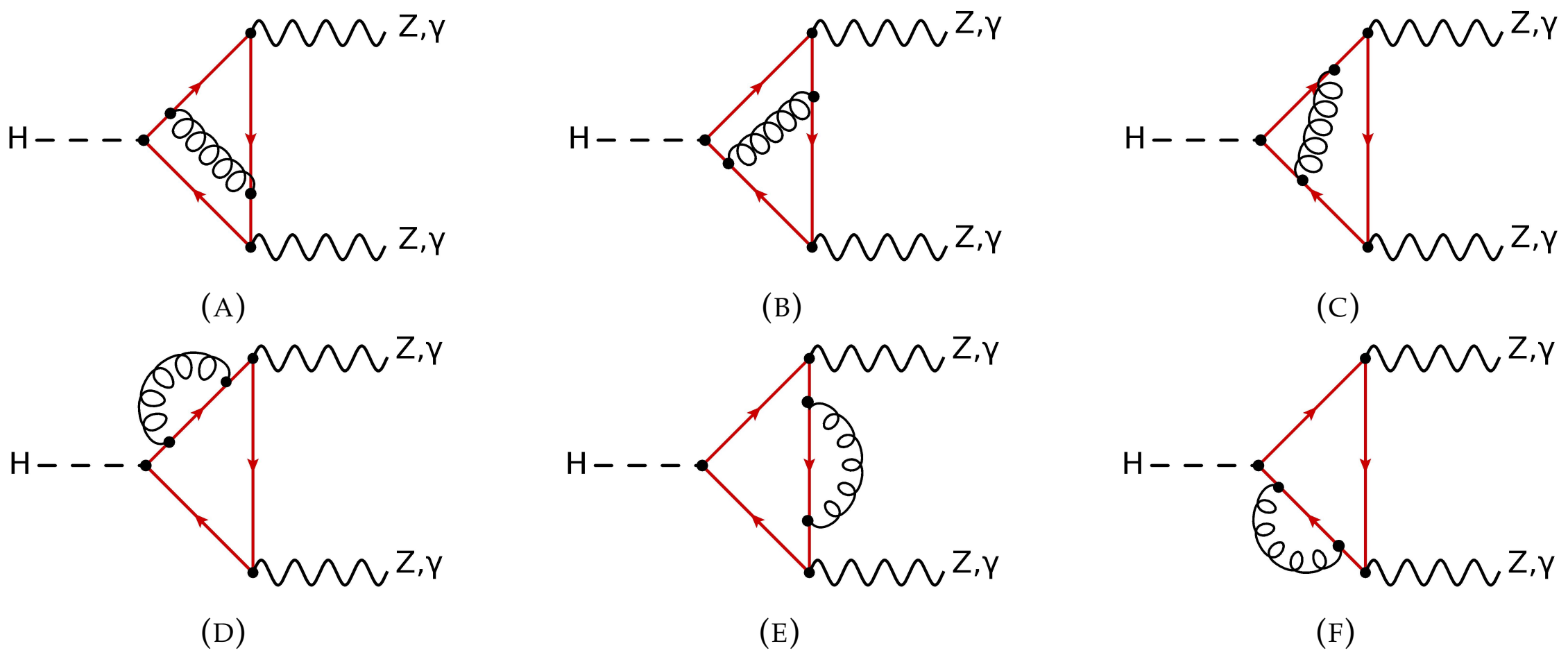

FIGURE 4.6: Representative two-loop triangle diagrams contributing to the bare amplitude of $H \to V_1V_2$ $(V_1, V_2 = Z, \gamma)$ with the top quark running in the loop. Diagrams with the reversed direction of the fermionic current are not shown.

and $V_2$ are taken off-shell, has the four contributions given as

$$\delta M^{\alpha\alpha_s}_{HV_1V_2} = \delta M^{\alpha\alpha_s}_{HZZ} + \delta M^{\alpha\alpha_s}_{HZ\gamma} + \delta M^{\alpha\alpha_s}_{H\gamma Z} + \delta M^{\alpha\alpha_s}_{H\gamma\gamma}. \tag{4.16}$$

In the above, the $HZ\gamma$ and $H\gamma Z$ contributions are written explicitly to take care of the fact that $e^+e^-$ can come from either $Z$ or $\gamma$. As the leptonic decay of the vector bosons is not affected by the QCD corrections to the $HV_1V_2$ vertex. Therefore, we can decompose the amplitude for the $HV_1V_2$ vertex corrections as

$$\delta M_{HV_1V_2} = M^{\mu\nu} J_\mu(p_1) J_\nu(p_2), \tag{4.17}$$

where, $M^{\mu\nu}$ is the two-loop amplitude for $H(q) \to V_1(p_1)V_2(p_2)$ decay; $J_\mu(p_1)$ and $J_\nu(p_2)$ are the fermionic currents corresponding to $V_1 \to e^+e^-$ and $V_2 \to \mu^+\mu^-$ respectively. Since the coupling of the top quark with the Higgs is the largest among all the quark flavors, we neglect the contributions from diagrams with any quark other than the top quark in the loop. There are in total 48 two-loop triangle diagrams contributing to the $H(q) \to V_1(p_1)V_2(p_2)$ decay at $\mathcal{O}(\alpha\alpha_s)$, out of which some representative diagrams are shown in figure 4.6.

Instead of using conventional methods, the *projector technique* (refer to [205] and references therein for details), based on Lorentz covariant tensor decomposition, has opted for a more systematic evaluation of the amplitude. In this technique, a general scattering amplitude can be expressed as a linear combination of scalar functions called *form-factors* multiplying all possible Lorentz and Dirac tensor structures. This decomposition incorporates physical constraints such as the Ward identities, transversality conditions, and symmetries of the problem under consideration.

Using the projector technique, the amplitude for $H \to V_1V_2$ can be expressed in its most general form using Lorentz covariance as

$$M^{\mu\nu} = (A\, g^{\mu\nu} + B\, p_1^{\nu}p_2^{\mu} + C\, \epsilon^{\mu\nu p_1 p_2} + D\, p_1^{\nu}p_1^{\mu} + E\, p_2^{\nu}p_2^{\mu} + F\, p_1^{\mu}p_2^{\nu}), \tag{4.18}$$

where $A$, $B$, $C$, $D$, $E$ and $F$ are the form-factors and $\epsilon^{\mu\nu p_1 p_2} = \epsilon^{\mu\nu\rho\sigma}p_{1\rho}p_{2\sigma}$. To extract these form factors, we construct suitable projectors $P^i_{\mu\nu}$ $(i = A, B, C, D, E, F)$ as a linear combination of the same basis tensors as $M^{\mu\nu}$ in the equation above. These projectors are constructed such that when applied to the two-loop amplitude $M^{\mu\nu}$ of any contributing diagram shown in figure 4.6, they project out a specific tensor structure. For example

$$P^A_{\mu\nu}M^{\mu\nu} = A. \tag{4.19}$$

The required projectors are formulated in $d$ dimensions and are expressed as:

$$\begin{aligned}
P^A_{\mu\nu} &= \frac{1}{d-2}\left(g_{\mu\nu} + \frac{p_2^2 p_{1\mu}p_{1\nu} + p_1^2 p_{2\mu}p_{2\nu} - (p_1.p_2)(p_{1\nu}p_{2\mu} + p_{1\mu}p_{2\nu})}{(p_1.p_2)^2 - p_1^2p_2^2}\right),\\
P^B_{\mu\nu} &= \frac{1}{(d-2)((p_1.p_2)^2 - p_1^2p_2^2)^2}\left((p_1.p_2)^3\left(\frac{p_1^2p_2^2}{(p_1.p_2)^2} - 1\right)g_{\mu\nu} + (d-1)(p_1.p_2)^2 p_{1\mu}p_{2\nu}\right.\\
&\quad \left. - (d-1)(p_1.p_2)(p_2^2 p_{1\mu}p_{1\nu} + p_1^2 p_{2\mu}p_{2\nu}) + \left((p_1.p_2)^2 + (d-2)p_1^2p_2^2\right)p_{1\nu}p_{2\mu}\right),\\
P^C_{\mu\nu} &= \frac{\epsilon_{\mu\nu p_1 p_2}}{(d-2)(d-3)((p_1.p_2)^2 - p_1^2p_2^2)},\\
P^D_{\mu\nu} &= \frac{1}{(d-2)((p_1.p_2)^2 - p_1^2p_2^2)^2}\left((p_1.p_2)^2 p_2^2\left(1 - \frac{p_1^2p_2^2}{(p_1.p_2)^2}\right)g_{\mu\nu} + (p_2.p_2)^2(d-1)p_{1\mu}p_{1\nu}\right.\\
&\quad \left. - (d-1)p_2^2(p_1.p_2)(p_{1\mu}p_{2\nu} + p_{1\nu}p_{2\mu}) + (p_1^2p_2^2 + (d-2)(p_1.p_2)^2)p_{2\mu}p_{2\nu}\right),\\
P^E_{\mu\nu} &= \frac{1}{(d-2)((p_1.p_2)^2 - p_1^2p_2^2)^2}\left((p_1.p_2)^2 p_1^2\left(1 - \frac{p_1^2p_2^2}{(p_1.p_2)^2}\right)g_{\mu\nu} + (p_1.p_1)^2(d-1)p_{2\mu}p_{2\nu}\right.\\
&\quad \left. - (d-1)p_1^2(p_1.p_2)(p_{1\mu}p_{2\nu} + p_{1\nu}p_{2\mu}) + (p_1^2p_2^2 + (d-2)(p_1.p_2)^2)p_{1\mu}p_{1\nu}\right),\\
P^F_{\mu\nu} &= \frac{1}{(d-2)((p_1.p_2)^2 - p_1^2p_2^2)^2}\left((p_1.p_2)^3\left(\frac{p_1^2p_2^2}{(p_1.p_2)^2} - 1\right)g_{\mu\nu} + (d-1)(p_1.p_2)^2 p_{2\mu}p_{1\nu}\right.\\
&\quad \left. - (d-1)(p_1.p_2)(p_1^2 p_{2\mu}p_{2\nu} + p_2^2 p_{1\mu}p_{1\nu}) + \left((p_1.p_2)^2 + (d-2)p_1^2p_2^2\right)p_{2\nu}p_{1\mu}\right).
\end{aligned}$$

The entire calculation of form-factors and contractions is performed in $d = 4 - 2\epsilon$ dimensions using *dimensional regularization*. While calculating the amplitude $M^{\mu\nu}$ for each contributing diagram, Larin's prescription given in [206] is used to handle traces involving the $\gamma^5$ matrices.

Furry's theorem prohibits the appearance of a single $\gamma^5$ in the trace over a closed fermionic loop due to charge invariance, which gives $C = 0$ on summing the contributions from all the triangle diagrams together. Additionally, since current conservation is associated with massless leptons in the final state, only the form-factors $A$ and $B$ contribute at the squared amplitude level. These form-factors can be written as linear combinations of scalar two-loop integrals of the type

$$I_{\{\nu_i\}}\left(d, p_1^2, p_2^2, m_t^2, \mu^2\right) = e^{2\gamma_E \epsilon}\left(\mu^2\right)^{\nu-d} \int \frac{d^d k_1}{i\pi^{\frac{d}{2}}} \frac{d^d k_2}{i\pi^{\frac{d}{2}}} \prod_{i=1}^{7} \frac{1}{P_i^{\nu_i}}. \tag{4.20}$$

Here, $k_1$ and $k_2$ are loop momenta, $d$ denotes the space-time dimension, $\mu$ represents an arbitrary scale introduced to maintain the dimensionlessness of the integral, $\{\nu_i\}$ = $\{\nu_1\nu_2\nu_3\nu_4\nu_5\nu_6\nu_7\}$ are powers of inverse propagators $P_i$, and $\nu = \sum_{i=1}^{7} \nu_i$. The inverse propagators $P_i$ are given by

$$\begin{aligned}
&P_1 = k_1^2 - m_t^2, && P_2 = k_2^2 - m_t^2, && P_3 = (k_1 - k_2)^2, \\
&P_4 = (k_1 - p_1)^2 - m_t^2, && P_5 = (k_2 - p_1)^2 - m_t^2, && P_6 = (k_1 - p_1 - p_2)^2 - m_t^2, \\
&P_7 = (k_2 - p_1 - p_2)^2 - m_t^2.
\end{aligned} \tag{4.21}$$

The set of two-loop integrals in these form-factors is reduced to a minimal set of integrals called master integrals (MIs) using integration-by-parts (IBP) [70, 71] and Lorentz-invariance (LI) [65] identities with the programs **`LiteRed`** [97, 98] combined with **`Mint`** [207] and **`FIRE`** [92–94]. We find in total 41 master integrals after IBP reduction. The choice of master integrals is not unique. The basis set $\vec{I}$ of the 41 master integrals we get is listed below.

$$\begin{aligned}
&\{I_{0000011}, I_{0000111}, I_{0001111}, I_{0010110}, I_{0020110}, I_{0100011}, I_{0100110}, I_{0100111}, I_{0101011}, I_{0101110}, \\
&I_{0101111}, I_{0110010}, I_{0110110}, I_{0111000}, I_{0111001}, I_{0111010}, I_{0111011}, I_{0111110}, I_{0112011}, I_{0112110}, \\
&I_{0120110}, I_{0121001}, I_{0121010}, I_{0121011}, I_{0121110}, I_{0210010}, I_{0210110}, I_{0211000}, I_{0211001}, I_{0211010}, \\
&I_{0211011}, I_{0211110}, I_{1100011}, I_{1100110}, I_{1100111}, I_{1101100}, I_{1101101}, I_{1110110}, I_{1120110}, I_{1210110}, \\
&I_{2110110}\}.
\end{aligned}$$

The involvement of two-loop integrals makes the evaluation of these form-factors highly non-trivial. The presence of multiple mass scales, particularly the inclusion of the heavy top-quark loop, presents significant challenges in analytically evaluating these integrals, and their exact analytical solutions remain unknown in the literature. The computation of these two-loop integrals is crucial for evaluating form-factors for $HV_1V_2$ vertex corrections, and one can choose between numerical or analytical techniques to compute these integrals. In the next chapter, we will present the first-ever full analytical computation of these two-loop master integrals using the *method of differential equations*.

5

# Two-loop Master Integrals for $HV_1V_2$ Vertex Corrections

IN this chapter, we will employ the method of differential equations introduced in chapter 3 to obtain the analytic results of two-loop master integrals that appear in the two-loop virtual QCD corrections to the $H \to ZZ^*$ decay. The material presented in this chapter is entirely based on the paper [66] and is done in collaboration with E. Chaubey and A. Shivaji. The analytic results obtained here also cover a more general decay $H \to Z^{(*)}Z^{(*)}$ as the number of mass scales remains the same and thus are directly applicable for the master integrals appearing in $\mathcal{O}(\alpha\alpha_s)$ corrections to the $HV_1V_2$ vertex discussed in chapter 4.

## 5.1 State Of The Art

Among five prominent decay modes of the Higgs, $H \to ZZ^*$ decay, where $Z^*$ is off-shell, is a rare one. When both the $Z$ and $Z^*$ further decay into a pair of charged leptons, it is also known as the “Golden decay channel”. This decay process holds significant phenomenological importance in studying the Higgs properties, necessitating precise predictions, including higher-order terms, in the perturbative calculation of the amplitude for the partial decay width of $H \to ZZ^*$. At the two-loop level, the amplitude for the $H \to ZZ^*$ decay receives contributions from virtual QCD corrections, which is an $\mathcal{O}(\alpha\alpha_s)$ effect. The computation of these corrections is challenging due to the presence of non-trivial two-loop Feynman integrals. The Feynman integrals governing these two-loop virtual QCD corrections belong to the two-loop three-point topology diagrams involving the massive top quark loop, as illustrated in Figure 5.1. A subset of these integrals also appear in the QCD corrections to the $H \to \gamma\gamma, \gamma Z$ decays. Due to their dependence on multiple mass scales along with massive internal propagators, the analytical evaluation of these integrals turns out to be a bottleneck.

The state-of-the-art for the analytic study of two-loop three-point functions, including massive propagators, is as follows. For two-loop master integrals which appear in the virtual QCD corrections of $H \to \gamma\gamma$, the analytic results in terms of harmonic polylogarithms have been known for quite some time [208]. Also, the analytic results of master integrals for massless two-loop 3-point functions relevant for QCD corrections to decay $H \to V^*V^*$ ($V = Z$ or $W$) with three off-shell legs have been obtained in terms of Goncharov polylogarithms [209]. The results for the master integrals relevant to NLO QCD corrections to $H \to Z\gamma$ with a massive top-quark loop are also available [210], and the results are expressed in terms of Goncharov polylogarithms. In [211], two-loop master integrals for QCD correction to neutral massive boson coupling to a pair of $W$ bosons were computed. The master integrals in these cases depend on two or three mass scales at maximum. The full analytic results for the master integrals appearing in $\mathcal{O}(\alpha\alpha_s)$ corrections to $HW^+W^-$ vertex are also known in terms of multiple polylogarithms [212]. A few examples of one-loop triangle diagrams with arbitrary mass dependence are given in [213–216]. A canonical set of master integrals applicable to $\mathcal{O}(\alpha\alpha_s)$ corrections in $H \to ZZ^*$ decay with four mass scale dependence was studied in [217], however the analytic results for these integrals are not public.

For the first time, we provide the full analytic result for all the two-loop canonical master integrals that contribute to $\mathcal{O}(\alpha\alpha_s)$ corrections to $H \to ZZ^*$ decay, keeping the full dependence on the mass of the top quark ($m_t$) in the loop employing the method of differential equations. The presence of multiple square roots in the system of differential equations poses technical challenges in the analytic computations of these master integrals. We construct ansätze to obtain a dlog-form of the system of differential equations with a minimal set of independent one-forms involving a non-rationalizable square root. Furthermore, we express the results of all the master integrals in terms of Chen's iterated integrals. In addition, we identify the hypersurface obtained during a simultaneous rationalization of all the square roots, occurring because of the multiple mass scales, to be a $CY_3$ manifold. This establishes a connection between the relevant Feynman integrals for this process and integrals that occur in more symmetric quantum field theories. Furthermore, we derive the analytic results of all the boundary constants along with a compact set of letters called the alphabet. These are important for phenomenological applications as a minimal set of letters is needed for a fast numerical evaluation, and analytic boundary constants help in evaluating the master integrals with higher precision. Lastly, we perform numerical checks on our results by matching them against those obtained using publicly available numerical tools. The choice of a set of canonical master integrals presented here is also useful for similar two-loop massive triangles with three off-shell legs and a massive internal loop. For instance, this set of master integrals, apart from contributing to $H \to ZZ^*$, is also useful for mixed electroweak-QCD corrections to Higgs production via the process $e^+e^- \to HZ$ and $e^+e^- \to H\mu^+\mu^-$, which will be relevant at future $e^+e^-$ colliders [217–219]. The results can also be used for precision studies in processes like $\mu^+\mu^- \to HZ, H\ell^+\ell^-$ at future muon collider [220].

This chapter is organized as follows: kinematic invariants and notations of the Feynman integrals are discussed in section 5.2. In section 5.3, we present all the pre-canonical as well as canonical master integrals. In section 5.3.2, we present the details of the square root(s) and discuss their rationalization. In section 5.3.3, we discuss all the one-forms present in the differential equations and present the alphabet. In section 5.4, we present the analytic results of all the boundary constants for the canonical master integrals and perform checks.

## 5.2 Set-up For Solving the Master Integrals

The representative Feynman diagrams contributing to $H(q) \to Z(p_1) + Z^*(p_2)$ at $\mathcal{O}(\alpha\alpha_s)$ are shown in figure 5.1. There are two independent external momenta $p_1$ and $p_2$ corresponding to $V_1$ and $V_2$ respectively, with $p_1^2 = m_Z^2$ and $p_2^2 = m_{Z^*}^2$. The external momenta corresponding to $Z^*$ is off-shell implying $p_1^2 \neq p_2^2$, in general. The Higgs boson is on-shell with $q^2 = (p_1+p_2)^2 = m_H^2$.

There are seven linearly independent scalar products involving loop momenta $k_1$ and $k_2$, and external momenta $p_1$ and $p_2$ ($k_1^2, k_2^2, k_1.k_2, k_1.p_1, k_1.p_2, k_2.p_1$ and $k_2.p_2$), which appear in our Feynman integrals. The relevant two-loop Feynman diagrams, however, have a maximum of six propagators. In order to express Feynman integrals in terms of master integrals, we introduce an auxiliary topology with seven propagators as shown in figure 5.2. The corresponding integral family is given by

$$I_{\nu_1\nu_2\nu_3\nu_4\nu_5\nu_6\nu_7}\left(d, p_1^2, p_2^2, m_t^2, \mu^2\right) = e^{2\gamma_E\epsilon}\left(\mu^2\right)^{\nu-d}\int \frac{d^dk_1}{i\pi^{\frac{d}{2}}}\frac{d^dk_2}{i\pi^{\frac{d}{2}}}\prod_{j=1}^{7}\frac{1}{P_j^{\nu_j}}. \tag{5.1}$$

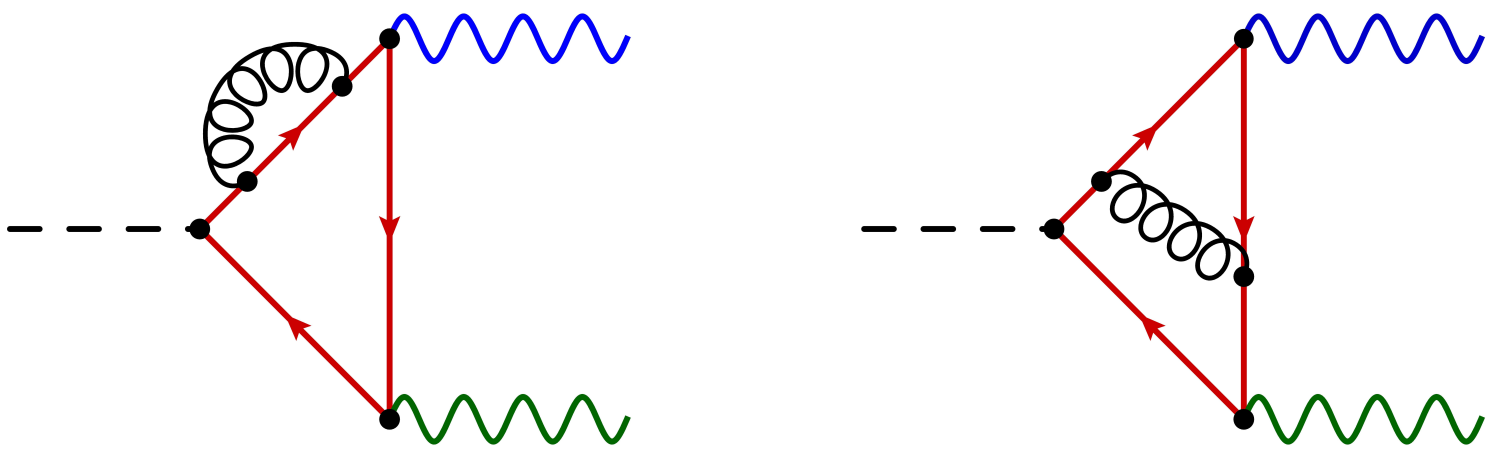

FIGURE 5.1: Examples of Feynman diagrams contributing to the $\mathcal{O}(\alpha\alpha_s)$-corrections to the decay $H \to ZZ^*$ via a top-quark loop. The Higgs boson is denoted by a dashed line, a top quark by a red line, $Z$ and $Z^*$ by blue and green wavy lines and a gluon by a curly line.

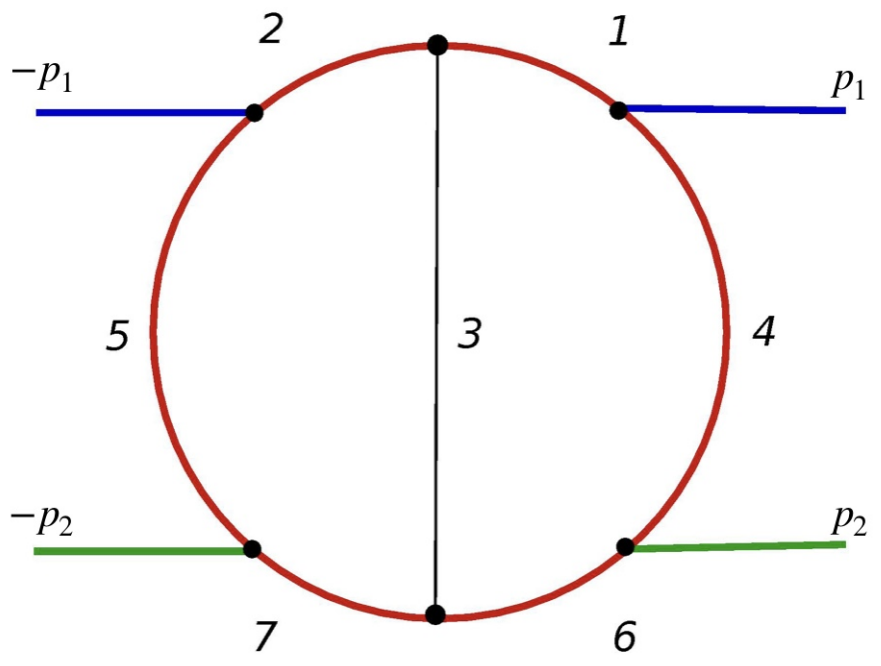

FIGURE 5.2: The auxiliary diagram for two-loop triangle diagrams in $H(q) \to Z(p_1)Z^*(p_2)$ decay at $\mathcal{O}(\alpha\alpha_s)$. The masses of propagators are encoded in the colours of the propagators; massless(light black), $m_t$(red), $m_Z$(blue), $m_{Z^*}$(green).

Here a factor of $e^{2\gamma_E\epsilon}$ is introduced to keep the final results for integrals free from the Euler-Mascheroni constant $\gamma_E$, $d$ is the space-time dimension, $\mu$ is an arbitrary scale introduced to render the Feynman integral dimensionless, the quantity $\nu$ is given by

$$\nu = \sum_{j=1}^{7}\nu_j. \tag{5.2}$$

The propagators $P_j$ are given by

$$P_1 = -k_1^2 + m_t^2, \qquad P_2 = -k_2^2 + m_t^2, \qquad P_3 = -\left(k_1 - k_2\right)^2,$$
$$P_4 = -\left(k_1 - p_1\right)^2 + m_t^2, \qquad P_5 = -\left(k_2 - p_1\right)^2 + m_t^2, \qquad P_6 = -\left(k_1 - p_1 - p_2\right)^2 + m_t^2,$$

$$P_7 = -(k_2 - p_1 - p_2)^2 + m_t^2. \tag{5.3}$$

We obtain the set of master integrals using the IBP program **`Fire`** [92–94] combined with **`LiteRed`** [97, 98], which gives us 41 master integrals. The set of 41 master integrals forms a basis which we denote by $\vec{I}$. For convenience, the master integrals are classified into sectors, in our case, some sectors have more than one master integral. The list of all 41 master integrals, classified sector-wise and following the notation in equation (5.1), is shown in column 2 of table 5.1. We may set $\mu^2 = m_t^2$, after which our master integrals depend kinematically on three dimensionless quantities defined by

$$\frac{m_{Z^*}^2}{m_t^2} = u, \quad \frac{m_H^2}{m_t^2} = v, \quad \frac{m_Z^2}{m_t^2} = w. \tag{5.4}$$

## 5.3 An Epsilon-form for the Master Integrals

We set up a linear system of differential equations of the form $d\vec{I} = A\vec{I}$ for all the master integrals of basis $\vec{I}$ by taking their derivatives with respect to the kinematic invariants ($u$, $v$ and $w$) and using IBP identities. The $41 \times 41$ matrix $A$ depends on kinematic invariants and space-time dimensions ($d$). The system of differential equations for the master integrals is solved iteratively at each order in the dimensional regularization parameter $\epsilon$ after switching from basis $\vec{I}$ to a canonical basis $\vec{J}$ such that the system of differential equations has an $\epsilon$-form [72] given by

$$d\vec{J} = \epsilon \tilde{A} \vec{J}. \tag{5.5}$$

Here matrix $\tilde{A}$ is independent of the dimensional regulator $\epsilon$. We aim to find a transformation matrix that brings the pre-canonical basis $\vec{I}$ to a canonical basis $\vec{J}$ leading to a differential system given in equation (5.5).

### 5.3.1 Finding the Canonical Basis

Arranging the master integrals sector-wise gives a block-diagonal form for their differential equations. We also use a "bottom-up" approach where we bring the differential equations of a lower sector to a canonical form using a coordinate system suitable to that sector before moving on to a higher sector. For some of the master integrals, it is convenient to work in $d = 2 - 2\epsilon$ dimensions. We use the dimension shifting operator $\mathbf{D}^-$ to shift the integrals from $4-2\epsilon$ to $2-2\epsilon$. We express such integrals as a linear combination of master integrals in $d = 4-2\epsilon$ dimensions using the dimensional shift relations [97]. To find a transformation matrix that brings our pre-canonical basis to a canonical form, we use the information of the maximal cuts of each of the sector [87, 119, 121, 221, 222]. Following the same method, as described in [125, 223, 224], we first construct an ansatz for a canonical basis using the information from the maximal cut of the sector in consideration. This fixes the diagonal part of the differential equations for that sector. We then include the contribution from the sub-sectors to complete the construction of a canonical basis for that sector. Finally, we repeat this process to reach the top sector and construct a canonical basis for the full system. Our canonical basis $\vec{J}$, in terms of dimensionless quantities $u, v$ and $w$, is given by the following transformations.

| number of propagators | master integrals basis $\vec{I}$ | master integrals basis $\vec{J}$ | kinematic dependence |
|---|---|---|---|
| 2 | $I_{0000011}$ | $J_1$ | – |
| 3 | $I_{0000111}$ | $J_2$ | $u$ |
| | $I_{0020120}, I_{0010220}$ | $J_4, J_5$ | $u$ |
| | $I_{0100011}$ | $J_6$ | $v$ |
| | $I_{0100110}$ | $J_7$ | $w$ |
| | $I_{0120020}, I_{0210020}$ | $J_{12}, J_{13}$ | $v$ |
| | $I_{0122000}, I_{0212000}$ | $J_{14}, J_{15}$ | $w$ |
| 4 | $I_{0001111}$ | $J_3$ | $u$ |
| | $I_{0100111}$ | $J_8$ | $u, v, w$ |
| | $I_{0101011}$ | $J_9$ | $u, v$ |
| | $I_{0101110}$ | $J_{10}$ | $u, w$ |
| | $I_{1100011}$ | $J_{16}$ | $v$ |
| | $I_{1100110}$ | $J_{17}$ | $v, w$ |
| | $I_{1101100}$ | $J_{19}$ | $w$ |
| | $I_{0110110}, I_{0120110}, I_{0110120}$ | $J_{21}, J_{22}, J_{23}$ | $u, v, w$ |
| | $I_{0111001}, I_{0121001}, I_{0112001}$ | $J_{24}, J_{25}, J_{26}$ | $u, v, w$ |
| | $I_{0111010}, I_{0121010}, I_{0211010}$ | $J_{27}, J_{28}, J_{29}$ | $u, v, w$ |
| 5 | $I_{0101111}$ | $J_{11}$ | $u, v, w$ |
| | $I_{1100111}$ | $J_{18}$ | $u, v, w$ |
| | $I_{1101101}$ | $J_{20}$ | $u, v, w$ |
| | $I_{0111011}, I_{0211011}, I_{0112011}, I_{0212011}$ | $J_{30}, J_{31}, J_{32}, J_{33}$ | $u, v, w$ |
| | $I_{0111110}, I_{0111120}, I_{0211110}, I_{0211120}$ | $J_{34}, J_{35}, J_{36}, J_{37}$ | $u, v, w$ |
| | $I_{1110110}, I_{1110210}, I_{1110120}, I_{1110220}$ | $J_{38}, J_{39}, J_{40}, J_{41}$ | $u, v, w$ |

TABLE 5.1: Overview of the set of master integrals for two-loop triangle diagrams in $H \to ZZ^*$ decay. The first column gives the number of non-zero propagators that appear in the master integrals, the second column lists the master integrals in the basis $\vec{I}$, the third column lists corresponding master integrals in the basis $\vec{J}$. The last column gives the kinematic dependence of master integrals in terms of dimensionless quantities defined in equation 5.4.

$$
\begin{aligned}
J_1 &= \epsilon^2 \ \mathbf{D}^- I_{0000011}, \\
J_2 &= \epsilon^2 \ \sqrt{-u \ (4-u)} \ \mathbf{D}^- I_{0000111}, \\
J_3 &= \epsilon^2 \ u \ (4-u) \ \mathbf{D}^- I_{0001111}, \\
J_4 &= \epsilon^2 \sqrt{-u \ (4-u)} \left[ I_{0020120} + \frac{1}{2} I_{0010220} \right], \\
J_5 &= \epsilon^2 \ u \ I_{0010220}, \\
J_6 &= \epsilon^2 \ \sqrt{-v \ (4-v)} \ \mathbf{D}^- I_{0100011}, \\
J_7 &= \epsilon^2 \ \sqrt{-w \ (4-w)} \ \mathbf{D}^- I_{0100110}, \\
J_8 &= \frac{\epsilon^2}{2\sqrt{\lambda}} \Big[ 2 (\lambda + u \ v \ w) \ \mathbf{D}^- I_{0100111} + w \ (u+v-w) \ \mathbf{D}^- I_{0100110} \\
&\quad + v(u-v+w) \ \mathbf{D}^- I_{0100011} + u \ (-u+v+w) \ \mathbf{D}^- I_{0000111} \Big], \\
J_9 &= \epsilon^2 \ \sqrt{u \ (4-u)} \ \sqrt{v \ (4-v)} \ \mathbf{D}^- I_{0101011}, \\
J_{10} &= \epsilon^2 \ \sqrt{u \ (4-u)} \ \sqrt{w \ (4-w)} \ \mathbf{D}^- I_{0101110}, \\
J_{11} &= \frac{\epsilon^2}{2} \sqrt{\frac{-u \ (4-u)}{\lambda}} \Big[ 2 (\lambda + u \ v \ w) \ \mathbf{D}^- I_{0101111} + w \ (u+v-w) \ \mathbf{D}^- I_{0101110} \\
&\quad + v \ (u-v+w) \ \mathbf{D}^- I_{0101011} + u \ (-u+v+w) \ \mathbf{D}^- I_{0001111} \Big], \\
J_{12} &= \epsilon^2 \ \sqrt{-v \ (4-v)} \left[ I_{0120020} + \frac{1}{2} I_{0210020} \right], \\
J_{13} &= \epsilon^2 \ v \ I_{0210020}, \\
J_{14} &= \epsilon^2 \ \sqrt{-w \ (4-w)} \left[ I_{0122000} + \frac{1}{2} I_{0212000} \right], \\
J_{15} &= \epsilon^2 \ w \ I_{0212000}, \\
J_{16} &= \epsilon^2 \ v \ (4-v) \ \mathbf{D}^- I_{1100011}, \\
J_{17} &= \epsilon^2 \ \sqrt{v \ (4-v)} \ \sqrt{w \ (4-w)} \ \mathbf{D}^- I_{1100110}, \\
J_{18} &= \frac{\epsilon^2}{2} \sqrt{\frac{-v \ (4-v)}{\lambda}} \Big[ 2 (\lambda + u \ v \ w) \ \mathbf{D}^- I_{1100111} + w \ (u+v-w) \ \mathbf{D}^- I_{1100110} \\
&\quad + v \ (u-v+w) \ \mathbf{D}^- I_{1100011} + u \ (-u+v+w) \ \mathbf{D}^- I_{0101011} \Big], \\
J_{19} &= \epsilon^2 \ w \ (4-w) \ \mathbf{D}^- I_{1101100}, \\
J_{20} &= \frac{\epsilon^2}{2} \sqrt{\frac{-w \ (4-w)}{\lambda}} \Big[ 2 (\lambda + u \ v \ w) \ \mathbf{D}^- I_{1101101} + w \ (u+v-w) \ \mathbf{D}^- I_{1101100}
\end{aligned}
$$

$$+ v\,(u - v + w)\,\mathbf{D}^{-} I_{1100110} + u\,(-u + v + w)\,\mathbf{D}^{-} I_{0101110}\Big],$$

$$J_{21} = \epsilon^3\,\sqrt{\lambda}\, I_{0120110},$$

$$J_{22} = \epsilon^3\,\sqrt{\lambda}\, I_{0110120},$$

$$J_{23} = \epsilon^2\,\frac{\sqrt{-w\,(4-w)}}{w\,(u+v-w)}\Big[(\lambda + u\,v\,w)\;\mathbf{D}^{-} I_{0110110} - 2\,\epsilon\,\lambda\,(I_{0120110} + \frac{1}{2} I_{0110120})$$

$$+ v\,(u - v + w)\,(I_{0120020} + \frac{1}{2} I_{0210020}) + u\,(-u + v + w)\,(I_{0020120} + \frac{1}{2} I_{0010220})\Big],$$

$$J_{24} = \epsilon^3\,\sqrt{\lambda}\, I_{0121001},$$

$$J_{25} = \epsilon^3\,\sqrt{\lambda}\, I_{0112001},$$

$$J_{26} = \epsilon^2\,\frac{\sqrt{-v\,(4-v)}}{v\,(-u+v-w)}\Big[(\lambda + u\,v\,w)\;\mathbf{D}^{-} I_{0111001} - 2\,\epsilon\,\lambda\,(I_{0121001} + \frac{1}{2} I_{0112001})$$

$$+ w\,(u + v - w)\,(I_{0122000} + \frac{1}{2} I_{0212000}) + u\,(-u + v + w)\,(I_{0020120} + \frac{1}{2} I_{0010220})\Big],$$

$$J_{27} = \epsilon^3\,\sqrt{\lambda}\, I_{0121010},$$

$$J_{28} = \epsilon^3\,\sqrt{\lambda}\, I_{0211010},$$

$$J_{29} = \epsilon^2\,\frac{\sqrt{-u\,(4-u)}}{u\,(-u+v+w)}\Big[(\lambda + u\,v\,w)\;\mathbf{D}^{-} I_{0111010} - 2\,\epsilon\,\lambda\,(I_{0121010} + \frac{1}{2} I_{0211010})$$

$$+ w\,(u + v - w)\,(I_{0122000} + \frac{1}{2} I_{0212000}) + v\,(u - v + w)\,(I_{0120020} + \frac{1}{2} I_{0210020})\Big],$$

$$J_{30} = \epsilon^4\,\sqrt{\lambda}\, I_{0111011},$$

$$J_{31} = \epsilon^3\,\sqrt{\lambda}\,\sqrt{-v\;(4-v)}\; I_{0211011},$$

$$J_{32} = \epsilon^3\,\sqrt{\lambda}\,\sqrt{-u\;(4-u)}\; I_{0112011},$$

$$J_{33} = \epsilon^2\,\Big[(\lambda + u\,v\,w)\;I_{0212011} + \epsilon\,u\,(-u + v + w)\;I_{0112011} + \epsilon\,v\,(u - v + w)\;I_{0211011}$$

$$+ (w + \frac{1}{2}\,u\,(-2 + v) - v)\,\mathbf{D}^{-} I_{0101011}\Big],$$

$$J_{34} = \epsilon^4\,\sqrt{\lambda}\, I_{0111110},$$

$$J_{35} = \epsilon^3\,\sqrt{\lambda}\,\sqrt{-u\;(4-u)}\; I_{0111120},$$

$$J_{36} = \epsilon^3\sqrt{\lambda}\,\sqrt{-w\;(4-w)}\; I_{0211110},$$

$$J_{37} = \epsilon^2\,\Big[(\lambda + u\,v\,w)\;I_{0211120} + \epsilon\,w\,(u + v - w)\;I_{0211110} + \epsilon\,u\,(-u + v + w)\;I_{0111120}$$

$$+ (v + \frac{1}{2}\,u(-2 + w) - w)\,\mathbf{D}^{-} I_{0101110}\Big],$$

$$J_{38} = \epsilon^4\,\sqrt{\lambda}\, I_{1110110},$$

$$J_{39} = \epsilon^3\,\sqrt{\lambda}\,\sqrt{-w\;(4-w)}\; I_{1110210},$$

$$J_{40} = \epsilon^3\,\sqrt{\lambda}\,\sqrt{-v\;(4-v)}\; I_{1110120},$$

$$
\begin{aligned}
J_{41} &= \epsilon^2 \Bigg[ (\lambda + u\, v\, w)\; I_{1110220} + \epsilon\, v\, (u - v + w)\; I_{1110120} + \epsilon\, w\, (u + v - w)\; I_{1110210} \\
&\quad + (u + \frac{1}{2}\, v\, (-2 + w) - w)\; \mathbf{D}^{-} I_{1100110} \Bigg].
\end{aligned}
$$

In the above, $\lambda(u, v, w)$ is the Källen function defined by

$$
\lambda(u, v, w) = u^2 + v^2 + w^2 - 2uv - 2vw - 2wu. \tag{5.6}
$$

Note that in terms of the coordinates of equation (5.4), the system of differential equations contain multiple square roots. In particular, the following square roots appear

$$
\sqrt{-u\,(-4+u)}, \quad \sqrt{-v\,(-4+v)}, \quad \sqrt{-w\,(-4+w)} \quad \text{and} \quad \sqrt{\lambda\,(u, v, w)}. \tag{5.7}
$$

A non-trivial task is to find a coordinate system that rationalizes all these square roots simultaneously. We introduce the following set of transformations that takes us from $(u, v, w)$ to $(x, y, z)$ and rationalizes the first three square roots in equation (5.7)

$$
\frac{m_{Z^*}^2}{m_t^2} = u = -\frac{(1-x)^2}{x}, \quad \frac{m_H^2}{m_t^2} = v = -\frac{(1-y)^2}{y}, \quad \frac{m_Z^2}{m_t^2} = w = -\frac{(1-z)^2}{z}. \tag{5.8}
$$

These transformations are ubiquitous in literature [225]. After using these set of transformations we are left with only one square root $\sqrt{\lambda\,(u, v, w)}$, which in the new coordinate system becomes

$$
r = \sqrt{\begin{aligned}&(y^2z^2 + x^4y^2z^2 - 2xyz(z + y(1 + z(-2 + y + z))) - 2x^3yz(z + y(1 + z(-2 + y + z))) \\ &+ x^2(-2y(-1+z)^2z - 2y^3(-1+z)^2z + z^2 + y^4z^2 + y^2(1 + z(4 + z(-6 + z(4 + z))))))\end{aligned}}. \tag{5.9}
$$

This square root $r$ cannot be rationalized by doing further coordinate transformations. We discuss the appearance of square roots and their rationalizability in more detail in the next section.

Sector-wise classification of the master integrals of canonical basis and their kinematic dependence is given in columns 3 and 4, respectively, of table 5.1. The master integrals from higher sectors have more propagators and are usually more complicated to solve than the master integrals from the lower sectors. We also look at the topology of each sector, where multiple sectors may correspond to the same topology. The form of canonical basis can be recycled within sectors belonging to the same topology since these sectors are usually obtained by a permutation of the external legs. In our case, we have 9 master topologies which are shown in figure 5.3. Apart from being a guiding principle in the construction of a canonical basis,

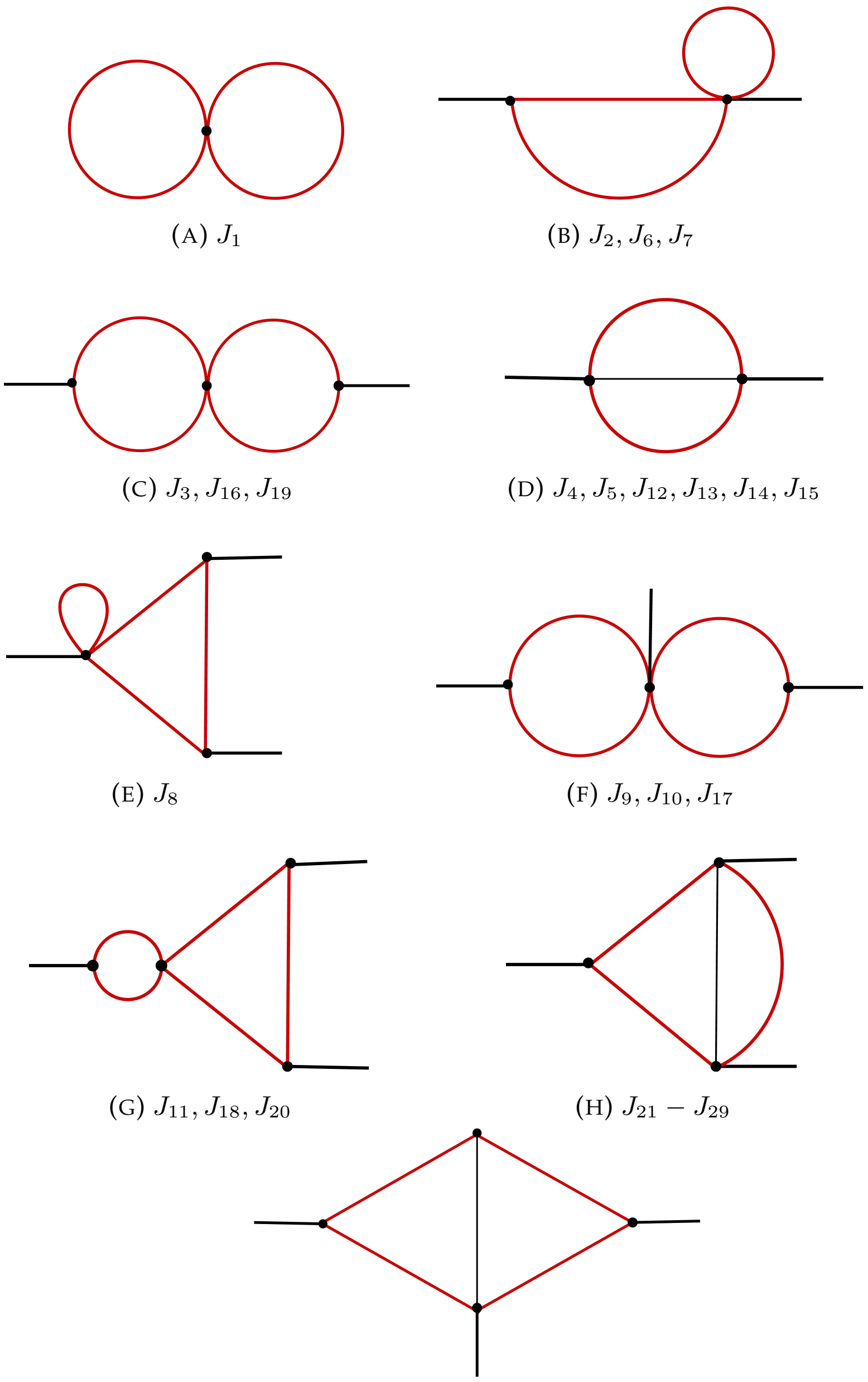

(A) $J_1$ (B) $J_2, J_6, J_7$

(C) $J_3, J_{16}, J_{19}$ (D) $J_4, J_5, J_{12}, J_{13}, J_{14}, J_{15}$

(E) $J_8$ (F) $J_9, J_{10}, J_{17}$

(G) $J_{11}, J_{18}, J_{20}$ (H) $J_{21} - J_{29}$

(I) $J_{30} - J_{41}$

FIGURE 5.3: Topologies relevant to two-loop three-point diagrams in $H \to ZZ^*$ decay at $\mathcal{O}(\alpha\alpha_s)$. Light black lines represent massless propagators. The caption of each topology denotes the set of canonical master integrals belonging to that topology.

the classification of integrals in terms of master topologies also helps in fixing the boundary constants while solving the differential equations. The boundary constants are discussed in section 5.4.

### 5.3.2 The Square Root

Due to the presence of many mass scales in our system, the differential equations contain 4 square roots in the coordinate system given in equation (5.4). However, our choice of variables given in equation (5.8) rationalizes 3 out of 4 square roots present in the system. The fourth square root is $r = \sqrt{P(x,y,z)}$ given in equation (5.9). This square root is present in the analytic expressions of the master integrals $J_8$, $J_{11}$, $J_{18}$, $J_{20} - J_{41}$. In general, one can try to perform a change of variables using

$$x = \psi_1(x_1), \quad y = \psi_2(x_2), \quad z = \psi_3(x_3), \tag{5.10}$$

with rational functions $\psi_1$, $\psi_2$ and $\psi_3$, such that $P(\psi_1(x_1), \psi_2(x_2), \psi_3(x_3))$ becomes a perfect square.

The literature studies indicate that it is simpler to find a transformation for rationalizing the square roots with one-variable dependence. The condition to find such a parametrization is that the degree of the polynomial involved must be less than or equal to 2. However, in some cases, the presence of a single square root of a multi-variable polynomial or a polynomial with a degree greater than 2 makes the subject of rationalizability more complicated [140–142, 145, 226].

Recent studies show that, for massive cases at higher orders in scattering amplitudes, the analytic computation of Feynman integrals involves non-rationalizable square roots which results in the appearance of complicated functions in their respective solutions. In many cases, the results are expressed as integrals over manifolds in higher dimensions, such as elliptic curves [123–126, 148, 150–169], $K3$ surfaces [139, 140, 144, 170, 171] and Calabi-Yau manifolds [172–176].

In order to study the square root $r$ and to understand its connection with a Calabi-Yau threefold variety, discussed before in the literature [175, 176], we proceed as follows. We start by noting that $P(x,y,z)$ in the square root $r$ we obtain after simultaneously rationalizing three out of the four square roots of equation (5.7)

$$
\begin{aligned}
P(x,y,z) &= r^2 \\
&= y^2z^2 + x^4y^2z^2 - 2xyz(z + y(1 + z(-2 + y + z))) - 2x^3yz(z + y(1 + z(-2 + y + z))) \\
&+ x^2(-2y(-1+z)^2z - 2y^3(-1+z)^2z + z^2 + y^4z^2 + y^2(1 + z(4 + z(-6 + z(4+z)))))
\end{aligned}
\tag{5.11}
$$

is a degree 8 in-homogeneous polynomial with no repeated roots. Then we assign weight one to all the three coordinates $x$, $y$ and $z$ to realize this surface as a projective variety. In the next step, we introduce an additional auxiliary coordinate '$l$' to homogenize $P(x,y,z)$. As a result, we obtain a homogeneous polynomial $P_8(x,y,z,l)$ of overall degree 8 in 4 variables. Writing this hypersurface as

$$Q(x,y,z,l,r) = r^2 - P_8^2(x,y,z,l) = 0, \tag{5.12}$$

we can identify it as a degree 8 hypersurface in 4-dimensional weighted projective space $\mathbb{WP}^{1,1,1,1,4}$, with weight 4 assigned to $r$. The sum of the weights of this weighted projective space is 8, which is exactly equal to the degree of the polynomial $Q$, known as a Calabi-Yau threefold $CY_3$ in $\mathbb{WP}^{1,1,1,1,4}$ [176][1]. With this identification, one can obtain the Hodge structure and Euler characteristics that characterize the Calabi-Yau manifold. After establishing this interesting connection, we proceed to bring the epsilon-factorized differential equation system into a dlog-form and express our results for the master integrals in terms of Chen's iterated integrals with dlog one-form kernels. Even though from a mathematical point of view, it might be possible to obtain a representation of these iterated integrals in terms of elliptic multiple polylogarithms (eMPLs) [148] with such kernels, we would like to keep this discussion for the future and proceed keeping the phenomenological applications of our results in mind. More details on one-forms and the the motivation behind using iterated integrals over dlog one-forms for phenomenological applications are provided in the following section.

### 5.3.3 dlog One-forms

We Taylor expand the canonical basis integrals $J_k$ around $\epsilon = 0$ as

$$J_k = \sum_{j=0}^{\infty} \epsilon^j J_k^{(j)}. \tag{5.13}$$

[1]We thank Matthias Wilhelm & Matthew von Hippel for discussions regarding this.

Putting it in the $\epsilon$-factorized differential equation of (5.5), allows us to express each $J_k^{(j)}$ in terms of Chen's iterated integrals up to a boundary constant. In order to present the analytic results in terms of iterated integrals over algebraic dlog one-forms, we need to bring the epsilon-factorized differential equation system to a dlog-form. In the dlog-form, the entries of the matrix $\tilde{A}$ are $\mathbb{Q}$-linear combination of dlog one-forms so that

$$\tilde{A}_{ij} = \sum_{k=1}^{31} \tilde{c}_{ijk} d \, \log(p_k(x,y,z)), \quad \tilde{c}_{ijk} \in \mathbb{Q}. \tag{5.14}$$

From the matrix $\tilde{A}$, a set of rational letters can be trivially obtained for the rational one-forms. For the one-forms with a dependence on $r$, getting a dlog one-form and, therefore, the list of non-rational letters, is not a trivial exercise. To construct the non-rational letters, we implement the algorithm from [143, 227] explained in the following:

1. We find the list of all the rational letters and the list of square roots. In our case, we have only one square root $r$ with a degree-8 polynomial.
2. We construct all possible monomials up to degree 8 using the rational letters, including $r^2$. Further, we identify those monomials which can be factorized as $(q_a + r)(q_a - r)$. This factorization gives the list of $q_a$'s.
3. Using these $q_a$'s, we construct ansätze $l_a$ for dlog one-forms as

$$l_a = \frac{q_a + r}{q_a - r}. \tag{5.15}$$

4. Using these ansätze, we fit all the non-rational one-forms, which gives us a minimal set of dlog one-forms $p_a$.

The complete alphabet with 31 letters $p_a$ that we obtain after the above steps is given by

$$
\begin{aligned}
p_1 &= x,\\
p_2 &= x-1,\\
p_3 &= x+1,\\
p_4 &= y,\\
p_5 &= y-1,\\
p_6 &= y+1,
\end{aligned}
$$

$$p_7 = z,$$

$$p_8 = z - 1,$$

$$p_9 = z + 1,$$

$$p_{10} = x^2y - x\left(y^2+1\right)z + yz^2,$$

$$p_{11} = -y\left(x^2z^2+1\right) + xy^2z + xz,$$

$$p_{12} = -x^2y^2z + xy\left(z^2+1\right) - z,$$

$$p_{13} = x^2yz - x\left(y^2z^2+1\right) + yz,$$

$$p_{14} = xy^2z - xyz + xz - yz^2 + yz - y,$$

$$p_{15} = x^2z - x\left(yz^2 - yz + y + z\right) + z,$$

$$p_{16} = x^2yz - x\left((y-1)z + z^2 + 1\right) + yz,$$

$$p_{17} = x^2yz - x\left(y^2 + y(z-1) + 1\right) + yz,$$

$$p_{18} = -y\left(x^2 + x(z-1) + 1\right) + xy^2z + xz,$$

$$p_{19} = \frac{-r + x^2yz - xy^2z - xyz^2 - xy + xz + yz}{r + x^2yz - xy^2z - xyz^2 - xy + xz + yz},$$

$$p_{20} = \frac{-r + x^2yz - 2x^2y + xy^2z - xyz^2 + xy + xz - yz}{r + x^2yz - 2x^2y + xy^2z - xyz^2 + xy + xz - yz},$$

$$p_{21} = \frac{-r + 2x^2y^2z - x^2yz - xy^2z - xyz^2 - xy + xz + yz}{r + 2x^2y^2z - x^2yz - xy^2z - xyz^2 - xy + xz + yz},$$

$$p_{22} = \frac{-r + x^2yz - 2xy^2z^2 + xy^2z + xyz^2 - xy - xz + yz}{r + x^2yz - 2xy^2z^2 + xy^2z + xyz^2 - xy - xz + yz},$$

$$p_{23} = \frac{-r + 2x^2yz^2 - x^2yz - xy^2z - xyz^2 + xy - xz + yz}{r + 2x^2yz^2 - x^2yz - xy^2z - xyz^2 + xy - xz + yz},$$

$$p_{24} = \frac{-r + x^2yz - 2xy^2z^2 + 3xy^2z - 2xy^2 + xyz^2 - 2xyz + xy - xz + yz}{r + x^2yz - 2xy^2z^2 + 3xy^2z - 2xy^2 + xyz^2 - 2xyz + xy - xz + yz},$$

$$p_{25} = \frac{-r + 2x^2y^3z - 2x^2y^2z + x^2yz - 2xy^2z^2 + xy^2z - 2xy^2 + xyz^2 + xy - xz + yz}{r + 2x^2y^3z - 2x^2y^2z + x^2yz - 2xy^2z^2 + xy^2z - 2xy^2 + xyz^2 + xy - xz + yz},$$

$$p_{26} = \frac{-r + 2x^2yz^3 - 2x^2yz^2 + x^2yz - 2xy^2z^2 + xy^2z + xyz^2 - xy - 2xz^2 + xz + yz}{r + 2x^2yz^3 - 2x^2yz^2 + x^2yz - 2xy^2z^2 + xy^2z + xyz^2 - xy - 2xz^2 + xz + yz},$$

$$p_{27} = \frac{-r + 2x^3y^2z - 2x^2y^2z - 2x^2yz^2 + x^2yz - 2x^2y + xy^2z + xyz^2 + xy + xz - yz}{r + 2x^3y^2z - 2x^2y^2z - 2x^2yz^2 + x^2yz - 2x^2y + xy^2z + xyz^2 + xy + xz - yz},$$

$$p_{28} = \frac{-r + 2x^3y - 2x^2y^2z + x^2yz - 2x^2y - 2x^2z + xy^2z + xyz^2 + xy + xz - yz}{r + 2x^3y - 2x^2y^2z + x^2yz - 2x^2y - 2x^2z + xy^2z + xyz^2 + xy + xz - yz},$$

$$p_{29} = \frac{-r + 2x^3yz^2 - 2x^2y^2z - 2x^2yz^2 + x^2yz - 2x^2z + xy^2z + xyz^2 + xy + xz - yz}{r + 2x^3yz^2 - 2x^2y^2z - 2x^2yz^2 + x^2yz - 2x^2z + xy^2z + xyz^2 + xy + xz - yz},$$

$$p_{30} = (xy - z)(-y + xz)(x - yz)(-1 + xyz),$$

$$p_{31} = r.$$

We also find that the matrix $\tilde{A}$ in equation (5.5) contains only 31 $\mathbb{Q}$-independent linear combinations of dlog one-forms. Out of these 31 independent one-forms, 16 are rational in the variables $x, y$ and $z$, whereas 15 contain the square root $r$.

In recent times, iterated integral representation with dlog one-forms has been shown to be very efficient for phenomenological applications, see for example [228, 229]. Iterated integrals with logarithmic one-forms have a clear branch-cut structure, and the results expressed in terms of these functions have a compact analytic form. For a numerical evaluation of these functions, we can use a local series expansion of the one-forms combined with path-decomposition property satisfied by iterated integrals, see for example [230], to integrate from one phase-space point to another. Logarithmic one-forms can be series expanded, and since they have a power-log expansion, a numerical evaluation is easy to implement. A representation in terms of iterated integrals is even useful for cases where public tools cannot be used due to an absence of power-log representation of the one-forms [231].

## 5.4 Results and Checks

In order to check the correctness of results, we numerically evaluate the iterated integrals using an in-house implementation in **`Mathematica`**. For the reader's convenience, we explain the steps that can be taken to do the numerical evaluation of iterated integrals in the following:

1. we parametrize the one-forms on a path,
2. we series expand the one-dimensional one-forms around a point on the path,
3. we perform the iterated integration of the expanded one-forms.

We might need multiple path segments if the phase-space point is far from the boundary point. In these cases, we perform the steps mentioned above on each path segment and use the path-decomposition formula [231, 232] to obtain the final numerical result. For phenomenological applications, the numerical evaluation of these functions can also be combined with several publicly available tools for fast evaluation. For example, one can set up a differential system just for the iterated integrals and evaluate them with generalized power series expansions [233] using tools like **`DiffExp`** [234] or parametrize the one-forms on a path and use **`GiNaC`** [184].

As explained in the previous section, the result of master integrals can be written as Taylor expansion in $\epsilon$. Since our results are applicable to calculations involving two-loop integrals, we calculate master integrals only up to $\mathcal{O}(\epsilon^4)$. We provide results for all the 41 master integrals of canonical basis in terms of the iterated integrals with the dlog one-forms in the supplementary material attached to the arXiv [66]. We want to emphasize once again that the choice of the coordinate system given in equation (5.8) allows us to write the iterated integrals appearing in $J_2 - J_7$, $J_9$, $J_{10}$, $J_{12} - J_{17}$, $J_{19}$ in terms of MPLs.

The complete results for the master integrals depend on the boundary terms. To obtain the analytic form of these boundary constants, we first evaluate the integrals in a regular limit [102, 223]. We then match these values against the functional part of the results by first evaluating the iterated integrals up to many (100) digits and then using **PSLQ** [73] to extract the analytic constants. The analytic expressions of all the boundary constants $B_i = (J_i)_{|(x=0,y=0,z=0)}$ in equation (5.13) upto $\mathcal{O}\left(\epsilon^4\right)$ are as follows:

$$
\begin{aligned}
B_1 &= 1 + \zeta_2\epsilon^2 - \frac{2}{3}\zeta_3\epsilon^3 + \frac{7}{4}\zeta_4\epsilon^4,\\
B_2 &= 2\zeta_2\epsilon^2 + 4\zeta_3\epsilon^3 + \frac{19}{2}\zeta_4\epsilon^4,\\
B_3 &= -10\zeta_4\epsilon^4,\\
B_4 &= -\zeta_2\epsilon^2 - 11\zeta_3\epsilon^3 - \frac{29}{2}\zeta_4\epsilon^4,\\
B_5 &= 6\zeta_3\epsilon^3 + \frac{13}{2}\zeta_4\epsilon^4,\\
B_6 &= 2\zeta_2\epsilon^2 + 4\zeta_3\epsilon^3 + \frac{19}{2}\zeta_4\epsilon^4,\\
B_7 &= 2\zeta_2\epsilon^2 + 4\zeta_3\epsilon^3 + \frac{19}{2}\zeta_4\epsilon^4,\\
B_8 &= 6i \text{ Im Li}_2\left(\frac{1}{2}(-1+i\sqrt{3})\right)\epsilon^2 - \frac{i}{2}\Bigg(16 \text{ Im Li}_3\left(\frac{i}{\sqrt{3}}\right) - 23 \text{ Im Li}_3\left(\frac{1}{2}(-1+i\sqrt{3})\right)\\
&\quad + 4 \text{ Im Li}_3\left(\frac{1}{2}(3-i\sqrt{3})\right)\Bigg)\epsilon^3 - \frac{i}{18}\Bigg(92\zeta_2 \text{ Im Li}_2\left(\frac{1}{2}(-1+i\sqrt{3})\right) + 92\zeta_2 \text{ Im Li}_2\left(\frac{1}{2}(3-i\sqrt{3})\right)\\
&\quad - 288 \text{ Im Li}_4\left(\frac{i}{\sqrt{3}}\right) + 207 \text{ Im Li}_4\left(\frac{1}{2}(-1+i\sqrt{3})\right) + 72 \text{ Im Li}_4\left(\frac{1}{2}(3-i\sqrt{3})\right)\Bigg)\epsilon^4,\\
B_9 &= -10\zeta_4\epsilon^4,\\
B_{10} &= -10\zeta_4\epsilon^4,\\
B_{11} &= 12i\zeta_2 \text{ Im Li}_2\left(\frac{1}{2}(-1+i\sqrt{3})\right)\epsilon^4,\\
B_{12} &= -\zeta_2\epsilon^2 - 11\zeta_3\epsilon^3 - \frac{29}{2}\zeta_4\epsilon^4,\\
B_{13} &= 6\zeta_3\epsilon^3 + \frac{13}{2}\zeta_4\epsilon^4,
\end{aligned}
$$

$$
\begin{aligned}
B_{14} &= -\zeta_2\epsilon^2 - 11\zeta_3\epsilon^3 - \frac{29}{2}\zeta_4\epsilon^4,\\
B_{15} &= 6\zeta_3\epsilon^3 + \frac{13}{2}\zeta_4\epsilon^4,\\
B_{16} &= -10\zeta_4\epsilon^4,\\
B_{17} &= -10\zeta_4\epsilon^4,\\
B_{18} &= 12i\zeta_2 \operatorname{Im}\operatorname{Li}_2\big(\frac{1}{2}(-1+i\sqrt{3})\big)\epsilon^4,\\
B_{19} &= -10\zeta_4\epsilon^4,\\
B_{20} &= 12i\zeta_2 \operatorname{Im}\operatorname{Li}_2\big(\frac{1}{2}(-1+i\sqrt{3})\big)\epsilon^4,\\
B_{21} &= -6i \operatorname{Im}\operatorname{Li}_2\big(\frac{1}{2}(-1+i\sqrt{3})\big)\epsilon^2 - \frac{i}{2}\bigg( -16 \operatorname{Im}\operatorname{Li}_3\big(\frac{i}{\sqrt{3}}\big) + 23 \operatorname{Im}\operatorname{Li}_3\big(\frac{1}{2}(-1+i\sqrt{3})\big)\\
&\quad - 4 \operatorname{Im}\operatorname{Li}_3\big(\frac{1}{2}(3-i\sqrt{3})\big)\bigg)\epsilon^3 + \frac{i}{9}\bigg(100\zeta_2 \operatorname{Im}\operatorname{Li}_2\big(\frac{1}{2}(-1+i\sqrt{3})\big) + 46\zeta_2 \operatorname{Im}\operatorname{Li}_2\big(\frac{1}{2}(3-i\sqrt{3})\big)\\
&\quad - 144 \operatorname{Im}\operatorname{Li}_4\big(\frac{i}{\sqrt{3}}\big) - 18 \operatorname{Im}\operatorname{Li}_4\big(\frac{1}{2}(-1+i\sqrt{3})\big) + 36 \operatorname{Im}\operatorname{Li}_4\big(\frac{1}{2}(3-i\sqrt{3})\big)\bigg)\epsilon^4,\\
B_{22} &= \frac{i}{2}\bigg(12\zeta_2 \operatorname{Im}\operatorname{Li}_2\big(\frac{1}{2}(-1+i\sqrt{3})\big) - 27 \operatorname{Im}\operatorname{Li}_4\big(\frac{1}{2}(-1+i\sqrt{3})\big)\bigg)\epsilon^4,\\
B_{23} &= \zeta_2\epsilon^2 + 10\zeta_3\epsilon^3 + \frac{1}{4}\bigg(47\zeta_4 - 36 \operatorname{Im}\operatorname{Li}_2\big(\frac{1}{2}(-1+i\sqrt{3})\big)^2\bigg)\epsilon^4,\\
B_{24} &= -6i \operatorname{Im}\operatorname{Li}_2\big(\frac{1}{2}(-1+i\sqrt{3})\big)\epsilon^2 - \frac{i}{2}\bigg( -16 \operatorname{Im}\operatorname{Li}_3\big(\frac{i}{\sqrt{3}}\big) + 23 \operatorname{Im}\operatorname{Li}_3\big(\frac{1}{2}(-1+i\sqrt{3})\big)\\
&\quad - 4 \operatorname{Im}\operatorname{Li}_3\big(\frac{1}{2}(3-i\sqrt{3})\big)\bigg)\epsilon^3 + \frac{i}{9}\bigg(100\zeta_2 \operatorname{Im}\operatorname{Li}_2\big(\frac{1}{2}(-1+i\sqrt{3})\big) + 46\zeta_2 \operatorname{Im}\operatorname{Li}_2\big(\frac{1}{2}(3-i\sqrt{3})\big)\\
&\quad - 144 \operatorname{Im}\operatorname{Li}_4\big(\frac{i}{\sqrt{3}}\big) - 18 \operatorname{Im}\operatorname{Li}_4\big(\frac{1}{2}(-1+i\sqrt{3})\big) + 36 \operatorname{Im}\operatorname{Li}_4\big(\frac{1}{2}(3-i\sqrt{3})\big)\bigg)\epsilon^4,\\
B_{25} &= \frac{i}{2}\bigg(12\zeta_2 \operatorname{Im}\operatorname{Li}_2\big(\frac{1}{2}(-1+i\sqrt{3})\big) - 27 \operatorname{Im}\operatorname{Li}_4\big(\frac{1}{2}(-1+i\sqrt{3})\big)\bigg)\epsilon^4,\\
B_{26} &= -\zeta_2\epsilon^2 - 10\zeta_3\epsilon^3 - \frac{1}{4}\bigg(47\zeta_4 - 36 \operatorname{Im}\operatorname{Li}_2\big(\frac{1}{2}(-1+i\sqrt{3})\big)^2\bigg)\epsilon^4,\\
B_{27} &= -6i \operatorname{Im}\operatorname{Li}_2\big(\frac{1}{2}(-1+i\sqrt{3})\big)\epsilon^2 - \frac{i}{2}\bigg( -16 \operatorname{Im}\operatorname{Li}_3\big(\frac{i}{\sqrt{3}}\big) + 23 \operatorname{Im}\operatorname{Li}_3\big(\frac{1}{2}(-1+i\sqrt{3})\big)\\
&\quad - 4 \operatorname{Im}\operatorname{Li}_3\big(\frac{1}{2}(3-i\sqrt{3})\big)\bigg)\epsilon^3 + \frac{i}{9}\bigg(100\zeta_2 \operatorname{Im}\operatorname{Li}_2\big(\frac{1}{2}(-1+i\sqrt{3})\big) + 46\zeta_2 \operatorname{Im}\operatorname{Li}_2\big(\frac{1}{2}(3-i\sqrt{3})\big)\\
&\quad - 144 \operatorname{Im}\operatorname{Li}_4\big(\frac{i}{\sqrt{3}}\big) - 18 \operatorname{Im}\operatorname{Li}_4\big(\frac{1}{2}(-1+i\sqrt{3})\big) + 36 \operatorname{Im}\operatorname{Li}_4\big(\frac{1}{2}(3-i\sqrt{3})\big)\bigg)\epsilon^4,\\
B_{28} &= \frac{i}{2}\bigg(12\zeta_2 \operatorname{Im}\operatorname{Li}_2\big(\frac{1}{2}(-1+i\sqrt{3})\big) - 27 \operatorname{Im}\operatorname{Li}_4\big(\frac{1}{2}(-1+i\sqrt{3})\big)\bigg)\epsilon^4,\\
B_{29} &= \zeta_2\epsilon^2 + 10\zeta_3\epsilon^3 + \frac{1}{4}\bigg(47\zeta_4 - 36 \operatorname{Im}\operatorname{Li}_2\big(\frac{1}{2}(-1+i\sqrt{3})\big)^2\bigg)\epsilon^4,\\
B_{30} &= \frac{27i}{2} \operatorname{Im}\operatorname{Li}_4\big(\frac{1}{2}(-1+i\sqrt{3})\big)\epsilon^4,\\
B_{31} &= -\frac{i}{2}\bigg( -12\zeta_2 \operatorname{Im}\operatorname{Li}_2\big(\frac{1}{2}(-1+i\sqrt{3})\big) - 27 \operatorname{Im}\operatorname{Li}_4\big(\frac{1}{2}(-1+i\sqrt{3})\big)\bigg)\epsilon^4,
\end{aligned}
$$

$$B_{32} = -\frac{i}{2}\Big( - 12\zeta_2 \text{ Im Li}_2\big(\frac{1}{2}(-1+i\sqrt{3})\big) - 27 \text{ Im Li}_4\big(\frac{1}{2}(-1+i\sqrt{3})\big)\Big)\epsilon^4,$$

$$B_{33} = -2\zeta_3\epsilon^3 + \frac{1}{2}\Big( - \zeta_4 - 36 \text{ Im Li}_2\big(\frac{1}{2}(-1+i\sqrt{3})\big)^2\Big)\epsilon^4,$$

$$B_{34} = \frac{27i}{2} \text{ Im Li}_4\big(\frac{1}{2}(-1+i\sqrt{3})\big)\epsilon^4,$$

$$B_{35} = -\frac{i}{2}\Big( - 12\zeta_2 \text{ Im Li}_2\big(\frac{1}{2}(-1+i\sqrt{3})\big) - 27 \text{ Im Li}_4\big(\frac{1}{2}(-1+i\sqrt{3})\big)\Big)\epsilon^4,$$

$$B_{36} = -\frac{i}{2}\Big( - 12\zeta_2 \text{ Im Li}_2\big(\frac{1}{2}(-1+i\sqrt{3})\big) - 27 \text{ Im Li}_4\big(\frac{1}{2}(-1+i\sqrt{3})\big)\Big)\epsilon^4,$$

$$B_{37} = -2\zeta_3\epsilon^3 + \frac{1}{2}\Big( - \zeta_4 - 36 \text{ Im Li}_2\big(\frac{1}{2}(-1+i\sqrt{3})\big)^2\Big)\epsilon^4,$$

$$B_{38} = \frac{27i}{2} \text{ Im Li}_4\big(\frac{1}{2}(-1+i\sqrt{3})\big)\epsilon^4,$$

$$B_{39} = -\frac{i}{2}\Big( - 12\zeta_2 \text{ Im Li}_2\big(\frac{1}{2}(-1+i\sqrt{3})\big) - 27 \text{ Im Li}_4\big(\frac{1}{2}(-1+i\sqrt{3})\big)\Big)\epsilon^4,$$

$$B_{40} = -\frac{i}{2}\Big( - 12\zeta_2 \text{ Im Li}_2\big(\frac{1}{2}(-1+i\sqrt{3})\big) - 27 \text{ Im Li}_4\big(\frac{1}{2}(-1+i\sqrt{3})\big)\Big)\epsilon^4,$$

$$B_{41} = -2\zeta_3\epsilon^3 + \frac{1}{2}\Big( - \zeta_4 - 36 \text{ Im Li}_2\big(\frac{1}{2}(-1+i\sqrt{3})\big)^2\Big)\epsilon^4.$$

To verify the correctness of our results, we numerically evaluate the master integrals at multiple phase-space points. For this, we parametrize the one-forms on a path, use a series expansion of the integrands around ($x = 0, y = 0, z = 0$), and integrate the iterated integrals at some phase-space point. We then match the results against the numerical values obtained from **pySecdec** [101] and **AMFlow** [102] at the same phase-space point and find good agreement ($\sim$ 100 digits). We present the numerical results of one of the master integrals from the top sectors $J_{41}$ up to $\mathcal{O}(\epsilon^4)$ evaluated at point $(x, y, z)$ = (0.5, 0.5, 0.5) up to 90 decimal places in table 5.2.

| $\mathcal{O}$ | $J_{41}$ |
|---|---|
| $\epsilon^0$ | 0 |
| $\epsilon^1$ | 0 |
| $\epsilon^2$ | 0.480453013918201424667102526326664971730552951594545586866864133623665382259834472199948263 |
| $\epsilon^3$ | 0.091897357209531231888638645571102756639040958784768669397674639926722578733737539395677303 |
| $\epsilon^4$ | 0.778715275596109429538387498220501297387117166408046400975462356038382264999189687267373506 |

TABLE 5.2: Numerical values of $J_{41}$ evaluated at $(x, y, z)$ = (0.5, 0.5, 0.5) for first five orders of $\epsilon$ using analytic results.

Note that the analytical results for master integrals presented in this work, expressed as iterated integrals over logarithmic kernels, are suitable for numerical computations in the Euclidean region, where these iterated integrals are real. However, for phenomenological applications, we require the computation of master integrals in the physical region, which necessitates additional work. Specifically, we need to perform an analytical continuation of the iterated integrals across the branch cuts and branch points appropriately, which is left for future investigation.

# 6

# Renormalization and Parameter Scheme

IN the previous chapter, we obtained the analytic results for the 41 master integrals appearing in the form-factors for the $HV_1V_2$ vertex correction amplitude $(\delta M^{\alpha\alpha_s}_{HV_1V_2})$ at $\mathcal{O}(\alpha\alpha_s)$, discussed in chapter 4. These two-loop form-factors contributing to the amplitude for the $HV_1V_2$ vertex corrections and the amplitude for the self-energy corrections $(\delta M^{\alpha\alpha_s}_{S.E.})$ can exhibit both ultraviolet (UV) and infrared (IR) divergences due to unconstrained loop momenta in the involved two-loop integrals. In order to make any physical interpretation using the total $\mathcal{O}(\alpha\alpha_s)$ two-loop amplitude for $H \to e^+e^-\mu^+\mu^-$, it is crucial to eliminate these divergences. In this chapter, we will discuss in detail how to eliminate these divergences to get finite results for the amplitude at $\mathcal{O}(\alpha\alpha_s)$.

To regulate both the UV and IR divergences simultaneously, we adopt the dimensional regularization scheme (discussed in section 2.2.1 of chapter 2), working in $d = 4 - 2\epsilon$ space-time dimensions. After regularization, the divergences are encoded in the two-loop amplitude as poles in $\epsilon$, with $\frac{1}{\epsilon^4}$ being the highest order of pole that can appear. Notably, the $\frac{1}{\epsilon^3}$ and $\frac{1}{\epsilon^4}$ poles are exclusively due to the IR singularities, while $\frac{1}{\epsilon^2}$ and $\frac{1}{\epsilon}$ poles can be due to both IR and UV singularities. According to the KLN (Kinoshita-Lee-Nauenberg) theorem [235–238], the IR singularities eventually get cancelled against the real emission Feynman diagrams to give IR safe observables. As mentioned previously in section 4.3 of chapter 4, since the amplitude at $\mathcal{O}(\alpha\alpha_s)$ does not receive contributions from any real emission diagram, the absence of real corrections thus demands the complete cancellation of virtual IR divergences among the contributing diagrams. Thus, the two-loop amplitude at $\mathcal{O}(\alpha\alpha_s)$ is expected to be completely free from the IR divergences in our case.

Since the contributing form-factors $A$ and $B$ for the $HV_1V_2$ vertex corrections are independent, at the two-loop, they should be separately free from $\frac{1}{\epsilon^4}$ and $\frac{1}{\epsilon^3}$ poles. This fact will provide one of the important checks on our calculation. Furthermore, since the form-factor $B$ is zero at the tree-level, and the first non-zero contribution to it arises at the one-loop, we expect that the two-loop form-factor $B$ does not have $\frac{1}{\epsilon^2}$ UV pole dependence. This can serve as another consistency check on our calculation.

As of now, due to the absence of IR divergences, we only need to address the elimination of UV divergences from the virtual two-loop amplitude at $\mathcal{O}(\alpha\alpha_s)$, which we will discuss in the following section. The procedure of removing UV poles is called *renormalization*, involving the redefinition of couplings, masses, and fields of the theory under consideration.

## 6.1 Renormalization of Virtual Corrections

In order to get rid of the UV divergences, we have employed the *counterterm approach*. This method involves the replacement of UV-divergent fields and parameters with their respective finite renormalized counterparts, along with introducing divergent renormalization constants called *counterterms*. The inclusion of these counterterms ensures that the divergences present in the unrenormalized amplitude are precisely cancelled out, resulting in a UV finite amplitude after renormalization. This process allows us to separate the renormalized finite virtual amplitude into two components: the divergent unrenormalized amplitude, referred to as the bare amplitude, and the corresponding counterterm amplitude. This division can be expressed as

$$M_{renor.} = M_{bare} + M_{CT}. \tag{6.1}$$

The $M_{CT}$ part of the amplitude receives contribution from the Feynman diagrams with modified vertices and propagators involving divergent counterterms. These counterterms are later fixed through renormalization conditions, which mainly relate the renormalized parameters to physical parameters and can be chosen arbitrarily. The choice of renormalization conditions essentially defines the *renormalization scheme*.

In our calculations, we have employed the standard *on-shell (OS) renormalization scheme*, first introduced by Ross and Taylor [74], for the renormalization of all electroweak parameters and fields involved. In this renormalization scheme, the counterterms are fixed in such a way that the finite renormalized parameters match the physical parameters at all perturbative orders. This renormalization scheme has advantages such as all the parameters in the renormalized theory have physical meaning, and can be directly measured in suitable experiments. Interested readers, for more details on the on-shell renormalization scheme and corresponding renormalization conditions, can refer to [75]. In the following subsections, we will discuss the renormalization of our bare two-loop amplitude for the $HV_1V_2$ vertex corrections and the self-energy corrections utilizing the OS renormalization scheme.

### 6.1.1 Renormalization of Amplitude for $HV_1V_2$ Vertex Corrections

To renormalize the divergent bare amplitude for $HV_1V_2$ vertex corrections ($M^{\alpha\alpha_s}_{HV_1V_2}$), the CT amplitude receives contribution from 48 one-loop triangle and 4 tree-level CT diagrams. The representative CT diagrams are shown in figure 6.1. The list of required counterterm Feynman rules for the involved quark propagator and vertices is provided in appendix A.

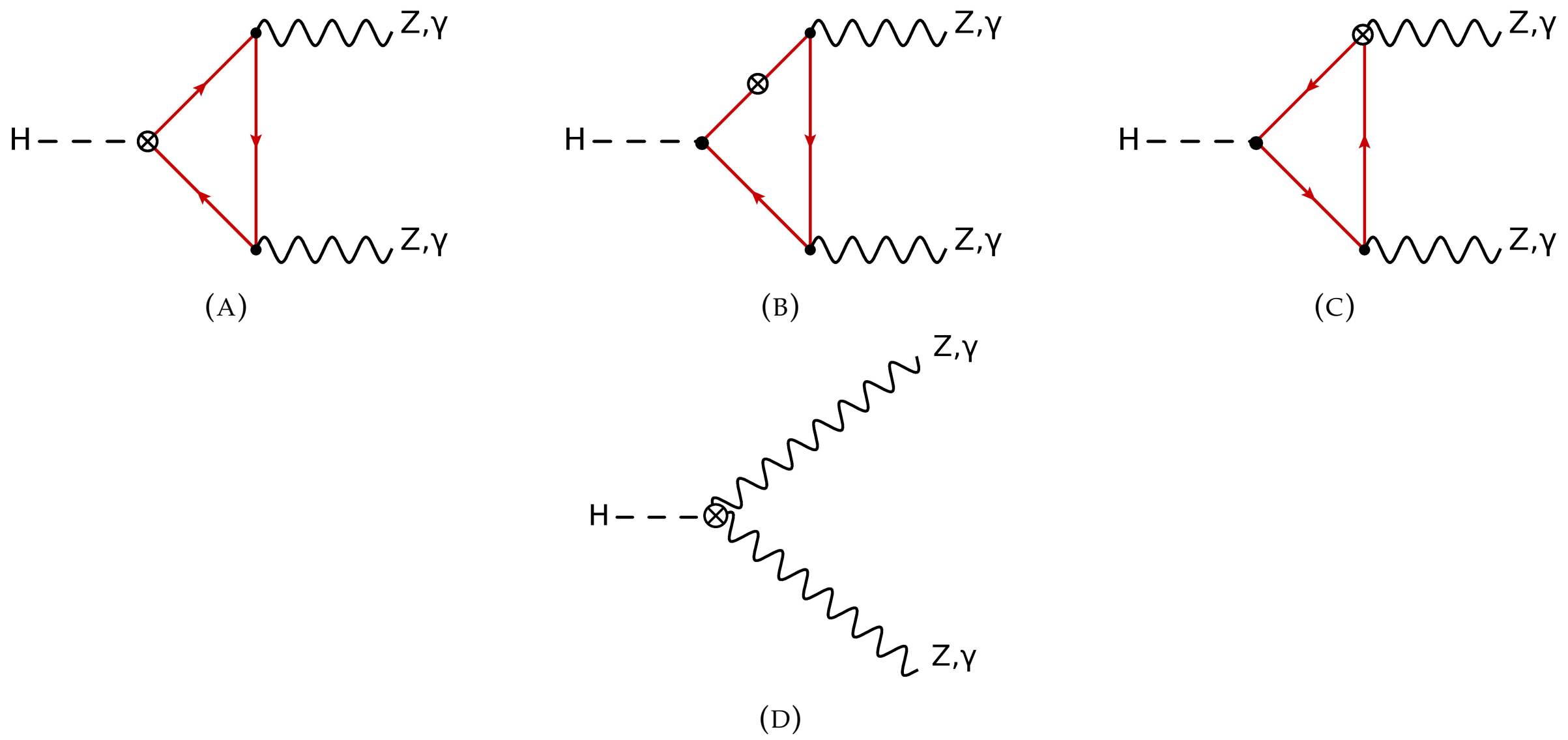

FIGURE 6.1: Representative one-loop triangle and tree-level CT diagrams. The counterterm vertex denoted by a crossed circle is proportional to $\alpha_s$ for the triangle CT diagrams and is proportional to $\alpha\alpha_s$ for the tree-level CT diagrams.

The one-loop triangle counterterm diagrams shown in figures 6.1 (A-C) mainly involve counterterm insertions on the $Vt\bar{t}$ and $Ht\bar{t}$ vertices, as well as counterterms for the top-quark mass and wave function. As indicated by the counterterm Feynman rules in appendix A, the renormalization of the quark wave function is related to the vertex renormalization. At the $\mathcal{O}(\alpha\alpha_s)$, the vertex counterterms involving quarks completely cancel out the contributions from the quark wave function counterterms. It is not surprising, as in our calculation the total amplitude does not depend on the quark wave function. Thus, the complete cancellation of the quark wave function counterterm provides another important check on our calculation. Therefore, at the one-loop level, we only need to evaluate diagrams with the top-quark mass counterterm, denoted by $\delta m_t$, insertions into the internal top-quark propagators, as well as into the $Ht\bar{t}$ vertex. In the OS scheme, mass counterterm $\delta m_t$ is fixed in such a way that the real part of the pole of the renormalized top-quark propagator appears at the physical or on-shell top-quark mass $(m_t)$ [239]. Specifically, the OS condition is

$$\frac{\delta m_t}{m_t} = \frac{1}{4m_t^2}\,\mathrm{Tr}(\not{p} + m_t)\ \mathrm{Re}\ \Sigma(p)\Big|_{p^2=m_t^2}, \tag{6.2}$$

where $p$ is the four-momentum of the top-quark and $\Sigma(p)$ denotes the self-energy of the top-quark due to a gluon exchange. A straightforward calculation in the OS scheme yields $\mathcal{O}(\alpha_s)$ expression for $\delta m_t$ with $d = 4 - 2\epsilon$ as [240]

$$\delta m_t = -m_t\,\Gamma(1+\epsilon)\left(\frac{4\pi\mu^2}{m_t^2}\right)\frac{C_F}{4}\frac{\alpha_s}{\pi}\frac{3-2\epsilon}{\epsilon(1-2\epsilon)}, \tag{6.3}$$

where $C_F = 4/3$ is the quadratic Casimir factor of the fundamental representation of $SU(3)$ and $\mu$ is the regularization mass scale introduced to make the coupling constant dimensionless.

On the other hand, the evaluation of the tree-level CT diagrams shown in figure 6.1 (D) requires the $\mathcal{O}(\alpha\alpha_s)$ contributions to the following renormalization constants: the gauge-boson mass renormalization constants $\delta M_W^2$, $\delta M_Z^2$, the gauge-boson field renormalization constants $\delta\mathcal{Z}_{V'V}$ $(VV' = ZZ, Z\gamma, \gamma Z, \gamma\gamma)$, the renormalization constant $\delta\sin\theta_W$ for the weak mixing angle, and the charge renormalization constant $\delta Z_e$. Utilizing the OS conditions, i.e., the position of the poles of the propagator equals to the square of the physical mass and the residue of the propagator equals 1, to fix these counterterms (for details refer to [75]), leads to the following expressions for the required renormalization constants. The $\mathcal{O}(\alpha\alpha_s)$ contributions to the required renormalization constants are expressed in terms of transverse part of the gauge-boson self-energies $\Sigma_T^{V'V}$ $(VV' = ZZ, Z\gamma, \gamma Z, \gamma\gamma, WW)$ obtained using the conventions of [75] by re-placing the one-loop gauge-boson self-energies with their two-loop $\mathcal{O}(\alpha\alpha_s)$ counterparts [204][1] (provided in appendix B).

$$\delta M^2_{W,(\alpha\alpha_s)} = \text{Re}\ \Sigma^{WW}_{\text{T},(\alpha\alpha_s)}(M_W^2), \tag{6.4}$$

$$\delta M^2_{Z,(\alpha\alpha_s)} = \text{Re}\ \Sigma^{ZZ}_{\text{T},(\alpha\alpha_s)}(M_Z^2), \tag{6.5}$$

$$\delta \mathcal{Z}_{ZZ,(\alpha\alpha_s)} = -\frac{\partial}{\partial s}\text{Re}\ \Sigma^{ZZ}_{\text{T},(\alpha\alpha_s)}(s)\Big|_{s=M_Z^2}, \tag{6.6}$$

$$\delta \mathcal{Z}_{\gamma\gamma,(\alpha\alpha_s)} = -\frac{\partial}{\partial s}\text{Re}\ \Sigma^{\gamma\gamma}_{\text{T},(\alpha\alpha_s)}(s)\Big|_{s=0}, \tag{6.7}$$

$$\delta \mathcal{Z}_{Z\gamma,(\alpha\alpha_s)} = \frac{2}{M_Z^2}\text{Re}\ \Sigma^{\gamma Z}_{\text{T},(\alpha\alpha_s)}(s)\Big|_{s=0}, \tag{6.8}$$

$$\delta \mathcal{Z}_{\gamma Z,(\alpha\alpha_s)} = -\frac{2}{M_Z^2}\text{Re}\ \Sigma^{\gamma Z}_{\text{T},(\alpha\alpha_s)}(s)\Big|_{s=M_Z^2}. \tag{6.9}$$

In the OS scheme, the weak mixing angles $\sin\theta_W = s_W$ and $\cos\theta_W = c_W$ are dependent parameters and are expressed in terms of renormalized masses of the gauge-bosons ($W$, $Z$) to all orders in perturbation theory as

$$s_W^2 = 1 - c_W^2 = 1 - \frac{M_W^2}{M_Z^2}. \tag{6.10}$$

Due to the above relation, the renormalization constants $\delta s_W$ and $\delta c_W$ for the weak mixing angles are related to the mass renormalization constants of $W$ and $Z$ bosons and their required expressions at $\mathcal{O}(\alpha\alpha_s)$ read as

$$\frac{\delta s_{W,(\alpha\alpha_s)}}{s_W} = -\frac{c_W^2}{s_W^2}\frac{\delta c_{W,(\alpha\alpha_s)}}{c_W} = -\frac{c_W^2}{2s_W^2}\left(\frac{\delta M^2_{W,(\alpha\alpha_s)}}{M_W^2} - \frac{\delta M^2_{Z,(\alpha\alpha_s)}}{M_Z^2}\right), \tag{6.11}$$

The above equation (6.11) can be further expressed in terms of gauge-boson self-energies utilizing the equations (6.4) and (6.5).

Lastly, we will briefly discuss the renormalization of electromagnetic charge $e$. In the OS scheme, $e$ is defined as the coupling of $e^+e^-\gamma$ vertex function in the Thomson limit, i.e. for zero photon momentum transfer. The corresponding charge renormalization constant $\delta Z_e$ is determined by requiring that all the corrections to this $ee\gamma$ vertex vanish for on-shell electrons and for zero momentum transfer [75, 241]. Using the Ward identities, the charge renormalization constant $\delta Z_e$ can be expressed in terms of gauge-boson self energies instead of vertex functions, as shown for the NLO case in [241]. The $\mathcal{O}(\alpha\alpha_s)$ expression for $\delta Z_e$ analogous to NLO [241],

[1] In these Refs. $\gamma$ is denoted by $A$

takes the form

$$\delta Z_{e,(\alpha\alpha_s)} = \frac{\delta e}{e} = -\frac{1}{2}\delta Z_{\gamma\gamma,(\alpha\alpha_s)} - \frac{s_W}{c_W}\frac{1}{2}\delta Z_{Z\gamma,(\alpha\alpha_s)},$$
$$= \frac{1}{2}\frac{\partial}{\partial s}\text{Re }\Sigma^{\gamma\gamma}_{T,(\alpha\alpha_s)}(s)\Big|_{s=0} - \frac{s_W}{c_W}\frac{\text{Re }\Sigma^{\gamma Z}_{T,(\alpha\alpha_s)}(s)}{M_Z^2}\Big|_{s=0}, \quad (6.12)$$

which simplifies at $\mathcal{O}(\alpha\alpha_s)$ to

$$\delta Z_{e,(\alpha\alpha_s)} = \frac{1}{2}\frac{\partial}{\partial s}\text{Re }\Sigma^{\gamma\gamma}_{T,(\alpha\alpha_s)}(s)\Big|_{s=0}, \quad (6.13)$$

because at $\mathcal{O}(\alpha\alpha_s)$ the contribution to $\Sigma^{\gamma Z}_{T,(\alpha\alpha_s)}(0)$ vanishes [203]. The on-shell electric charge counterterm $\delta e$ at $\mathcal{O}(\alpha^2)$ and $\mathcal{O}(\alpha\alpha_s)$ is discussed in [242]. Note that the above expression for $\delta Z_e$ is applicable for the $\alpha(0)$ scheme only, where $\alpha$ assumes its Thomson-limit value. Care should be taken with $\delta Z_e$ when switching to a different input parameter scheme for $\alpha$, as discussed later in section 6.3.

### 6.1.2 Renormalization of Amplitude for Self-energy Corrections

The CT amplitude for the renormalization of the bare amplitude for the self-energy corrections ($\delta M_{S.E.}$) receives contributions from 96 one-loop self-energy and 4 tree-level CT diagrams. The representative CT diagrams are shown in figure 6.2.

The one-loop self-energy diagrams, depicted in figures 6.2 (A-C), involve counterterm insertions on the $Vq\bar{q}$ $(q = u, d, c, s, b, t)$ vertex and the quark propagator. Due to the QED-like Ward identity, the counterterms of quark vertex and wavefunction cancel each other out completely [203], and we only need to consider the diagrams with quark mass counterterm insertions. On the other hand, for the evaluation of the tree-level counterterm diagrams shown in figure 6.2 (D), one needs $\mathcal{O}(\alpha\alpha_s)$ expressions for the renormalization constants $\delta Z_{ZZ}$, $\delta M_Z^2$, $\delta Z_{\gamma Z}$ and $\delta Z_{Z\gamma}$, which can be deduced from the gauge-boson self-energies given in [203, 240], as discussed previously in section 6.1.1.

Note that, in our calculations, we have not explicitly accounted for contributions from self-energy insertions on the Higgs leg. This omission is because we consider the Higgs boson as on-shell. Therefore, all the self-energy corrections to the external Higgs leg can be taken into account via its wave-function renormalization, in accordance with the Lehmann, Symanzik and Zimmerman (LSZ) formalism [4]. In our calculations, this contribution is incorporated by inserting a factor of $\frac{1}{2}\delta\mathcal{Z}_H$ in the $HZZ$ counter vertex Feynman rule (see appendix A), where $\delta\mathcal{Z}_H$ denotes the Higgs wave-function renormalization constant. In the OS scheme, the $\mathcal{O}(\alpha\alpha_s)$ contribution to the $\delta\mathcal{Z}_H$ is given by [75],

$$\delta\mathcal{Z}_{H,(\alpha\alpha_s)} = -\frac{\partial}{\partial s}\mathrm{Re}\,\Sigma^H_{(\alpha\alpha_s)}(s)\Big|_{s=M_H^2}, \tag{6.14}$$

where $\Sigma^H(s)$ represents the Higgs boson self-energy. The $\mathcal{O}(\alpha\alpha_s)$ contributions to $\Sigma^H(s)$ (provided in appendix B) are available in [240] in terms of scalar function denoted by $\Pi^H(s)$.

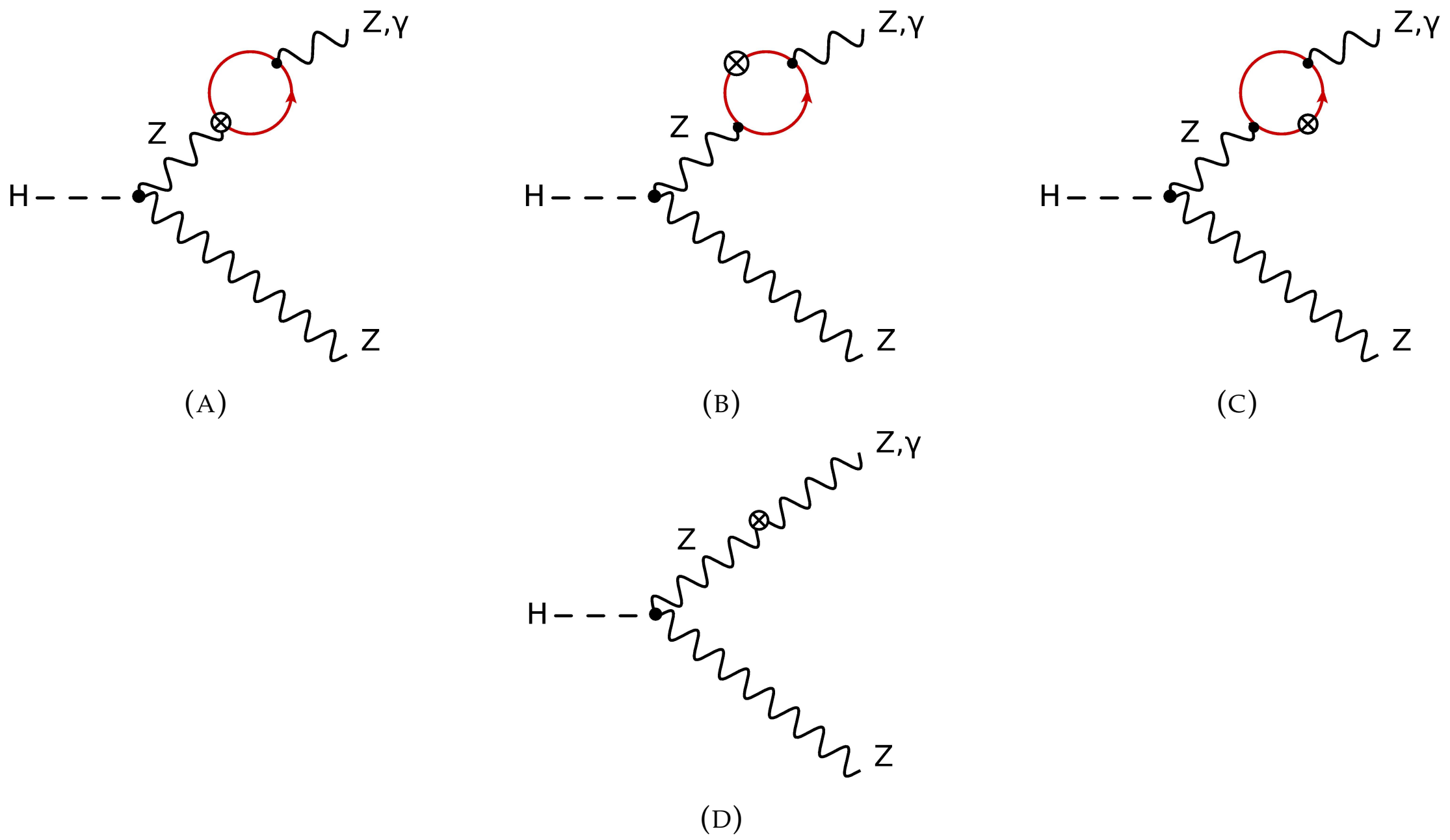

FIGURE 6.2: Representative one-loop self-energy and tree-level CT diagrams. The counterterm vertex denoted by a crossed circle is proportional to $\alpha_s$ for the one-loop self-energy CT diagrams and is proportional to $\alpha\alpha_s$ for the tree-level CT diagrams. The diagrams with counterterm insertions on the lower leg are not shown.

In addition to the renormalization of electroweak parameters (fields) involved in our $\mathcal{O}(\alpha\alpha_s)$ amplitude calculation, we must also address the renormalization of the strong coupling constant $\alpha_s$. We have opted the *Modified Minimial Subtraction ($\overline{MS}$) scheme* for the renormalization of $\alpha_s$. In this scheme, the strong coupling constant becomes scale-dependent, denoted by $\alpha_s(Q)$. In our perturbative calculation of the amplitude for $H \to e^+e^-\mu^+\mu^-$ up to the two-loop level, $\alpha_s$ appears for the first time at the two-loop level. Therefore, with respect to the $\alpha_s$, the $\mathcal{O}(\alpha\alpha_s)$ two-loop corrections we are interested in are at the LO. Hence, it is sufficient to consider the one-loop running of the $\alpha_s$, given by the expression

$$\alpha_s(Q) = \frac{\alpha_s(Q_R)}{1 + \alpha_s(Q_R)\frac{\beta_0}{4\pi}\log\left(\frac{Q^2}{Q_R^2}\right)}; \qquad \beta_0 = \frac{11}{3}C_A - \frac{4}{3}T_R n_q. \tag{6.15}$$

Here $n_q$ is the number of active quark flavors below the scale $Q$, $\beta_0$ is the one-loop $\beta$-function coefficient. $C_A = 3$ is the color factor associated with gluon emission from a gluon, and $T_R = 1/2$ is the color factor for a gluon splitting into a $q\bar{q}$ pair. $Q_R$ is some reference scale at which the value of $\alpha_s(Q)$ is known with accuracy. The most common choice is to take $Q_R$ equal to the mass of the $Z$-boson, $Q_R = M_Z \approx 91.1876$ GeV, at which $\alpha_s(Q_R) \approx 0.118$ [37].

Next, the two-loop diagrams for the process under consideration involve unstable particles, such as the $Z$-bosons and the top quark, in the propagators. Their presence introduces complications in perturbative calculations beyond the leading order. In the following section, we will discuss these issues and explore potential solutions.

## 6.2 Unstable Particles and the Complex-Mass Scheme

In this section, we will address the challenges posed by the introduction of finite decay width in the propagators of internal unstable particles during the calculation of matrix elements. Additionally, we will introduce the *complex-mass scheme* [78–80], which we have used for the calculations presented in this work.

Treating an intermediate massive unstable particle (Mass $M$) as stable can lead to divergences in the matrix elements when its propagator $(1/(p^2 - M^2))$ goes on-shell i.e. when for some momentum transfer $p$, we have $p^2 = M^2$. These divergences can be removed by at least summing the imaginary part of the self-energy corrections of the particle near the pole, known as *Dyson resummation* [241]. Denoting the renormalized self-energy of the particle by $\Sigma_R(p^2)$, the Dyson summation modifies the particle's LO propagator as follows

$$\frac{1}{p^2 - M^2} \to \frac{1}{p^2 - M^2 + \Sigma_R(p^2)}. \tag{6.16}$$

The *optical theorem*, a consequence of the unitarity of the S-matrix, establishes a relationship between the decay width $(\Gamma)$ of an unstable particle and the imaginary part of the renormalized self-energy (see references [4, 243]). In one-loop approximation, this relationship is given by

$$\Gamma = \frac{\text{Im}\Sigma_R(M^2)}{M}, \tag{6.17}$$

Moreover, since $\text{Re}\Sigma_R(M^2) = 0$ for the on-shell renormalized mass $M$, the propagator for the unstable particle near the pole behaves like $\dfrac{1}{p^2 - M^2 + i\Gamma M}$.

As the perturbative summation of self-energies includes higher-order terms, the introduction of finite widths in the propagators of unstable particles represents a higher-order effect and can lead to several issues, such as the violation of gauge invariance [76, 77] and gauge dependence in fixed-order calculations. Various schemes have been introduced to address these issues related to unstable particles, and an overview of possible solutions can be found in [241].

In this work, we have adopted the gauge-invariant *complex-mass scheme* (CMS), first introduced in [244] and generalized to one-loop calculations in [78], to address these problems related to unstable particles. In the CMS, we analytically continue the masses of the weak gauge bosons to the complex momentum ($k$) plane, denoting them as $\mu_V^2$, and define them as the locations of the propagator poles ($k^2 = \mu_V^2$) as follows

$$\mu_V^2 = M_V^2 - iM_V\Gamma_V \quad (V = W, Z), \tag{6.18}$$

where $M_V$ and $\Gamma_V$ are the pole masses and widths which differ from their on-shell definitions denoted by $M_V^{OS}$ and $\Gamma_V^{OS}$ at the one-loop and beyond. The relation between the two sets of values is given by [1, 241]

$$M_V = \frac{M_V^{OS}}{\sqrt{1 + \left(\frac{\Gamma_V^{OS}}{M_V^{OS}}\right)^2}}, \quad \Gamma_V = \frac{\Gamma_V^{OS}}{\sqrt{1 + \left(\frac{\Gamma_V^{OS}}{M_V^{OS}}\right)^2}} \quad (V = W, Z). \tag{6.19}$$

To maintain gauge invariance, complex masses must be consistently used in all the Feynman rules, including the definition of the weak mixing angle, which in the CMS becomes

$$c_W^2 = 1 - s_W^2 = \frac{\mu_W^2}{\mu_Z^2} = \frac{M_W^2 - iM_W\Gamma_W}{M_Z^2 - iM_Z\Gamma_Z}. \tag{6.20}$$

In the CMS, complex masses are introduced not only for the gauge-bosons but also for other unstable particles, such as the Higgs boson and the top quark. However, in our $\mathcal{O}(\alpha\alpha_s)$ calculations, we take the Higgs mass to be 125 GeV, which is below the $t\bar{t}$ threshold. Therefore, there is no need to introduce the finite width for the top quark running in the loop. Additionally, since the Higgs is taken to be on-shell, there is no need to introduce complex mass for the Higgs boson either.

Since the renormalization conditions for stable and unstable particles remain the same. Thus the CMS is just a generalization of the on-shell scheme where we replace the masses and counterterms with their corresponding complex counterparts. Therefore, for our calculations in the CMS, the required complex gauge-boson mass renormalization constants $\delta\mu_W^2$, $\delta\mu_Z^2$, and the gauge-boson field renormalization constants $\delta\mathcal{Z}_{V'V}$ at $\mathcal{O}(\alpha\alpha_s)$ take a similar form as in the on-shell scheme, expressed in terms of self-energies [245] but without taking the real part, as follows

$$\delta\mu^2_{W,(\alpha\alpha_s)} = \Sigma^{WW}_{T,(\alpha\alpha_s)}(\mu_W^2), \tag{6.21}$$

$$\delta\mu^2_{Z,(\alpha\alpha_s)} = \Sigma^{ZZ}_{T,(\alpha\alpha_s)}(\mu_Z^2), \tag{6.22}$$

$$\delta\mathcal{Z}_{ZZ,(\alpha\alpha_s)} = -\left.\frac{\partial\Sigma^{ZZ}_{T,(\alpha\alpha_s)}(k^2)}{\partial k^2}\right|_{k^2=\mu_Z^2}, \tag{6.23}$$

$$\delta\mathcal{Z}_{\gamma\gamma,(\alpha\alpha_s)} = -\left.\frac{\partial\Sigma^{\gamma\gamma}_{T,(\alpha\alpha_s)}(k^2)}{\partial k^2}\right|_{k^2=0}, \tag{6.24}$$

$$\delta\mathcal{Z}_{Z\gamma,(\alpha\alpha_s)} = \left.\frac{2}{\mu_Z^2}\Sigma^{\gamma Z}_{T,(\alpha\alpha_s)}(k^2)\right|_{k^2=0}, \tag{6.25}$$

$$\delta\mathcal{Z}_{\gamma Z,(\alpha\alpha_s)} = -\left.\frac{2}{\mu_Z^2}\Sigma^{\gamma Z}_{T,(\alpha\alpha_s)}(k^2)\right|_{k^2=\mu_Z^2}. \tag{6.26}$$

Similarly, for the complex cosine and sine of the weak mixing angle, the renormalization constants $\delta c_W$ and $\delta s_W$ are given by

$$\frac{\delta s_{W,(\alpha\alpha_s)}}{s_W} = -\frac{c_W^2}{s_W^2}\frac{\delta c_{W,(\alpha\alpha_s)}}{c_W} = -\frac{c_W^2}{2s_W^2}\left(\frac{\delta\mu^2_{W,(\alpha\alpha_s)}}{\mu_W^2} - \frac{\delta\mu^2_{Z,(\alpha\alpha_s)}}{\mu_Z^2}\right). \tag{6.27}$$

Finally, the charge renormalization constant $\delta Z_{e,(\alpha\alpha_s)}$ is fixed via the following condition in the CMS

$$\delta Z_{e,(\alpha\alpha_s)} = \frac{1}{2}\left.\frac{\partial\Sigma^{AA}_{T,(\alpha\alpha_s)}(k^2)}{\partial k^2}\right|_{k^2=0} - \frac{\sin\theta_W}{\cos\theta_W}\frac{\Sigma^{AZ}_{T,(\alpha\alpha_s)}(0)}{\mu_Z^2}. \tag{6.28}$$

In the next section, we will introduce the input parameter scheme used for the numerical calculations presented in this work.

## 6.3 Electroweak Input Parameter Scheme

In higher-order calculations, to guarantee the gauge independence and consistency in the results, it is important to stick to a set of independent parameters. For the input parameter set, well-defined and precisely known parameters are preferred. Among the SM parameters, an obvious choice for the input parameter set includes the electromagnetic coupling constant ($\alpha$), the strong coupling ($\alpha_s$) and the masses of the weak gauge-bosons $M_W$ and $M_Z$, the mass of the Higgs boson ($M_H$), the masses of the fermions $m_f$, and the Cabibbo-Kobayashi-Maskawa (CKM) matrix $V$.

Depending on the choice for the input value of the electromagnetic coupling constant $\alpha$, three different input-parameter schemes are introduced [241]:

1. $\boldsymbol{\alpha(0)}$ **scheme:** In this scheme, the fine-structure constant $\alpha$ is renormalized in the Thomson limit, namely at the scale of zero photon momentum transfer, leading to the renormalized value $\alpha = \alpha(0) \approx 1/137$. This scheme is suitable for processes involving external photons in the initial or final state. However, for processes at energies comparable to or exceeding the masses of the electroweak gauge-bosons, it results in large logarithmic corrections depending on the ratios of the masses of the light fermions and the EW gauge-bosons.

2. $\alpha(m_Z)$ **scheme:** This parameter scheme involves the effective electromagnetic coupling at the scale of the $Z$ mass i.e., $\alpha(m_Z) \approx 1/129$, obtained by evolving the $\alpha(0)$ from the scale $Q = 0$ to $M_Z$ via renormalization group inspired equation:

$$\alpha(M_Z) = \frac{\alpha(0)}{1 - \Delta\alpha(M_Z)}, \tag{6.29}$$

   where the quantity $\Delta\alpha(M_Z)$ in the denominator comes from the resummation of large light-fermion logarithms. The corresponding charge renormalization constant in this scheme gets modified as

$$\delta Z_e|_{\alpha(M_Z)} = \delta Z_e|_{\alpha(0)} - \frac{1}{2}\Delta\alpha(M_Z) \tag{6.30}$$

   Switching from the $\alpha(0)$ scheme to the $\alpha(M_Z)$ scheme effectively subtracts the $\Delta\alpha(M_Z)$ terms from the EW corrections, canceling all the large light-fermion logarithms appearing in charge renormalization. Therefore, for high-energy processes without external photons, the $\alpha(M_Z)$ scheme is preferred over the $\alpha(0)$ scheme.

3. $G_\mu$ **scheme:** In this scheme, instead of choosing the fine-structure constant $\alpha$, the Fermi constant $G_\mu$, most precisely measured in the muon decay, is chosen as the input for the EW coupling strength. The electromagnetic coupling constant $\alpha$ is derived using the relation

$$\alpha_{G_\mu} = \frac{\sqrt{2}G_\mu M_W^2}{\pi}\left(1 - \frac{M_W^2}{M_Z^2}\right). \tag{6.31}$$

The charge renormalization constant in this scheme is related to the one in $\alpha(0)$ scheme as follows

$$\delta Z_e|_{G_\mu} = \delta Z_e|_{\alpha(0)} - \frac{1}{2}\Delta r, \tag{6.32}$$

where $\Delta r$ that quantifies all higher-order radiative corrections to muon decay [246, 247] can be written as

$$\Delta r = \Delta\alpha(M_Z) - \frac{c_W^2}{s_W^2}\Delta\rho + \delta r, \tag{6.33}$$

here $\Delta r$, in addition to $\Delta\alpha(M_Z)$, also contains $\Delta\rho$ which receives isospin-violating corrections induced by the heavy top-quark mass. Additionally, it contains $\delta r$, called finite remainder involving corrections not considered in $\Delta\alpha(M_Z)$ and $\Delta\rho$ (refer to [241] and references therein for details). Therefore, using the $G_\mu$ scheme leads to the elimination of sensitivity of the corrections to the universal large logarithms from the light fermion loops and some non-logarithmic large terms arising from the heavy top quark loop.

In this work, we have employed the $G_\mu$ scheme for the calculations. For the mixed QCD-electroweak corrections under consideration, the $\mathcal{O}(\alpha\alpha_s)$ contributions to $\Delta r$, denoted as $\Delta r_{(\alpha\alpha_s)}$, are simplified because the finite remainder $\delta r$ receives no contribution at this order [204, 245]. Thus, the contributions to $\Delta r_{(\alpha\alpha_s)}$ come entirely from the fermion-loop contributions to the gauge-boson self-energies, given in the on-shell scheme as

$$\begin{aligned}\Delta r_{(\alpha\alpha_s)} =& \left.\frac{\partial\Sigma^{\gamma\gamma}_{T,(\alpha\alpha_s)}(k^2)}{\partial k^2}\right|_{k^2=0} - \frac{c_W^2}{s_W^2}\left(\frac{\Sigma^{ZZ}_{T,(\alpha\alpha_s)}(M_Z^2)}{M_Z^2} - \frac{\Sigma^{WW}_{T,(\alpha\alpha_s)}(M_W^2)}{M_W^2}\right)\\ &+ \frac{\Sigma^{WW}_{T,(\alpha\alpha_s)}(0) - \Sigma^{WW}_{T,(\alpha\alpha_s)}(M_W^2)}{M_W^2},\end{aligned} \tag{6.34}$$

where we have utilized the fact that at $\mathcal{O}(\alpha\alpha_s)$ contribution to $\Sigma^{\gamma Z}_{T,(\alpha\alpha_s)}(0)$ vanishes, as stated earlier. The $\mathcal{O}(\alpha\alpha_s)$ corrections to $\Delta r$ were presented in [203, 248]. The $\mathcal{O}(\alpha\alpha_s)$ expression for $\Delta r$ in the CMS scheme can be obtained by replacing the masses and renormalization constants with their complex counterparts.

This chapter provided a detailed discussion on the renormalization of our $\mathcal{O}(\alpha\alpha_s)$ amplitude for the $HV_1V_2$ vertex corrections and self-energy corrections using the complex-mass scheme. The renormalization process results in finite amplitudes for the $HV_1V_2$ vertex and the self-energy corrections. In the next chapter, we will discuss the numerical results of our calculations, obtained by utilizing these $\mathcal{O}(\alpha\alpha_s)$ UV finite amplitudes.

# 7

# Numerical Implementation and Results

IN the previous chapter, we discussed the renormalization of our $\mathcal{O}(\alpha\alpha_s)$ amplitude for the $HV_1V_2$ vertex corrections and the self-energy corrections using the complex-mass scheme. This chapter focuses on the numerical implementation of our calculation and the checks performed to ensure its correctness, along with the presentation and discussion of the numerical results obtained. The results presented in this chapter have been published in [67].

Section 7.1 provides a detailed discussion on the implementation of our UV finite $\mathcal{O}(\alpha\alpha_s)$ amplitude into a publicly available event generation code **`Hto4l`** [46]. Section 7.2 presents the numerical results provided by our implementation in the **`Hto4l`** code, including various differential distributions for the final state leptons.

## 7.1 Numerical Implementation and Checks

We combine the $\mathcal{O}(\alpha\alpha_s)$ UV finite amplitudes obtained after renormalization for the $HV_1V_2$ vertex and self-energy corrections with the fermionic currents to obtain the two-loop amplitude for $H \to e^+e^-\mu^+\mu^-$. Adding the finite $Z\ell\bar{\ell}$ contribution to it, we get the total renormalized two-loop amplitude, denoted by $\widehat{M}_2^{\alpha\alpha_s}$. In the perturbative expansion of amplitude-squared up to two-loop, this matrix element interferes with the LO amplitude as

$$|M|^2 = |M_0|^2 + |M_1^{(\alpha)}|^2 + 2\,\mathrm{Re}\left[\widehat{M}_2^{\alpha\alpha_s}(M_0)^*\right]. \tag{7.1}$$

Where, the interference term is organized in terms of two-loop form-factors and renormalization constants using the symbolic manipulation program **FORM** [201, 202], and a **FORTRAN** output is obtained for the numerical evaluation.

To obtain the partial decay width, we perform phase-space integration over the final state leptons. This is done using the publicly available **Hto4l**[1] code developed by Boselli et al. [46]. This code is a Monte Carlo program that generates unweighted events, for the channels $H \to \ell^+\ell^-\ell'^+\ell'^-$ (with $\ell, \ell' = e, \mu$), up to NLO electroweak accuracy matched with QED Parton Shower (PS), and in the presence of dimension-six operators [46, 249]. It can be used for calculating the partial width of the on-shell decay of the Higgs boson into 4 charged leptons or along with any event generator that provides events for the production of the Higgs boson in the Les Houches standard format.

We implement the **FORTRAN** output of our two-loop interference term given in equation (7.1) into this **Hto4l** code to obtain the improved predictions for the partial decay width of $H \to e^+e^-\mu^+\mu^-$ channel with an accuracy of $\mathcal{O}(\alpha\alpha_s)$. In order to achieve good accuracy on the observables of interest, a huge number of phase-space points must be sampled. The main computational bottleneck in our implementation arises from calculating the two-loop form-factors of the $HV_1V_2$ vertex correction amplitude for each phase point, which involves 41 non-trivial two-loop master integrals. While we have derived analytic results for these integrals in chapter 5, they are not yet optimized for numerical efficiency across the entire phase-space region relevant for the $H \to e^+e^-\mu^+\mu^-$ channel. To make these results suitable for phenomenological applications, further efforts are needed, including a thorough analysis of the integrals' properties and their relations, as well as the search for specific path choices that enhance efficient numerical evaluations. These tasks require time and are left for future investigations.

To ensure efficient numerical evaluations for each phase point, we find it more convenient to prepare a two-dimensional grid covering all the relevant phase-space points for the form-factors

[1] For detailed information regarding the code, visit: http://www.pv.infn.it/hepcomplex/hto4l.html

$A$ and $B$. This grid is constructed by numerically evaluating 41 master integrals. To perform these numerical evaluations, we utilize our in-house code based on the sector-decomposition method [250, 251] implementing a quasi-Monte Carlo method associated with CUDA/GPU technique – a reliable computational technique is detailed in [252].

This grid is parametrized in terms of $p_1^2$ and $p_2^2$ values, where $p_1$ and $p_2$ denote the virtualities of the gauge-bosons $V_1$ and $V_2$, using the input parameters given in section 7.2.1. The grid is prepared with an accuracy of $\mathcal{O}(10^{-3})$, which is sufficient for phenomenological studies but can be increased as needed.

The grid is then used to estimate the form-factors at random phase-space points with the help of a linear interpolation code developed in-house. It is important to note that any change in the input parameter set would require a new grid.

We interface the squared matrix elements, the grid of the form-factors, and the interpolation code with the **`Hto4l`** code to perform the phase-space integration. This allows us to obtain the partial decay width and kinematic distributions for the final state leptons at $\mathcal{O}(\alpha\alpha_s)$. In order to prove the reliability of our implementation, we have performed the following checks:

1. To good numerical accuracy, we find that the $\frac{1}{\epsilon^4}$ and $\frac{1}{\epsilon^3}$ poles cancel in both the form-factors $A$ and $B$. In the form-factor $B$, the $\frac{1}{\epsilon^2}$ pole also vanishes. The UV poles in the form-factors $A$ and $B$ cancel after adding the CTs, and the result does not depend on the choice of the scale $\mu$ in the dimensional regularization. These checks have been performed for several phase-space points.

2. By taking the gluon propagator in the $R_\xi$ gauge, we find that the two-loop form-factors, and consequently the two-loop amplitude, are gauge-parameter independent. Additionally, the Ward identity for the $HV_1V_2$ vertex demands $A + p_1.p_2\ B + p_1^2\ D = 0$, and we have verified this relation for the two-loop amplitude at $\mathcal{O}(\alpha\alpha_s)$.

3. As mentioned earlier, the two-loop diagrams for $H \to e^+e^-\mu^+\mu^-$ are closely related to the ones appearing in the production process $e^+e^- \to ZH$. In reference [218] analytical expressions for the contributing form-factors are given up to order $m_t^0$ after series expanding them in powers of $\frac{1}{m_t}$. In order to check the accuracy of the grid prepared for the form-factors, we produced the grid for $e^+e^- \to ZH$ taking a very large value of the top-quark mass ($m_t$). Further, we matched the numerical values of the form-factors from the grid with those given in [218]. We found excellent agreement between the two for different values of the center-of-mass energies.

4. The correctness of our numerical implementation is checked via reproducing the results for the mixed QCD-electroweak corrections for the $e^+e^- \to ZH$ process given in Ref. [253] in the $G_\mu$ and $\alpha(0)$ schemes. We performed this check by implementing our calculation in **`MadGraph`** [254], and we found that the calculated corrections matched the available results in both the schemes with a relative error of less than $1\%$.

## 7.2 Numerical Results

In this section, we will provide the improved predictions for the partial decay width of $H \to e^+e^-\mu^+\mu^-$ with an accuracy of $\mathcal{O}(\alpha\alpha_s)$ obtained from our calculations implemented in the **`Hto4l`** code. We also discuss the impact of these $\mathcal{O}(\alpha\alpha_s)$ corrections compared to the LO predictions and NLO corrections on the kinematical and angular distributions of interest for the final state leptons. The angular distributions in section 7.2.4 are defined in the rest frame of the Higgs boson.

### 7.2.1 Input Parameters and Setup

We use the following set of input parameters,

$$G_\mu = 1.1663787 \times 10^{-5}\ \mathrm{GeV}^{-2}, \qquad M_Z^{OS} = 91.1876\ \mathrm{GeV}, \quad M_W^{OS} = 80.379\ \mathrm{GeV},$$
$$\Gamma_Z^{OS} = 2.4952\ \mathrm{GeV}, \qquad \Gamma_W^{OS} = 2.141\ \mathrm{GeV}, \qquad M_H = 125\ \mathrm{GeV},$$
$$m_t = 173\ \mathrm{GeV}, \qquad \alpha_s(M_Z) = 0.1185.$$

We work in the $G_\mu$ scheme and derive the electromagnetic coupling constant $\alpha$ from the Fermi constant using the equation (6.31). As $\alpha$ is renormalized on-shell, it is scale-independent.

In the complex-mass scheme, from the above list of input parameters, the on-shell values of the masses and widths ($M_V^{OS}, \Gamma_V^{OS}$) of the $W$ and $Z$ bosons are transformed into their corresponding *pole values* denoted by $M_V$ and $\Gamma_V$ according to the equation (6.19), resulting in

$$M_Z = 91.1535\ \mathrm{GeV}, \qquad M_W = 80.3505\ \mathrm{GeV},$$

$$\Gamma_Z = 2.49427 \text{ GeV}, \qquad \Gamma_W = 2.1402 \text{ GeV}. \tag{7.2}$$

It is worth mentioning that the difference between using $(M_V^{OS}, \Gamma_V^{OS})$ or $(M_V, \Gamma_V)$ is hardly visible in the numerical results presented here. The strong coupling constant is renormalized in the $\overline{MS}$ scheme and is scale-dependent. Thus, the results for the mixed QCD-electroweak corrections are obtained using both fixed and running $\alpha_s(Q)$. We take $Q = M_Z$ to obtain results with the fixed value of the QCD coupling. For the running of $\alpha_s(Q)$ using equation (6.15), we have used transverse momentum of the lepton $p_T(\ell^-)$ and the invariant mass of the $\ell^+\ell^-$ pair $(M_{\ell^+\ell^-})$ as the running scale.

### 7.2.2 Results for Partial Decay Width

We define $\delta_i = \Gamma_{\textit{two-loop}}/\Gamma_i$ $(i =$ LO, NLO) to indicate the relative two-loop correction with respect to the LO contribution and NLO EW correction. In Table 7.1, we report the relative contribution of the mixed QCD-electroweak correction to the leading order partial decay width of $H \to e^+e^-\mu^+\mu^-$ for $M_H = 125$ GeV. The table shows results for different scale choices $Q$ for the strong coupling constant.

| $\Gamma_{\text{LO}}$ (KeV) | $\delta_{\text{LO}}[\%]$ | | |
|---|---|---|---|
| | $\alpha_s(M_Z)$ | $\alpha_s(M_{\ell^+\ell^-})$ | $\alpha_s(p_T(\ell^-))$ |
| 0.23726831 | 0.27 | 0.30 | 0.35 |

TABLE 7.1: Values of the LO partial decay width, and of the $\%$ mixed QCD-electroweak correction with respect to the leading order contribution, $\delta_{\text{LO}}$, at $M_H = 125$ GeV for different scale choices for the strong coupling $\alpha_s(Q)$.

We find that the mixed QCD-electroweak correction to the partial decay width is around 0.27% of the LO contribution for $\alpha_s$ at $Q = M_Z$[2]. With the running coupling, it becomes 0.30% for $Q = M_{\ell^+\ell^-}$ and 0.35% for $Q = p_T(\ell^-)$. To draw a comparison, we note that the inclusive two-loop QCD corrections in $H \to Z\gamma$ decay have been found around 0.22% of the LO [51, 210, 255]. The two-loop QCD corrections in the $H \to \gamma\gamma$ decay lie in the range of 1-2% for the intermediate Higgs mass below the $t\bar{t}$ threshold [256]. With respect to the NLO EW correction, the mixed QCD-electroweak correction amounts to 18% for the fixed and 21% (24%) for the running QCD coupling with $Q = M_{\ell^+\ell^-}$ $(Q = p_T(\ell^-))$.

[2]For any fixed scale choice $Q$ other than $M_Z$, the % correction can be estimated by multiplying 0.27 with a factor of $\alpha_s(Q)/\alpha_s(M_Z)$.

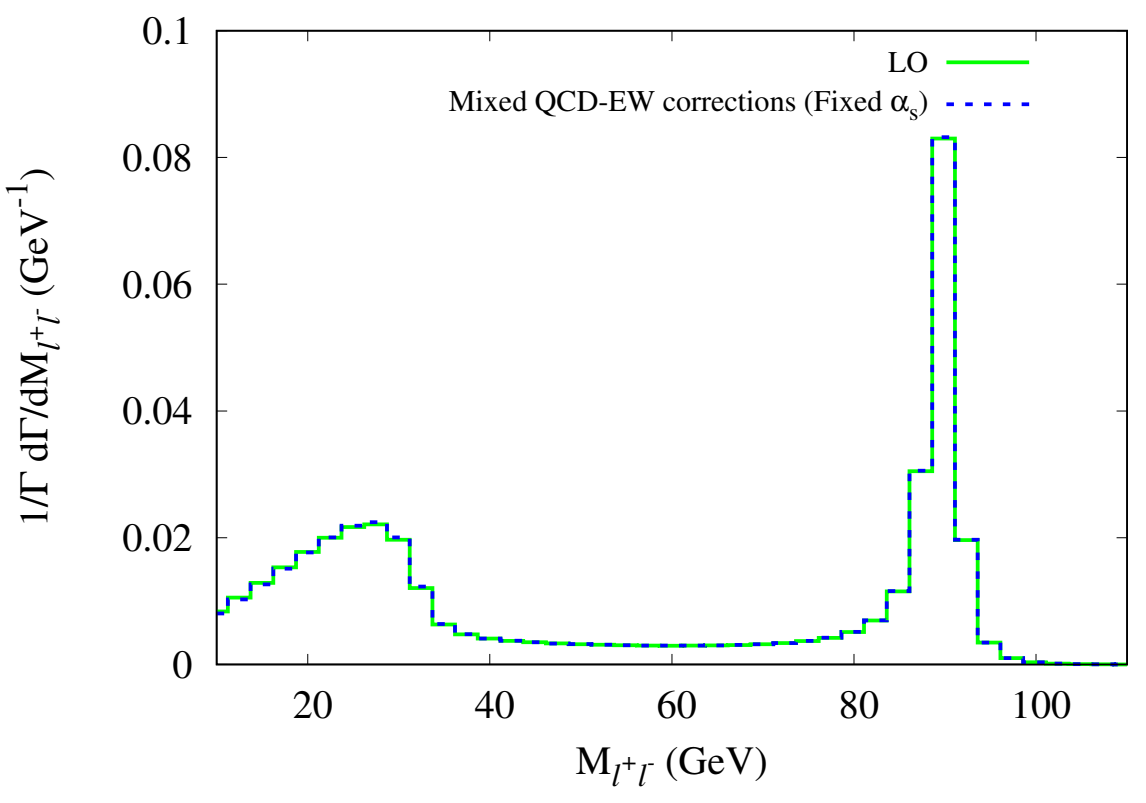

(A)

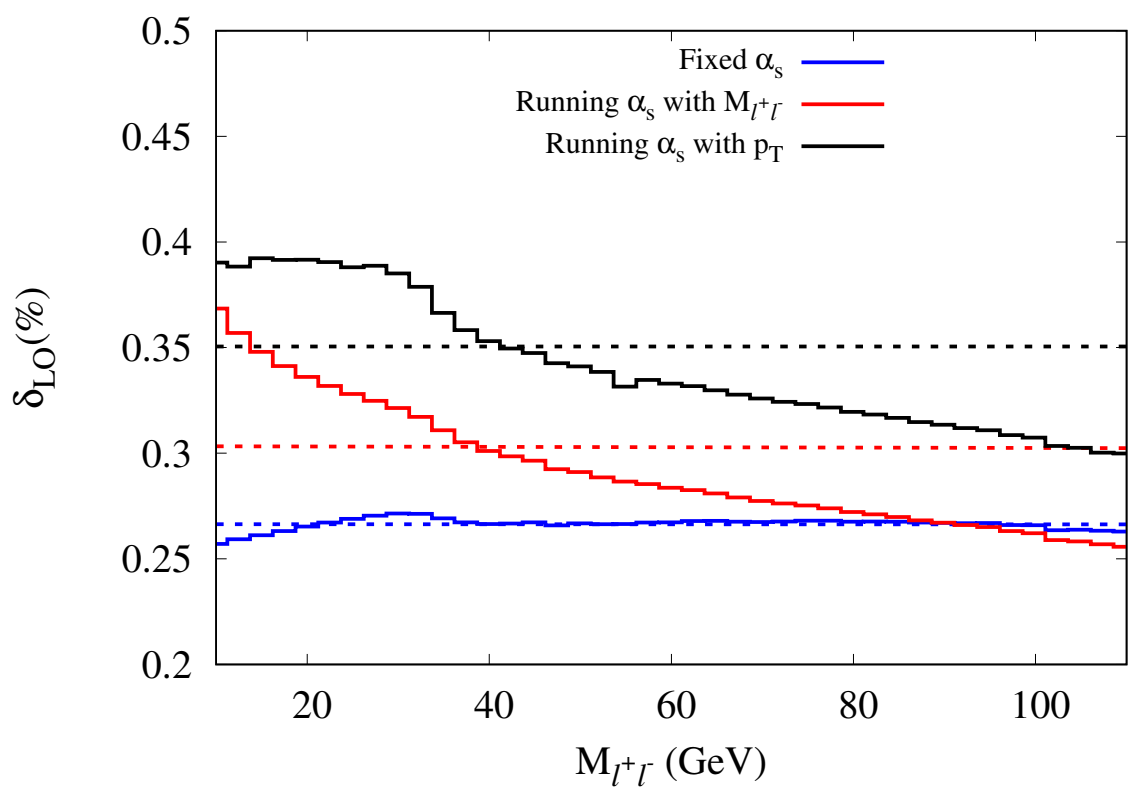

(B)

FIGURE 7.1: (A) Normalized differential distribution in the invariant mass of the final state lepton pair $\ell^+\ell^-$ $(\ell = e, \mu)$ for the LO contribution and the mixed QCD-electroweak corrections for fixed $\alpha_s$. (B) Effect of the mixed QCD-electroweak corrections relative to the LO predictions denoted by $\delta_{\text{LO}}$ on the invariant mass distribution of the final state lepton pair $\ell^+\ell^-$. The dotted straight lines mark the results at the inclusive level.

### 7.2.3 Invariant Mass Distributions

It is well known that the higher-order corrections are sensitive to the kinematics of the events. In figures 7.1 and 7.2, we investigate the impact of two-loop corrections with respect to the LO predictions and NLO EW corrections on the invariant mass distribution of the final state lepton pair.

For the fixed $\alpha_s$, the mixed QCD-electroweak corrections relative to the LO are roughly the same in all the bins and are of the order of 0.27%, as seen in figure 7.1 (B). This suggests that the nature of the events for the LO and two-loop corrections is kinematically similar, shown explicitly via the normalized differential distribution in figure 7.1 (A). The reason behind this is the absence of real corrections at the $\mathcal{O}(\alpha\alpha_s)$. Additionally, as seen in figure 7.1 (B), the relative two-loop corrections for the running $\alpha_s$ differ in each bin and are higher in the lower mass bins. It can go beyond 0.35% in the lower bins, depending on the choice of the running scale. For the invariant mass above the $Z$ pole, the corrections for the running $\alpha_s$ with $Q = M_{\ell^+\ell^-}$ become slightly smaller than the corrections for the fixed $\alpha_s$. This behavior is dictated by the one-loop running of $\alpha_s$.

In figure 7.2 (A), we plot the two-loop corrections with respect to the NLO EW corrections in the upper panel. Since the electroweak corrections are also sensitive to the kinematics of the events, we note that in certain bins, between 20 GeV and 40 GeV, $\delta_{\text{NLO}}$ becomes quite large, independent of the choice of the scale $Q$. The relative effect of the NLO EW corrections with respect to the LO contributions in each bin, as shown in the lower panel, can be referred to understand the features of the distribution in the upper panel of figure 7.2 (A). Significantly large values of $\delta_{\rm NLO}$ in certain bins are simply due to the fact that the NLO EW corrections are negligible in those bins. This observation can be useful for the bin-wise analysis of the data. The two-loop corrections appear flat in the bins between 40 GeV and 80 GeV. However, a closer look reveals that $\delta_{\rm NLO}$ decreases in higher mass bins due to larger NLO EW corrections in those bins. We have shown this in figure 7.2 (B). The inverse behavior of the distributions in the upper and lower panels can be attributed to the kinematic similarity between the two-loop events and the LO events noted earlier.

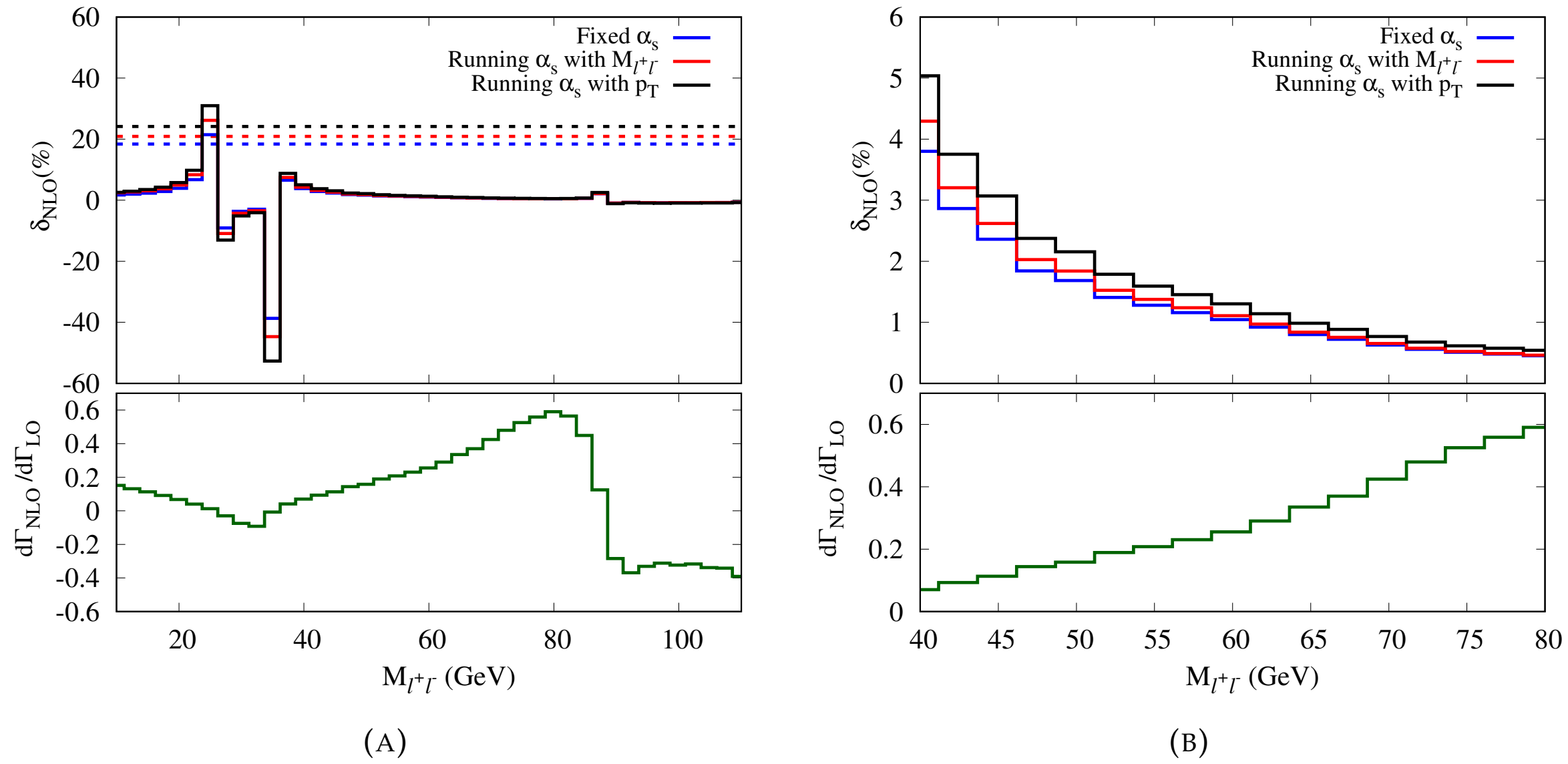

FIGURE 7.2: Effect of the mixed QCD-electroweak corrections on the invariant mass distribution of the final state lepton pair $\ell^+\ell^-$ relative to the NLO EW corrections denoted by $\delta_{\rm NLO}$ in the upper panels of (A) and (B). The quantity $\mathrm{d}\Gamma_{\rm NLO}/\mathrm{d}\Gamma_{\rm LO}$ in the lower panels of (A) and (B) gives the ratio of the NLO EW correction and the LO contribution. For clarity, plot (B) displays the information of plot (A) in the region between 40 GeV and 80 GeV. The distributions in blue and red (black) are the results obtained using the fixed scale $Q = M_Z$ and the running scale $Q = M_{\ell^+\ell^-}(p_T(\ell^-))$ for $\alpha_s$, respectively. The dotted straight lines mark the results at the inclusive level.

### 7.2.4 Angular Distributions

Apart from the invariant mass distribution, angular distributions are also effective in studying the Higgs properties. Therefore, we study the effect of the mixed QCD-electroweak corrections in the $\phi$ distribution, which is one of the most sensitive observables for the BSM studies. It is defined as the angle between the decay planes of the intermediate $Z$-bosons in the rest frame of the Higgs boson. This angle $\phi$ is the main observable for spin-parity assignment of the Higgs boson [257–263]. We have plotted the $\phi$ dependence of the relative corrections $\delta_i$ in figures 7.3 and 7.4.

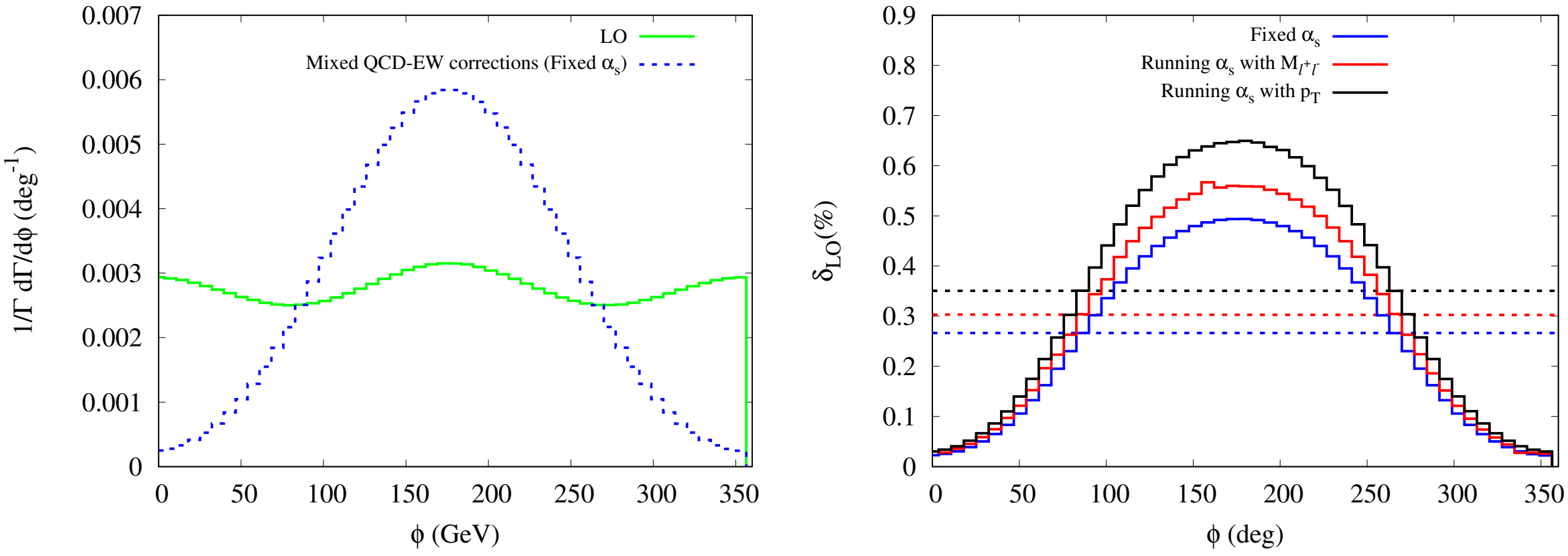

(A) (B)

FIGURE 7.3: (A) Normalized differential distribution in the angle $\phi$ between the decay planes of the intermediate $Z$-bosons for the LO contribution and the mixed QCD-electroweak corrections for fixed $\alpha_s$. (B) Effect of the mixed QCD-electroweak corrections relative to the LO predictions denoted by $\delta_{\rm LO}$ on the distribution for the angle $\phi$. The dotted straight lines mark the results at the inclusive level.

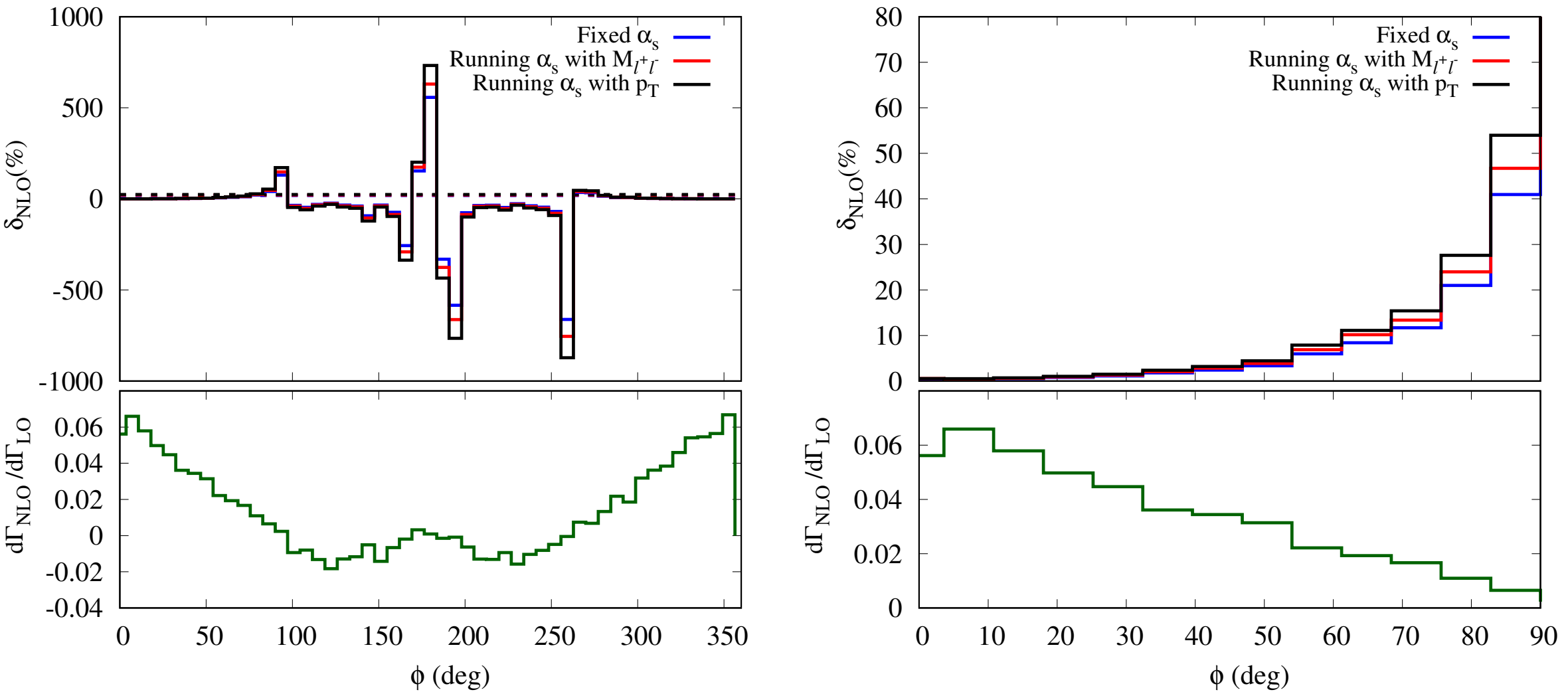

(A) (B)

FIGURE 7.4: Effect of the mixed QCD-electroweak corrections on the distribution for the angle $\phi$ between the decay planes of the intermediate $Z$-bosons relative to NLO EW corrections denoted by $\delta_{\rm NLO}$ in the upper panels of (A) and (B). The quantity $\mathrm{d}\Gamma_{\rm NLO}/\mathrm{d}\Gamma_{\rm LO}$ in the lower panels of (A) and (B) gives the ratio of the NLO EW correction and the LO contribution. For clarity, plot (B) displays the information of plot (A) in the region between 0 and $\frac{\pi}{2}$. The distributions in blue and red (black) are the results obtained using the fixed scale $Q = M_Z$ and the running scale $Q = M_{\ell^+\ell^-}(p_T(\ell^-))$ for $\alpha_s$, respectively. The dotted straight lines mark the results at the inclusive level.

In figure 7.3 (B), we see that the mixed QCD-electroweak corrections relative to the LO do not exhibit a flat behavior in the $\phi$-distribution for the fixed $\alpha_s$. This is in contrast to what we see in the case of the invariant mass distribution in figure 7.1 (B). We observe a $(1 - \cos\phi)$ dependence in the shape of the $\phi$-distribution due to the mixed QCD-electroweak corrections. Also, the shape is independent of the scale choice for $\alpha_s$. The stated dependence is different from the LO behavior, which follows a $\cos^2\phi$ dependence [263, 264], shown explicitly in figure 7.3 (A). The difference can be attributed to the change in the effective $HZZ$ coupling due to the two-loop corrections. Additionally, as shown in figure 7.3 (B), the two-loop corrections with respect to the LO are insignificant at the edges and large in the central region. The relative correction $\delta_{\rm LO}$ peaks at $\phi = \pi$. It is 0.49% for the fixed and 0.56% (0.64%) for the running QCD coupling with $Q = M_{\ell^+\ell^-}(p_T(\ell^-))$. Compared to the fixed $\alpha_s$, relative corrections are higher for the running $\alpha_s$ across all the bins.

In the upper panel of figure 7.4 (A), we have shown the relative effect of two-loop corrections with respect to the NLO EW corrections. In the lower panel of the figure, the NLO EW corrections with respect to the LO are displayed. In the mid-region of the distribution where the NLO EW corrections are negligible, more prominent peaks for two-loop corrections can be seen. The numerical values of these peaks should not be taken very seriously. It is just that the two-loop corrections are more relevant in the bins with peaks, and they should be taken into account in the bin-wise analysis of the data aimed at BSM searches. $\delta_{\rm NLO}$ for $\phi$ looks flat near the edges. However, it is indeed not the case as shown in figure 7.4 (B) for the bins between 0 and $\frac{\pi}{2}$. In this range, $\delta_{\rm NLO}$ rises as $\mathrm{d}\Gamma_{\rm NLO}/\mathrm{d}\Gamma_{\rm LO}$ goes down with increasing $\phi$. Similarly, in the region beyond $250^\circ$ as $\mathrm{d}\Gamma_{\rm NLO}/\mathrm{d}\Gamma_{\rm LO}$ increases, $\delta_{\rm NLO}$ decreases (not shown explicitly) with an increasing value of $\phi$. These features are independent of the scale choice for $\alpha_s$.

# 8

# Summary and Outlook

SINCE the momentous discovery of the Higgs boson – *the God particle* in 2012, the Higgs sector of the Standard Model has remained in the spotlight. Small discrepancies between experimental measurements and theoretical predictions within this sector hold the potential to unlock the mysteries of the universe. To detect these subtle discrepancies, alongside achieving the percent-level accuracy in measurements at the future Higgs factories, the pursuit of highly precise theoretical predictions of the Higgs-related observables is imperative. The precision of these theoretical predictions within the perturbative framework is mainly controlled by the higher-order corrections.

Given the importance of $H \to 4\ell$ channel in precision studies of the Higgs properties, in this thesis work we have provided the precise theoretical predictions for the partial decay width of $H \to ZZ^{(*)} \to e^+e^-\mu^+\mu^-$ channel, including $\mathcal{O}(\alpha\alpha_s)$ mixed QCD-electroweak corrections in its perturbative calculation [67]. The conventional Feynman diagram approach is employed to evaluate all the contributing matrix elements. Interestingly, matrix elements of $\mathcal{O}(\alpha\alpha_s)$ corrections to $HV_1V_2$ $(V_1, V_2 = Z, \gamma)$ vertex and $Z/\gamma$ self-energy receive contributions solely from virtual two-loop diagrams involving a quark-loop along with gluon dressing. For the $HV_1V_2$ vertex correction diagrams, where the Higgs boson directly couples with the quark loop, we solely consider the contribution from the top quark, anticipating that contributions from lighter quarks are negligible due to their weak interaction with the Higgs boson (described by Yukawa coupling). We have developed general-purpose codes for the systematic evaluation of amplitudes at $\mathcal{O}(\alpha\alpha_s)$ in terms of form-factors. The presence of divergent two-loop integrals poses a significant challenge in computing these amplitudes. Exploiting the relations between these integrals, we decompose the amplitudes into a linear combination of master integrals.

In this work, we have obtained the first-ever full analytic results for these master integrals specifically related to the $\mathcal{O}(\alpha\alpha_s)$ corrections to $HV_1V_2$ vertex by solving their respective differential equations while considering the full dependence of all the mass scales involved [66]. Remarkably, despite the presence of non-rationalizable square roots in the system of differential equations, the results for the master integrals are expressed using well-defined mathematical structures known as Chen's iterated integrals with logarithmic kernels along with analytic integration constants. The results presented are straightforwardly applicable to any production or decay process involving two-loop three-point Feynman diagrams with four mass scales.

Furthermore, to obtain the improved predictions for partial decay width of $H \to e^+e^-\mu^+\mu^-$ channel with an accuracy of $\mathcal{O}(\alpha\alpha_s)$, the contributing matrix elements after renormalization are implemented in a publicly available code **Hto4l** [46] to perform the phase-space integration. A pre-computed grid for the form factors, obtained through numerical evaluation of the contributing two-loop integrals using in-house codes, is interfaced with **Hto4l** code to ensure fast and efficient evaluation. The reliability of this implementation is validated by successfully reproducing existing results for mixed QCD-electroweak corrections for the $e^+e^- \to ZH$ process [253]. The final code is then used to obtain the partial decay width with an accuracy of $\mathcal{O}(\alpha\alpha_s)$ along with kinematic differential distributions for the $H \to e^+e^-\mu^+\mu^-$ process.

This work reveals that in the $G_\mu$ scheme, the mixed QCD-electroweak correction increases the partial decay width of Higgs decaying into four charged leptons ($e^+e^-\mu^+\mu^-$) by about $0.27\%$ with respect to the leading order prediction and by about $18\%$ with respect to the next-to-leading order electroweak correction. Interestingly, these $\mathcal{O}(\alpha\alpha_s)$ corrections do not significantly affect the invariant mass distribution of the final state leptons, for fixed $\alpha_s$, but do influence specific angular variables, such as the relative angle between the decay planes of the intermediate $Z$-bosons. Our findings are useful for the precision studies to probe BSM physics in the Higgs sector at future colliders. Needless to say, our computational framework also allows predictions for $H \to \gamma\gamma,\ \gamma Z,\ \gamma\ell^+\ell^-, Z\ell^+\ell^-$ decays. In addition to that, we can also use the ingredients calculated in this work to predict $\mathcal{O}(\alpha\alpha_s)$ corrections for the $H \to \ell^+\ell^-\ell^+\ell^-$ $(\ell = e, \mu)$.

The thesis presents improved predictions for the partial decay width of $H \to e^+e^-\mu^+\mu^-$, incorporating $\mathcal{O}(\alpha\alpha_s)$ corrections to the $HV_1V_2$ vertex solely from the top-quark loop diagrams. However, for the completeness of the study, it should be extended to include contributions from the bottom quark loop as well. While these bottom quark contributions are expected to be smaller compared to those from the top quark, they are still phenomenologically crucial and valuable for validating the previous findings and improving the accuracy of predictions. Extending the study to include bottom quark loop contributions would not only enhance the accuracy of predictions but also provide a more comprehensive understanding of mixed QCD-electroweak corrections at $\mathcal{O}(\alpha\alpha_s)$ to the golden decay channel ($H \to e^+e^-\mu^-\mu^+$) of the Higgs boson.

Moreover, to broaden the scope of our study, it is essential to compute $\mathcal{O}(\alpha\alpha_s)$ corrections to the partial decay width of $H \to 4f$, encompassing all possible final states (fully leptonic, semi-leptonic, and hadronic) at this order. The master integrals evaluated in chapter 5 are directly applicable for these final states as well. Therefore, it is imperative to perform the analytic continuation of the results obtained to access the entire relevant phase-space region. Additionally, dedicating sufficient time to obtaining a more compact form of these analytic results for the master integrals, utilizing the properties of the iterated integrals, would be advantageous, enabling faster numerical calculations with enhanced precision. With these modifications, the results can be used directly in event generation codes for the Higgs production and decay to carry out phenomenological studies and data analysis for future collider experiments such as $e^+e^-$, $\mu^+\mu^-$ and FCC-hh colliders.

In the long run, implementing the full calculation of the partial decay width of the Higgs boson decaying into four fermions, encompassing all possible final states, with an accuracy of $\mathcal{O}(\alpha\alpha_s)$ in a public code would be highly beneficial. This implementation would contribute to the wider scientific community by providing a valuable resource for high-precision calculations in the Higgs sector.

# Thesis Workflow

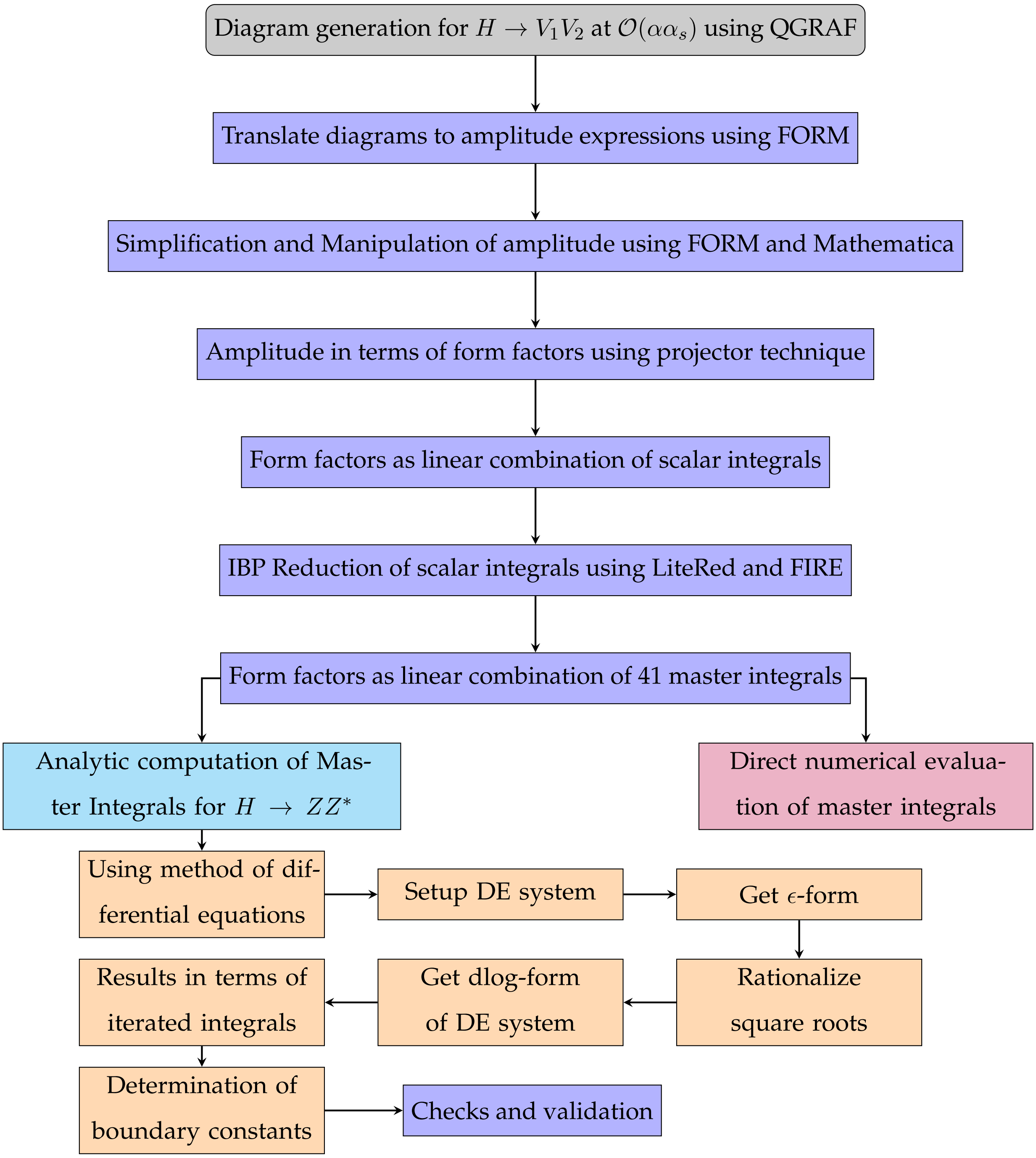

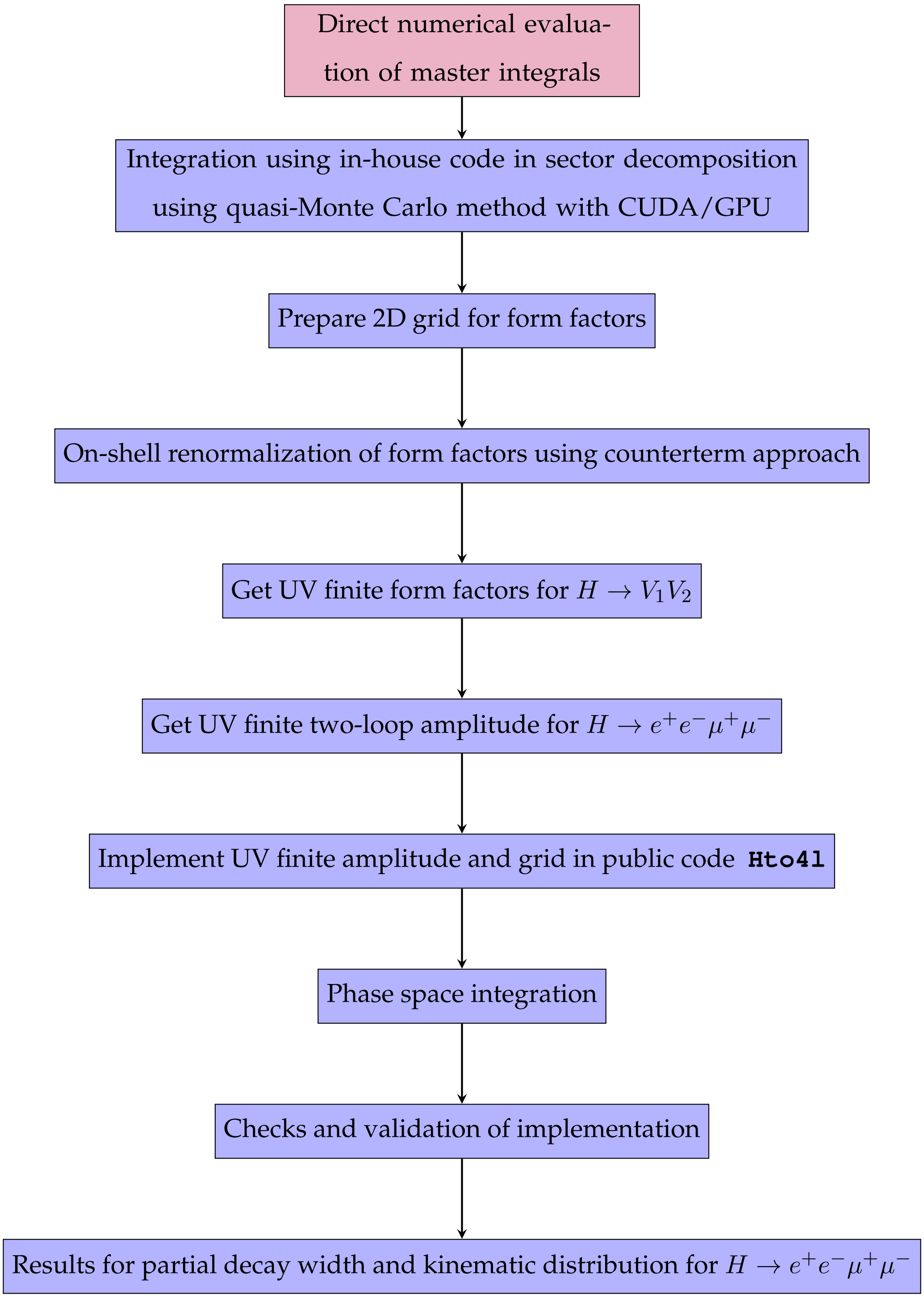
Direct numerical evaluation of master integrals
Integration using in-house code in sector decomposition using quasi-Monte Carlo method with CUDA/GPU
Prepare 2D grid for form factors
On-shell renormalization of form factors using counterterm approach
Get UV finite form factors for $H \to V_1 V_2$
Get UV finite two-loop amplitude for $H \to e^+e^-\mu^+\mu^-$
Implement UV finite amplitude and grid in public code **Hto4l**
Phase space integration
Checks and validation of implementation
Results for partial decay width and kinematic distribution for $H \to e^+e^-\mu^+\mu^-$

# Appendices

# A

# Feynman Rules

In this appendix, we list the selected Feynman rules of the Standard Model in the general $R_\xi$ gauge along with the counterterm rules which are relevant for the process under consideration in this thesis work i.e., $H \to Z^{(*)}Z^{(*)} \to e^+e^-\mu^+\mu^-$.

## Propagators:

1. Quark propagator

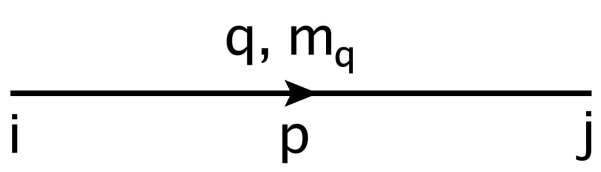

$$: \quad i\delta_{ij}\frac{\not{p} + m_q}{p^2 - m_q^2 + i\varepsilon}$$

2. Gluon propagator

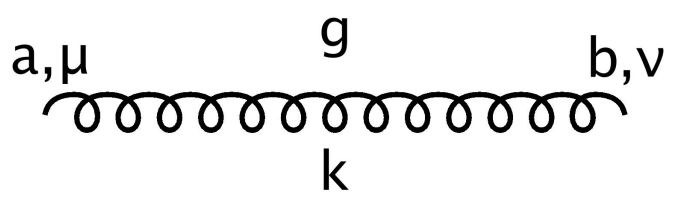

$$: \quad -i\delta_{ab}\frac{\left(g_{\mu\nu} - (1-\xi)\dfrac{k_\mu k_\nu}{k^2}\right)}{k^2 + i\varepsilon}$$

3. Photon propagator

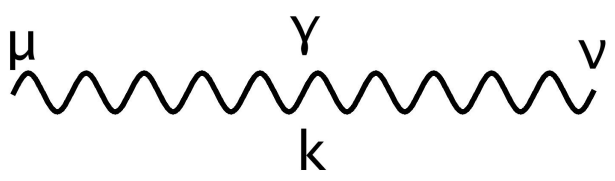

$$: \qquad -i\frac{\left(g_{\mu\nu} - (1-\xi)\frac{k_\mu k_\nu}{k^2}\right)}{k^2 + i\varepsilon}$$

4. Weak $Z$-boson propagator

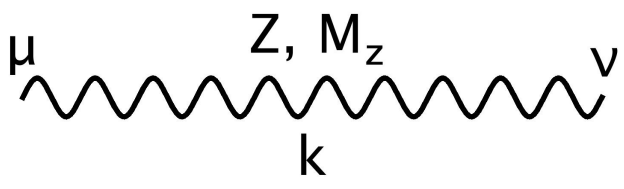

$$: \qquad -i\frac{\left(g_{\mu\nu} - (1-\xi)\frac{k_\mu k_\nu}{k^2 - \xi M_Z^2}\right)}{k^2 - M_Z^2 + i\varepsilon}$$

The two most common gauge choices are Feynman gauge, $\xi = 1$ and Landau gauge $\xi = 0$.

## Vertices:

1. Quark-gluon vertex

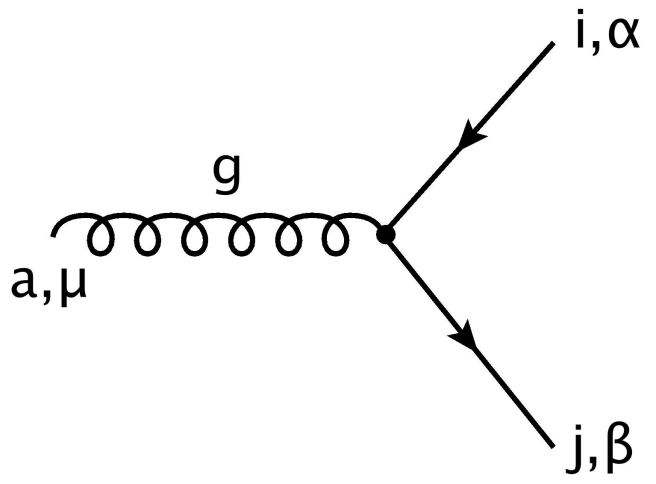

$$: \qquad -ig_s(\gamma_\mu)_{\beta\alpha}(\mathbf{T}_a)_{ji}$$

Here $\mathbf{T}_a$ are the generators of the $SU(3)$ gauge group.

2. Fermion-photon vertex

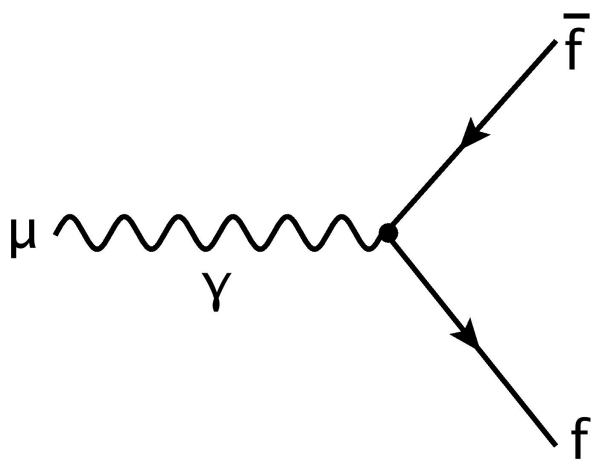

$$: \qquad -ieQ_f\gamma_\mu$$

3. Fermion-Higgs vertex

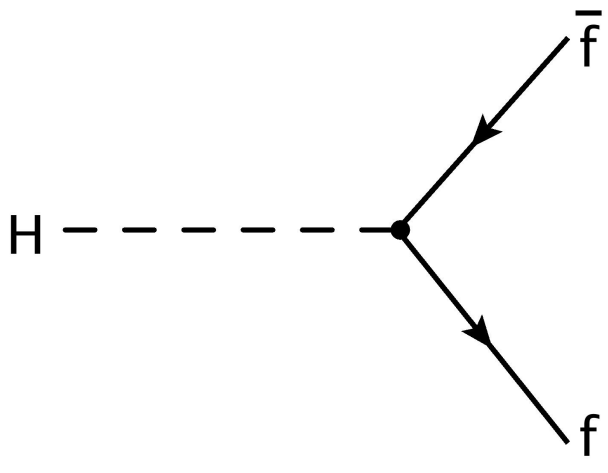

$$: \qquad -i\frac{m_f}{v}$$

4. Fermion-$Z$ boson vertex

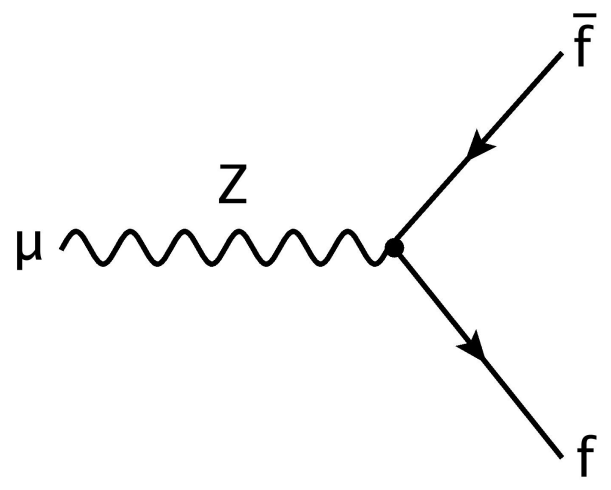

$$: \quad -ie\gamma_\mu(g_V^f \mathbb{1} - g_A^f \gamma_5)$$

$$g_V^f = \frac{1}{2\sin\theta_W \cos\theta_W}(I^3_{W,f} - 2Q_f \sin^2\theta_W)$$
$$g_A^f = \frac{1}{2\sin\theta_W \cos\theta_W} I^3_{W,f}.$$

5. $HZZ$ vertex

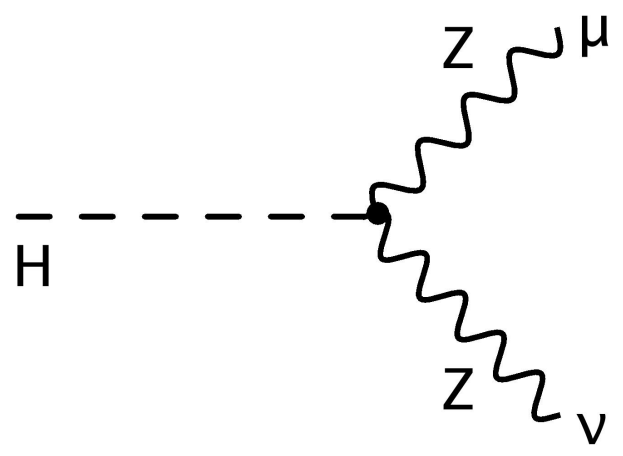

$$: \quad \frac{2iM_Z^2}{v} g_{\mu\nu}$$

## Counterterm Feynman Rules:

1. Quark propagator with counterterm insertion

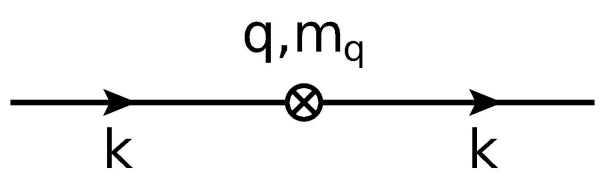

$$: \quad i\left[\delta Z_q\,(\not{k} - m_q) - \delta m_q\right]$$

Here $\delta Z_q$ and $\delta m_q$ are the quark wave function and mass counterterms.

2. $Z$-boson propagator counterterm

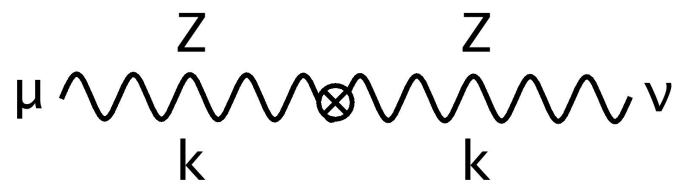

$$: \quad -ig_{\mu\nu}\left[\delta Z_{ZZ}(k^2 - M_Z^2) - \delta M_Z^2\right]$$

3. $Z - \gamma$ propagator mixing counterterm

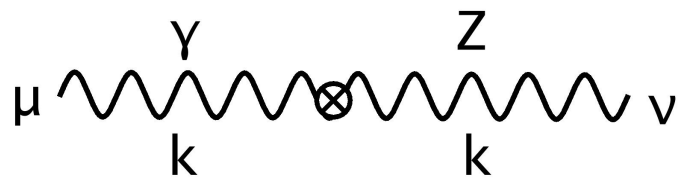

$$: \quad -ig_{\mu\nu}\left[-\frac{1}{2}\delta Z_{\gamma Z}k^2 - \frac{1}{2}\delta Z_{Z\gamma}(k^2 - M_Z^2)\right]$$

4. Quark-photon vertex counterterm

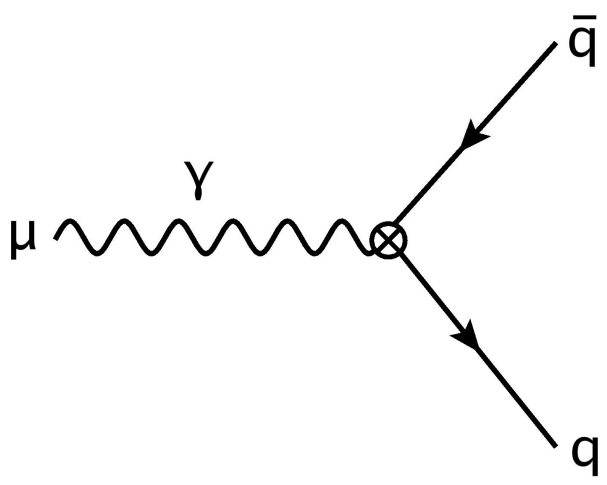

$$: \qquad -ieQ_f\gamma_\mu \, C_{\gamma qq}$$

For pure QCD correction to quark-photon vertex $C_{\gamma qq}$ receives contribution from $\mathcal{O}(\alpha)$ quark wave function counter term, i.e. $C_{\gamma qq} = \delta Z_q$.

5. Quark-Higgs vertex counterterm

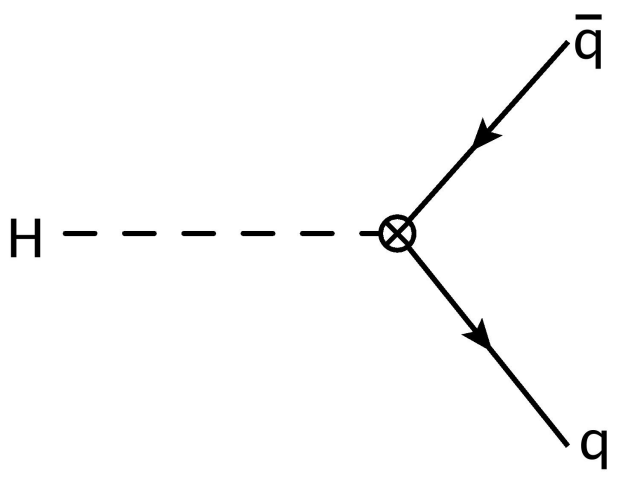

$$: \qquad -i\frac{m_q}{v} \, C_{Hqq}$$

Here the renormalization constant $C_{Hqq}$ which we need at order $\alpha_s$ consists of the quark mass and wave-function counterterms, i.e. $C_{Hqq} = \delta Z_q - \dfrac{\delta m_q}{m_q}$.

6. $HZZ$ vertex counterterm

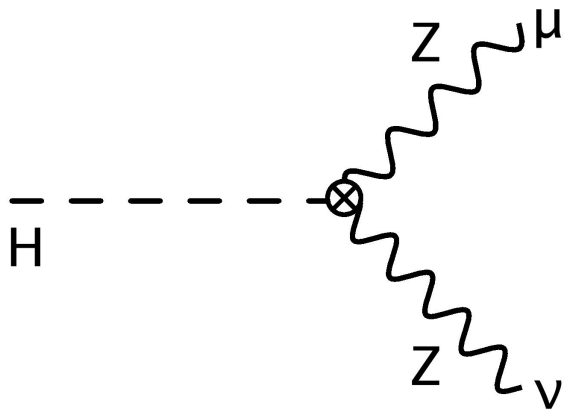

$$: \; -ig_{\mu\nu}\frac{2M_Z^2}{v}\left[\frac{\delta M_Z^2}{M_Z^2} - \frac{\delta v}{v} + \frac{1}{2}\delta Z_H + \delta Z_{ZZ}\right]$$

Here $\dfrac{\delta v}{v}$ is the counterterm for the vacuum expectation value. In $G_\mu$ scheme it has the following form

$$\frac{\delta v}{v} = \frac{\delta M_W^2}{2M_W^2} + \frac{\delta s_W}{s_W} - \delta Z_e. \tag{A.1}$$

7. $HZ\gamma$ vertex counterterm

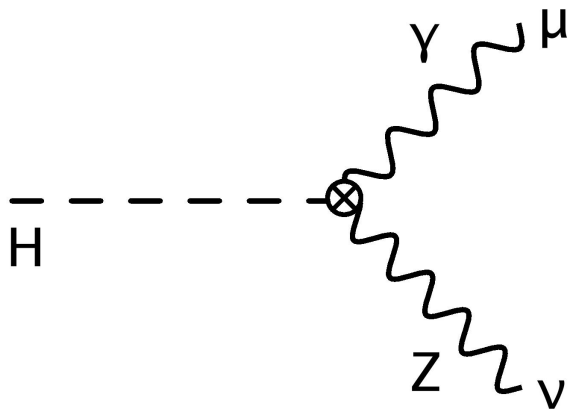

$$: \qquad -ig_{\mu\nu}\frac{M_Z^2}{v}\delta Z_{Z\gamma}$$

The above counterterm vanish at $\mathcal{O}(\alpha\alpha_s)$.

8. $Z\ell\bar{\ell}$ vertex counterterm

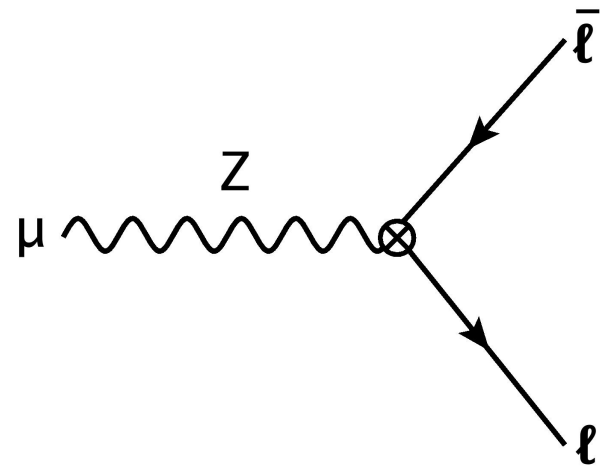

$$: \; ie\gamma_\mu(g_V^\ell \mathbb{1} - g_A^\ell \gamma_5)\left[\delta Z_e + \frac{s_W^2 - c_W^2}{c_W^2}\frac{\delta s_W}{s_W} + \frac{1}{2}\delta Z_{ZZ}\right] - \frac{2s_W}{c_W}Q_\ell \frac{\delta s_W}{s_W} - \frac{Q_\ell}{2}\delta Z_{\gamma Z}$$

9. $Zq\bar{q}$ vertex counterterm

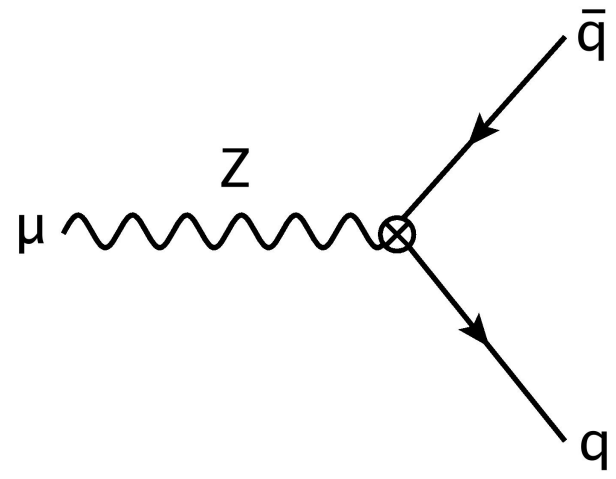

$$: \qquad -ie\gamma_\mu(g_V^f \mathbb{1} - g_A^f \gamma_5)\; C_{Zqq}$$

For QCD correction to $Zq\bar{q}$ vertex $C_{Zqq}$ receives contribution solely from $\mathcal{O}(\alpha)$ quark wave function counter term, i.e. $C_{Zqq} = \delta Z_q$.

# B

# Self-energies

In this appendix, we provide the required expressions for the transverse part of the gauge-boson self-energies $\Sigma_{\mathrm{T}}^{V'V}$ ($VV' = ZZ, \gamma Z, \gamma\gamma, WW$) and the self-energy of the Higgs boson $\Sigma^H$ at $\mathcal{O}(\alpha\alpha_s)$ in the on-shell renormalization scheme following the conventions of references [203, 204, 240]. The results presented are the sum of two-loop gauge (Higgs) boson self-energy, and one-loop counterterm gauge (Higgs) boson self-energy.

## Vector-Boson Self-Energies

We treat all quarks except the top quark as massless. We denote the momentum of the vector-boson propagator as $q$, and use $s = q^2$ for the self-energy expressions. The transverse parts of the vector-boson self-energies can be expressed in terms of scalar functions $\Pi_{\mathrm{T}}^{V,A}$, where the superscript denotes the vector and axial-vector parts of the self-energy, respectively, as follows:

$$\Sigma^{WW}_{\mathrm{T},(\alpha_s\alpha)}(s) = \frac{\alpha_s\alpha}{8\pi^2 s_W^2}\left[2(\Pi_{\mathrm{T}}^V(s,0,0)+\Pi_{\mathrm{T}}^A(s,0,0)) + (\Pi_{\mathrm{T}}^V(s,m_t^2,0)+\Pi_{\mathrm{T}}^A(s,m_t^2,0))\right],$$

$$\begin{aligned}\Sigma^{ZZ}_{\mathrm{T},(\alpha_s\alpha)}(s) = \frac{\alpha_s\alpha}{4\pi^2 s_W^2 c_W^2}\Bigg[&\left(\frac{44}{9}s_W^4 - \frac{14}{3}s_W^2 + \frac{5}{4}\right)\Pi_{\mathrm{T}}^V(s,0,0) + \frac{5}{4}\Pi_{\mathrm{T}}^A(s,0,0)\\ &+\left(\frac{1}{2}-\frac{4}{3}s_W^2\right)^2\Pi_{\mathrm{T}}^V(s,m_t^2,m_t^2) + \frac{1}{4}\Pi_{\mathrm{T}}^A(s,m_t^2,m_t^2)\Bigg],\end{aligned}$$

$$\Sigma^{\gamma\gamma}_{\mathrm{T},(\alpha_s\alpha)}(s) = \frac{\alpha_s\alpha}{\pi^2}\left[\frac{11}{9}\Pi^V_{\mathrm{T}}(s,0,0) + \frac{4}{9}\Pi^V_{\mathrm{T}}(s,m_t^2,m_t^2)\right],$$
$$\Sigma^{\gamma Z}_{\mathrm{T},(\alpha_s\alpha)}(s) = -\frac{\alpha_s\alpha}{2\pi^2 s_W c_W}\left[\left(\frac{7}{6}-\frac{22}{9}s_W^2\right)\Pi^V_{\mathrm{T}}(s,0,0) + \left(\frac{1}{3}-\frac{8}{9}s_W^2\right)\Pi^V_{\mathrm{T}}(s,m_t^2,m_t^2)\right]. \quad \text{(B.1)}$$

In order to write the expressions of $\Pi^{V,A}_{T,L}(s,m_a,m_b)$ in a compact way, we define following variables

$$t_a = -\frac{m_a^2}{s}, \qquad \rho_a = \ln\frac{m_a^2}{\mu^2}$$
$$x_a = \frac{2t_a}{1+t_a+t_b+\sqrt{\lambda}}, \qquad \lambda = 1+2t_a+2t_b+(t_a-t_b)^2. \quad \text{(B.2)}$$

Here $\mu$ denotes the renormalization mass scale. In our case both the masses $m_a$ and $m_b$ must be either the mass of top quark or zero. Therefore the corresponding scalar functions are given as

1. When both the masses are non-zero, and are equal to mass of top quark $m_t$, for example for the case of $Z/\gamma$ self-energy with top-quark loop. In this case the variables defined above reduce to $t_a = t_b = t$, $\rho_a = \rho_b = \rho$, $\lambda = 1+4t$, $x_a = x_b = x = \frac{4t}{(1+\sqrt{1+4t})^2}$. With these variables we have

$$\Pi^V_T(s,m_t^2,m_t^2) = s\Bigg[\frac{1}{2\epsilon} - \rho + \frac{55}{12} - \frac{26}{3}t + \sqrt{1+4t}(1-6t)\ln x - \frac{2}{3}t(4+t)\ln^2 x$$
$$+\frac{2}{3}(4t^2-1)[F(1)+F(x^2)-2F(x)] - \frac{4}{3}(1-2t)\sqrt{1+4t}[G(x^2)-G(x)]\Bigg],$$

$$\Pi^A_T(s,m_t^2,m_t^2) = s\Bigg[-\frac{6t}{\epsilon^2} + (1+24t\rho-22t)\frac{1}{2\epsilon} - (1+12t\rho-22t)\rho + \frac{55}{12}$$
$$-\frac{19}{6}t + 4t^2 - t\pi^2 + (1+12t+4t^2)\sqrt{1+4t}\ln x + \frac{2}{3}t(5+11t+6t^2)\ln^2 x$$
$$-\frac{2}{3}(1+2t)(1+4t)[F(1)+F(x^2)-2F(x)] - \frac{4}{3}(1+4t)^{3/2}[G(x^2)-G(x)]\Bigg],$$
(B.3)

where the functions $F$ and $G$ are given by[1]

$$F(x) = \int_0^x dy\left(\frac{\ln y}{1-y}\right)^2 \ln\frac{x}{y} = 6Li_3(x) - 4Li_2(x)\ln x - \ln^2 x\ln(1-x), \quad \text{(B.4)}$$
$$G(x) = x\frac{\partial F(x)}{\partial x} = \int_0^x dy\left(\frac{\ln y}{1-y}\right)^2 = 2Li_2(x) + 2\ln x\ln(1-x) + \frac{x}{1-x}\ln^2 x. \quad \text{(B.5)}$$

[1]When evaluating $F(x_a x_b)$ and $G(x_a x_b)$, the $\ln(x_a x_b)$ terms should be evaluated as $\ln x_a + \ln x_b$.

2. When one of the quark masses is zero, for example, $W$ boson self-energy with $tb$ quark loop. In this case, $t_b = 0$, $x_b = 0$ , $t_a = t$, $\rho_a = \rho$, $\lambda = (1+t_a)^2$, $x_a = x = \frac{t_a}{1+t_a}$. For this case we have

$$\Pi_T^{V,A}(s, m_t^2, 0) = s\Big[ -\frac{3t}{2\epsilon^2} + \Big(1 - \frac{11}{2}t + 6t\rho\Big)\frac{1}{2\epsilon} - \Big(1 - \frac{11}{2}t + 3t\rho\Big)\rho + \frac{55}{12} - \frac{71}{24}t - \frac{\pi^2}{4}t - \frac{5}{6}t^2 + (1+t)\Big(1 + \frac{3}{2}t - \frac{5}{6}t^2\Big)\ln x + \frac{t}{6}(1-3t)(1+t)\ln^2 x - \frac{1}{3}(2+t)(t-1)G(x) + \frac{1}{3}(t-2)(1+t)^2[F(1) - F(x)]\Big]. \tag{B.6}$$

3. When both the masses are zero, i.e., $m_a = m_b = 0$ for a massless quark loop, we have

$$\Pi_T^{V,A}(s, 0, 0) = s\left[\frac{1}{2\epsilon} + \ln\frac{\mu^2}{-s} + \frac{55}{12} - 4\zeta(3)\right]. \tag{B.7}$$

For $s > 0$, the real part is given as

$$\text{Re}\,\Pi_T^{V,A}(s, 0, 0) = s\left[\frac{1}{2\epsilon} + \ln\frac{\mu^2}{s} + \frac{55}{12} - 4\zeta(3)\right]. \tag{B.8}$$

## Higgs Boson Self-Energy

The expression for the scalar function of Higgs boson self-energy in terms of similar notations used for the gauge-boson self-energy when $m_a = m_b = m_t$ is given by

$$\Sigma^H(s, m_t, m_t) = \frac{\alpha_s}{4\pi^3}\frac{m_t^2}{v^2}\, s\Big[ -\frac{3}{2\epsilon^2}(1+12t) - \frac{1}{\epsilon}\Big(\frac{11}{4} - 3\rho + 42t - 36t\rho\Big) + \frac{11}{2}\rho - 3\rho^2 + 84t\rho + \frac{3}{8} - 73t - \frac{\pi^2}{4}(1+12t) + \frac{3}{2}\sqrt{1+4t}(14t+3)\ln x + \Big(\frac{3}{2} + 14t + 29t^2\Big)\ln^2 x - 36t\rho^2 - (1+2t)(1+4t)[F(1) + F(x^2) - 2F(x)] - 2(1+4t)^{3/2}[G(x^2) - G(x)].\Big] \tag{B.9}$$

Where $s = q^2$ is the squared four-momentum transfer and $v$ is the vacuum expectation value.

C

# The Tadpole Integral

In this appendix, I present the evaluation of one-loop tadpole integral $J_{20}(\epsilon, x)$, defined in equation (3.25) as

$$J_{20}(\epsilon, x) = \epsilon I_{20}(\epsilon, x), \tag{C.1}$$

where $I_{20}$, using equation (3.12), is given as

$$I_{20}(d, x) = e^{\epsilon\gamma_E}(m^2)^{2-d/2} \int \frac{d^d k}{i\pi^{d/2}} \frac{1}{(-k^2 + m^2)^2} \tag{C.2}$$

Now, performing a Wick rotation of the above integral to the Euclidean space and performing the angular integration using spherical coordinates, we get

$$I_{20}(d, x) = e^{\epsilon\gamma_E}(m^2)^{2-d/2} \int_0^\infty \frac{d^d K}{\pi^{d/2}} \frac{1}{(K^2 + m^2)^2}, \tag{C.3}$$

$$= e^{\epsilon\gamma_E}(m^2)^{2-d/2} \frac{2\pi^{d/2}}{\Gamma\left(\frac{d}{2}\right)} \int_0^\infty \frac{dK}{\pi^{d/2}} \frac{K^{d-1}}{(K^2 + m^2)^2}. \tag{C.4}$$

In the next step, we will manipulate the integrand to put the integral in the form of a beta function to perform the integration over $K$.

$$I_{20}(d,x) = e^{\epsilon\gamma_E}(m^2)^{2-d/2}\frac{2}{\Gamma\left(\frac{d}{2}\right)}\int_0^\infty \frac{1}{2}dK^2\frac{(K^2)^{d/2-1}}{(K^2+m^2)^2}. \tag{C.5}$$

Taking the mass dependence outside the integral and putting $K^2/m^2 = t$, we get

$$I_{20}(d,x) = e^{\epsilon\gamma_E}(m^2)^{2-d/2}\frac{1}{\Gamma\left(\frac{d}{2}\right)}(m^2)^{d/2-2}\int_0^\infty dt\, t^{d/2-1}(t+1)^{-2}. \tag{C.6}$$

Using definition of the beta function given in equation (2.23), we obtain

$$I_{20}(d,x) = \frac{e^{\epsilon\gamma_E}}{\Gamma\left(\frac{d}{2}\right)}\frac{\Gamma\left(\frac{d}{2}\right)\Gamma\left(2-\frac{d}{2}\right)}{\Gamma(2)} = e^{\epsilon\gamma_E}\Gamma\left(2-\frac{d}{2}\right). \tag{C.7}$$

Where in the last step we have used $\Gamma(2) = 1$. Taking $d = 4-2\epsilon$, we get

$$I_{20}(d,x) = e^{\epsilon\gamma_E}\Gamma(\epsilon). \tag{C.8}$$

Using this result in equation (C.1) gives

$$J_{20} = \epsilon e^{\epsilon\gamma_E}\Gamma(\epsilon). \tag{C.9}$$

# D

# Overview of Computational Tools

1. **LiteRed:** **`LiteRed`** [97] is a **`Mathematica`** based Integration-by-Parts (IBP) reduction package, that is heavily used in multi-loop calculations. It performs the reduction by finding symbolic reduction rules through heuristic methods. Apart from generating symbolic IBP relations, **`LiteRed`** is also employed for finding symmetry relations, deriving differential equation systems and obtaining dimensional-shift relations for scalar Feynman integrals.

2. **FIRE:** **`FIRE`** [92] is a publicly available algorithm whose name stands for Feynman Integral REduction. It is widely used for performing IBP reduction of Feynman integrals to an independent set of master integrals. **`FIRE`** basically utilize the Laporta algorithm [91], the s-bases algorithm [265], and various other strategies to systematically solve the IBP relations. The latest version of FIRE is primarily written in C++ with a smaller part in **`Mathematica`**, making the reduction process much faster. This program uses **`LiteRed`** to identify additional symmetry rules while performing the reduction.

3. **Mint:** **`Mint`** [207] is an automated **`Mathematica`** package often used along with **`LiteRed`** to find the number of master integrals for a given integral family.

4. **KIRA:** **`KIRA`** [95, 266] is a powerful C++ program used for the reduction of scalar Feynman integrals based on an advanced implementation of the Laporta algorithm. It utilizes finite field method [267] with the help of **FireFly** [268, 269] to efficiently solve systems of equations derived from IBP and Lorentz Invariance identities. **KIRA** supports parallelization and user-defined systems of equations, making it both efficient and flexible for reducing amplitudes, solving linear system of equations, and finding relations among Feynman integrals in complicated multi-loop calculations.

5. **Reduze:** **Reduze** [96] is an open-source C++ program which employs a variant of Laporta algorithm to reduce Feynman integrals to master integrals. It uses the **GiNAC** [270] or, **Fermat** libraries for simplifying algebraic coefficients in system of IBP equations. **Reduze** supports parallel computation and can handle multi-scale problems. It provides the flexibility to deal with multiple integral families within a single calculation and can provide the relations among integrals from different sectors of the same or different integral families. Additionally, it can also be used for generating system of differential equations for Feynman integrals, and for shifting loop momenta of **QGRAF** [200] generated Feynman diagrams to map sectors of integral families.

6. **FIESTA:** **FIESTA** (Feynman Integral Evaluation by a Sector decompositiOn Approach) [100] is a software package written in **Mathematica** and C++ designed for the numerical evaluation of the Feynman integrals using sector decomposition. Given the set of loop momenta, propagators, and indices along with kinematic substitutions, the program returns the epsilon-expansion of the corresponding integral. Additionally, **FIESTA** supports parallel computations and the asymptotic expansion of the Feynman integrals.

7. **pySecDec:** **pySecDec** [101] is a python package primarily used for numerical evaluation of dimensionally regularized multi-loop parametric Feynman integrals. It employs sector-decomposition approach for handling divergences. Apart from efficiently evaluating individual integrals the latest version of **pySecDec** [271] also features improvements for automating the evaluation of multi-loop amplitudes and performing asymptotic expansions of Feynman integrals.

The evaluation of Feynman integrals using the method of differential equation becomes simpler if the dependence on dimensional regularization parameter "epsilon" is factorized. The epsilon-factorized form, also called the canonical form of the differential equation system can be obtained by choosing an appropriate integral basis. Several automatic tools are publicly available to determine the same. Here we discuss a few of them:

8. **Fuchsia:** **`Fuchsia`** [127] is a public implementation of the Lee algorithm [272]. It reduces the given systems of one-variable differential equations of the Feynman integrals into the epsilon-factorized form. By leveraging the properties of Feynman integrals, it guarantees that solutions posses only regular singularities.

9. **Canonica:** **`Canonica`** is a **`Mathematica`** package [128] based on the algorithm described in [273]. It is the first automatic tool for obtaining the $\epsilon$-factorized form of differential equations involving multiple kinematic scales. The package takes the matrix $A$ and kinematic invariants as input and provides the corresponding canonical form for $A$, along with the transformation matrix for the basis of master integrals. A limitation of this package is that it only works for cases where rational transformations exist to convert the system of differential equations into the epsilon-factorized form.

10. **Libra:** **`Libra`** [129] is an automated **`Mathematica`** tool used to obtain the epsilon factorized form of the first-order differential equation systems. Unlike Canonica's black-box approach, Libra performs a semi-manual reduction of these systems. One important feature of **`Libra`** is its capability to fix boundary conditions for the given differential equation system. Similar to Canonica, it also works well for the multivariant problems.

11. **INITIAL:** A different approach for transforming systems of differential equations to canonical form, developed by Hoschele [274], is generalized in the **`Mathematica`** package **`INITIAL`** (an INitial InTegral ALgorithm)) [130]. This algorithm requires a single uniform transcendental weight integral from the top sector as input and then algorithmically obtains the full basis of canonical integrals, including the provided UT integral, if it exists. The algorithm works well for both univariate and multivariate cases. In the latest release [275], the applicability of this algorithm has been extended from polylogarithmic cases to those involving elliptic functions.

In multi-loop problems with multiple mass scales, directly evaluating Feynman integrals is a highly non-trivial task. Therefore, in such cases, one often opts for their study in various asymptotic limits. For the asymptotic expansion of these integrals, it is necessary to identify relevant regions in the integration space. There are some automated packages available for the identification of these regions, which are discussed below:

12. **asy.m:** A publicly available **`Mathematica`** algorithm, **`asy.m`** [188], automates the task of finding the relevant regions for the asymptotic expansion of Feynman integrals using alpha representation. The algorithm employs a geometric approach based on determining the convex hull of the integrand exponents. This code works well for Sudakov-type limits but initially lacked detection of potential and Glauber regions. The updated version, **asy2.m** [189], resolves these issues and identifies all relevant regions for a given integral, including the Glauber and potential regions.

13. **ASPIRE:** This algorithm introduces a novel method to identify the relevant regions for a given Feynman integral in the Method of Regions framework by studying its singularity structure and the associated Landau equations in the Alpha-representation. **ASPIRE** [190] employs Power Geometry techniques to expedite the process of finding the relevant regions. The algorithm is effective in identifying the potential and Glauber regions.

Some additional open-source packages used in multi-loop calculations are discussed below:

14. **QGRAF: QGRAF** is a FORTRAN-based program [200], developed for the automatic generation of Feynman diagrams. It generates the symbolic representation of all possible diagrams for a specified process at a given loop order within a defined model. **QGRAF** supports various gauge theory models and provides output that serve as the starting point for writing the amplitude for given process. Its output can be further processed by other software packages for algebraic manipulation and numerical evaluation.

15. **RationalizeRoots: RationalizeRoots** [141] is a **Mathematica** and **Maple** software package used to rationalize square roots through appropriate variable transformations, which appear during the analytic computation of multi-loop Feynman integrals. The implementation is based on an algorithm that generalizes the method presented in [140]. For a given set of square roots, the software finds a suitable variable transformations to rationalize all the square roots simultaneously through the parametrization of algebraic hypersurfaces associated to these roots using families of lines.

16. **DiffExp: DiffExp** [234] is a publicly available **Mathematica** package used for solving Feynman integral in terms of one-dimensional series expansions along set of line segments, given the differential equation systems for their integral families. These expansions are truncated at a specified order in the line parameter. The implementation of this package is based on the series expansion strategies discussed in [233] to solve Feynman integrals. For a given system of differential equations and suitable boundary conditions of an integral family in dimensional regularization, **DiffExp** computes highly accurate numerical results at a given point in phase-space. It is particularly effective for integral families where the prefactors in the basis are rational functions of kinematic invariants or involve square roots at most.

17. **GiNaC:** **GiNaC** is an iterated and recursive acronym for **GiNaC** is Not a CAS, where CAS stands for Computer Algebra System [276]. It is an open-source C++ library, allowing symbolic manipulations, and along with the CLN library, it supports numerical computations with arbitrary precision. It is designed to efficiently handle the multivariate polynomials, Clifford and color algebras and special functions needed for complex computations in quantum field theory and is thus extensively used for numerical calculation of Feynman integrals to very high precision.

18. **MadGraph:** **MadGraph** [254] is a Monte Carlo event generator that provides essential ingredients for studying the Standard Model and Beyond the Standard Model phenomenology. Specifically, it is used for automated cross-section computations, event generation, and their matching with event generators for experimental analysis, as well as matrix elements calculations. It supports a wide range of models, including the Standard Model, Supersymmetric (SUSY), extra dimensions (e.g., ADD, RS models), and dark matter models. It integrates seamlessly with Pythia and Delphes for parton showering, hadronization, and detector-level simulations, enabling final event simulations. The latest version, MadGraph5_aMC@NLO, enhances next-to-leading order (NLO) computations for a broader range of processes and offers improved integration with loop-induced interactions.